\documentclass[apl, graphicx, twocolumn]{revtex4-1}
\usepackage{xcolor}
\usepackage{graphicx}

\draft % marks overfull lines with a black rule on the right

\begin{document}

%\preprint{}

\title{Perspective: Phonon polaritons for infrared optoelectronics} %Title of paper



%NOTE - We can discus the author list further.

\author{Christopher R. Gubbin}
%\thanks{}
\affiliation{School of Physics and Astronomy, University of Southampton, Southampton, SO17 1BJ, United Kingdom}

\author{Simone De Liberato}
\email{s.de-liberato@soton.ac.uk}
%\thanks{}
\affiliation{School of Physics and Astronomy, University of Southampton, Southampton, SO17 1BJ, United Kingdom}

\author{Thomas G. Folland}
\email[]{thomas-folland@uiowa.edu}
%\thanks{}
\affiliation{Department of Physics and Astronomy, The University of Iowa}


\date{\today}

\begin{abstract}
In recent years there has been significant fundamental research into phonon polaritons, owing to their ability to compress light to extremely small dimensions, low-losses, and ability to support anisotropic propagation. In this perspective, after briefly reviewing the present state of mid-infrared optoelectronics, we will assess the potential of phonon-polariton based nanophotonics for infrared  (3-100$\mathrm{\mu}$m) light sources, detectors and modulators. These will operate in the Reststrahlen region where conventional semiconductor light sources become ineffective. Drawing on the results from the past few years, we will sketch some promising paths to create such devices and we will evaluate their practical advantages and disadvantages when compared to other approaches to infrared optoelectronics.
\end{abstract}
https://www.overleaf.com/project/60353a652cf72a3371a75a62
\pacs{}% insert suggested PACS numbers in braces on next line

\maketitle %\maketitle must follow title, authors, abstract and \pacs

\section{Introduction}
%\label{}
Achieving optimal performance from an optoelectronic device depends on controlling both light and matter degrees of freedom. Key is the strength of light-matter interaction which can be enhanced through careful tailoring of the electromagnetic environment via the Purcell effect \cite{Purcell1946,Ballarini2019}. In recent decades advances in optical engineering have allowed for control of light on length scales below the diffraction limit,  permitting tailoring of an optoelectronic devices photonic environment and increasing efficiencies \cite{Tan2018,Baranov2019,Kim2021,Stewart2021}. Nanophotonic concepts are of particular interest in the mid-to-far infrared ($3-100 \mu m$), where technologies are less mature, operate below theoretical efficiency limits \cite{Rogalski2012}, and the increasing mismatch between photon wavelength and matter degrees of freedom results in naturally weaker light-matter interaction.
Light can be confined deep below the diffraction limit by hybridisation with charge carriers. In the visible and near-infrared spectral regions this can be achieved utilising metallic elements to form hybrid surface plasmon polariton modes,  morphological excitations whose properties are characterised by the system's geometry and the plasma frequency of the free electron gas \cite{Maier2007}. The plasma frequency can occur from the UV (4eV, 0.3$\mathrm{\mu}$m for noble metals) through to the infrared and terahertz part of the electromagnetic spectrum (0.01eV, 120$\mathrm{\mu}$m for heavily doped semiconductors \cite{Naik2013,Taliercio2019}). As both noble metals and doped semiconductors are already ubiquitous in optoelectronic devices, surface plasmon polaritons became a natural nanophotonic platform. Since the surface plasmon laser was demonstrated in the infrared two decades ago \cite{Tredicucci2000}, and in the visible a decade ago \cite{Azzam2020}, there have been significant developments in this field. Such concepts were integral to the creation of the first THz quantum cascade lasers \cite{Kohler2002} and have been demonstrated in both light sources \cite{Sirtori2013,Briggs2020,Nordin2020} and detectors \cite{Tan2018,Palaferri2018,Goldflam2016}. Plasmons though, have intrinsic limitations. The electrons supporting the mode have relatively rapid scattering rates, which lead to modes with broad linewidths and short lifetimes \cite{Khurgin2015}. The presence of metals or free carriers can also be counterproductive to device operation, for example through diffusion of metals into active device layers degrading device performance \cite{Piotrowska1983} and enhanced non-radiative damping through electronic interactions \cite{Khurgin2014}.\\

Recently there has been growing interest in alternative mechanisms of light confinement, utilising the optical phonon modes of polar dielectric crystals, in modes termed surface phonon polaritons (SPhPs). 
These modes can exist in the Reststrahlen region of polar dielectrics, between their longitudinal optical (LO) and transverse optical (TO) frequencies. They thus naturally cover the mid-infrared (0.2eV in calcite \cite{Breslin2021} and hexagonal boron nitride \cite{Caldwell2014b, Dai2014, Dai2015, Li2015}) and terahertz (10meV in InSb  \cite{Hartstein1975}) regions. Although they lack some of the frequency tuneability of plasmons, the wide variety of available polar materials including SiC \cite{Caldwell2013, Chen2014, Wang2013}, GaN \cite{Feng2015}, SiO$_2$  \cite{Hafeli2011}, Al$_2$O$_3$ \cite{Berte2018}, GaAs \cite{Vassant2010}, among others \cite{Caldwell2015, Caldwell2016, Ratchford2018} (See Fig.~\ref{fig:Overview}) and their significantly reduced optical losses make them an strong alternative to plasmonic systems. The wavelength of a SPhP mode within the Restrahlen band can be set through the creation of patterned resonators into a crystal of the above materials \cite{Caldwell2013}. Additional dynamic wavelength tuneability can also be engineered into SPhP devices through the injection of free carriers \cite{Ratchford2018,Dunkelberger2020}, intersubband transitions \cite{Vassant2012}, or alternatively through the use of phase change materials \cite{Li2016,Folland2018,Chaudhary2019}. Both doping and phase change based tuning can be engineered through either optical or electrical stimulus to the SPhP device, opening up the potential for active devices. Many polar dielectrics are already utilised in mature optoelectronic devices \cite{Caldwell2015, Dubrovkin2020} and interactions of charge with optical phonons already plays a key part in intersubband based optoelectronics \cite{Rattunde2006,DeLiberato2012,Manceau2018}. These facts, together with recent high-profile research results from around the world \cite{Li2016,Tamagnone2020,Gubbin2019,Dunkelberger2020,Dai2014,Ma2018,Hu2020,Wu2020, Greffet2002, Hillenbrand2002,Feng2015b}, highlight phonon polaritons as a uniquely promising platform for infrared optoelectronics. 

This perspective will assess opportunities and challenges presented by SPhP-based optoelectronic devices. To begin, we outline the current state-of-the-art in infrared optoelectronics. We will then proceed to examine opportunities for light sources, light detectors and future phononic devices utilising the unique properties of SPhPs. Finally we outline emerging materials challenges. We hope to convey to the reader our excitement for this blossoming field and our belief that it is now the time to start putting together the proof-of-concept works and the know-how developed over the past decade to empower a novel generation of pioneering infrared optoelectronic devices. 

\subsection{Infrared Optoelectronics}
\begin{figure*}
    \centering
    \includegraphics[width=17cm]{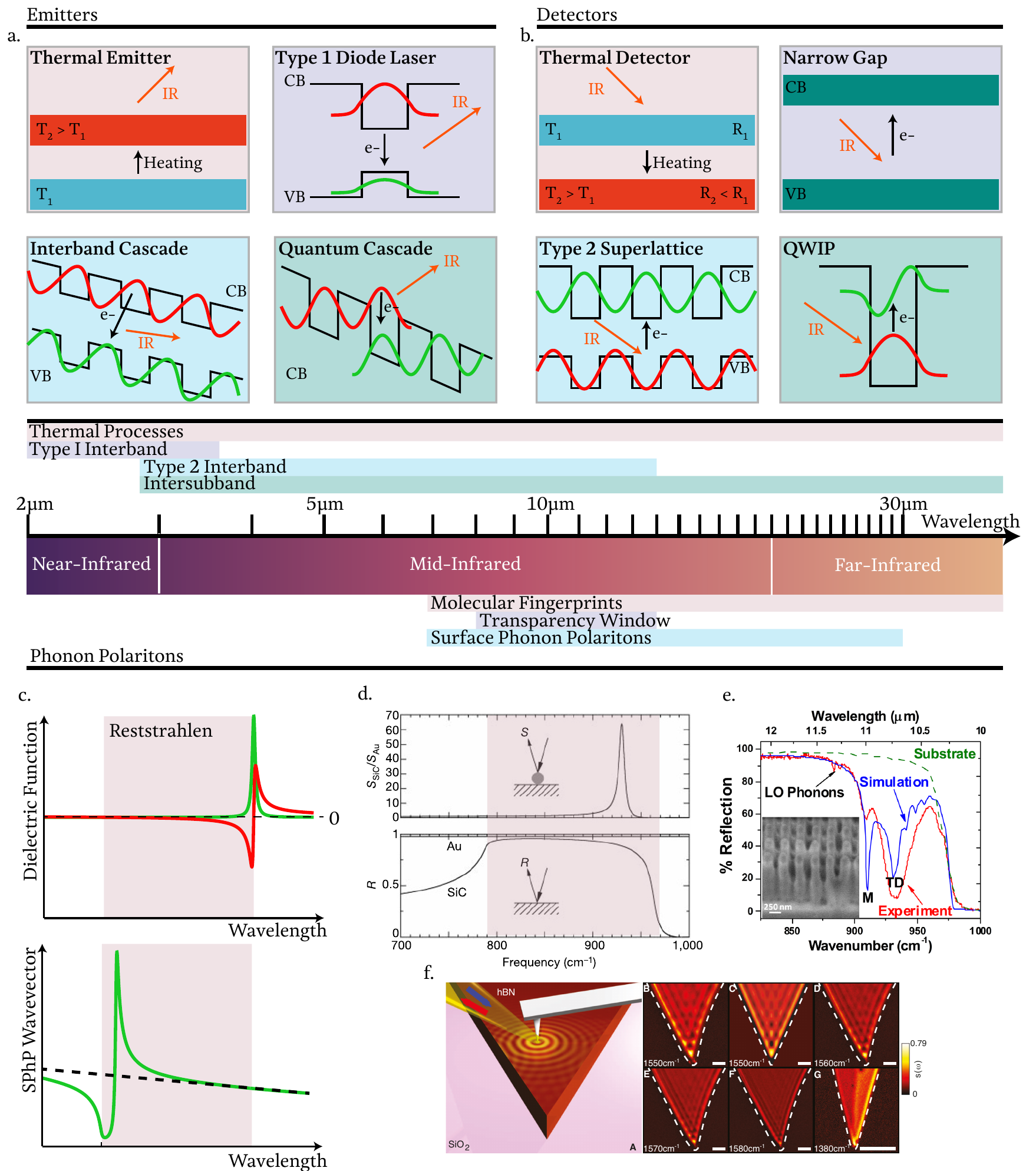}
    \caption{A comparison of mid-infrared optoelectronic a. emitter and b. detector technologies. The mid-infrared spectral region (centre) contains many technologically interesting regions, particularly the atmospheric transparency window between 8 and 14$\mathrm{\mu}$m. Upper panels detail common mid-infrared optoelectronic emitters and detectors and the spectral regions they typically operate in. In the lower panel the operation regions of common polar dielectric materials supporting surface phonon polaritons are shown. c. In the Reststrahlen region $\lambda_{\mathrm{LO}} < \lambda < \lambda_{\mathrm{TO}}$ these materials have negative dielectric function, are highly reflective and support bound excitations. d. Lower: reflectance from a 3C-SiC surface. Upper: Scattering from a metallic tip above a 3C-SiC surface. Reproduced from Ref. \cite{Hillenbrand2002}. e. Reflectance of a 3C-SiC nanopillar array. Reproduced from Ref. \cite{Caldwell2013}. f. Near field imaging of propagating SPhPs on a hBN flake, reproduced from Ref. \cite{Dai2014}.}
    \label{fig:Overview}
\end{figure*}

Infrared light plays a critical role in passive thermal imaging, exploiting black body radiation emitted by all objects. The infrared region also contains spectroscopic fingerprints for different chemical species relevant for healthcare, environmental sensing, and security. For example ammonia, NH$_3$, has spectral lines in the 10-12$\mu$m range and is used as a biomarker for predicting kidney diseases \cite{Chen2020}. A multitude of other environmental pollutants possess absorption bands at even longer wavelengths, including trichloroethylene and other volatile organics \cite{Lu2016}.
Furthermore, the reduced atmospheric scattering cross section for long wavelength infrared light and transparency in the 3-5 and 8-12 $\mu m$ mid-infrared windows, could be relevant for free-space laser ranging and communication \cite{Sobrino2016}. However, infrared light is challenging to generate and detect due to the low energy of photons and high absorption associated with many materials in the spectral region \cite{Folland2019}. Present-day optoelectronics utilises four main approaches to generate or detect mid-infrared light: thermal mechanisms, excitation of a narrow band gap semiconductor, excitation of a narrow band gap super lattice, or through intersubband transitions, as illustrated in the upper panel Fig.~\ref{fig:Overview}. The operation ranges of devices based on these approaches is also illustrated in Fig.~\ref{fig:Overview}. In this section we will discuss briefly the merits and drawbacks of each of these approaches.

When considering different mechanisms for light emission and detection it is worth highlighting specific figures of merit with respect to each mechanism. For emitters perhaps the most significant figure of merit is  the wall-plug efficiency of the device, which is simply the ratio of optical power output to electrical power input. This is often more relevant for applications where battery operation or portability is critical. A high wall-plug efficiency ensures good conversion from electrical to photonic energy, and reduces heating, improving reliability \cite{Day2013}. For detectors, the primary limitations are responsivity as well as the level of current noise in the device at a given speed. The specific detectivity ($D^*$) of the device, acts as a combined figure of merit encompassing response speed, responsivity (voltage or current output per unit optical power), and noise, with higher values indicating better performance. The noise signal can generally be split into noise due to thermal excitation of carriers, the background flux of photons, and defect-induced events in the material. The $D^*$ is defined in terms of responsivity ($R$), the area of the device ($A$), the electrical response frequency $\Delta f$, and the noise signal $I$ in voltage or current as
\begin{equation}
    D^* = \frac{R \sqrt{A \Delta f}}{I}\label{eq:Dstar}.
\end{equation}
The specific detectivity $D^*$ is a useful metric for comparing between detectors, however it is worth noting that other metrics apply - such as the background limited operating temperature, and maximum response speed depending on application. A more complete discussion of different metrics can be found in comprehensive reviews such as \cite{Rogalski2012,Tan2018}. Infrared detectors can also be operated as focal plane arrays, where multiple small detector elements (typically only $10 \mu$m across) can be used to form an infrared camera. This is a key application space, as it forms the basis for all infrared thermal imaging and hyperspectral imaging technology.

%moved 
\subsubsection{Thermal Emitters}
\begin{figure*}
\begin{center}
	\includegraphics[width=14cm]{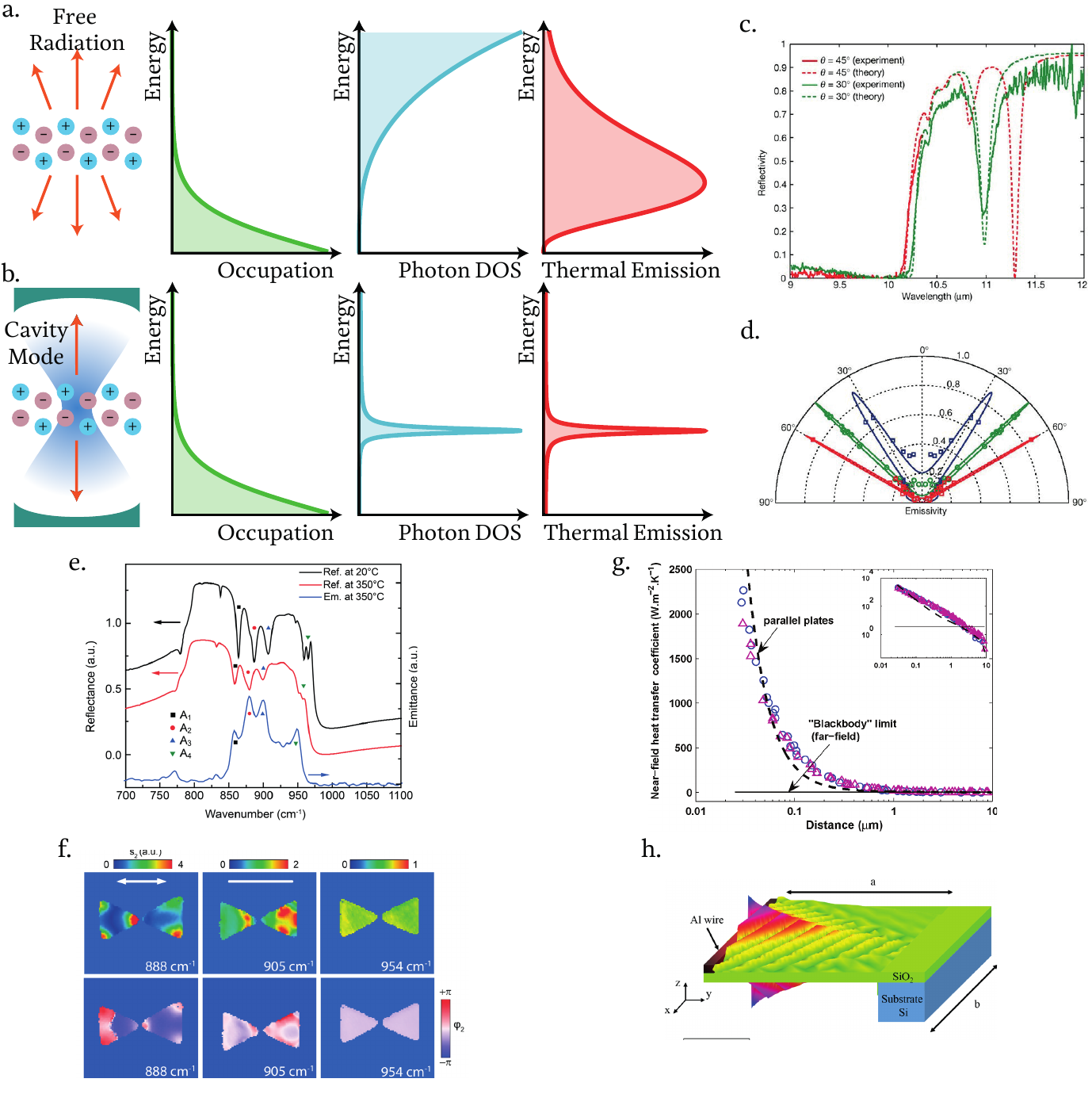}
    \caption{\label{fig:Figthermal} a. An illustration of thermal emission. The oscillations of charged media radiate photons. These are emitted and re-absorbed according to the environment density of photonic states, so the system remains at thermal equilibrium. In a non-resonant environment (upper panel) the photonic density of states increases monotonously, yielding a blackbody spectrum. b. In a resonant one (lower panel) the photonic density of states can peak sharply, leading to narrowband thermal emission. c. Reflection from a lameller 3C-SiC grating \cite{Greffet2002}. d. Directional thermal emission from the same 3C-SiC grating \cite{Greffet2002}. e. Comparison of reflectivity and thermal emission measured for arrays of SiC bowtie nano-antennas \cite{Wang2017} f. Near field at the corresponding mode frequencies \cite{Wang2017} g. Near field heat transfer coefficient between a polar dielectric sphere and a planar surface as a function of their spacing \cite{Shen2009} h. Illustration of SPhP heat transfer on a nanoscale SiO$_2$ film, reproduced from Tranchant {\it et al.} \cite{Tranchant2019}}
\end{center}
\end{figure*}
The simplest method of generating mid-infrared light is to exploit the natural thermal motion of charged particles, illustrated schematically in the upper panel of Fig.~\ref{fig:Overview}. Thermal emitters are useful for their simplicity, and their wide range of operation. Fluctuating currents generated by moving charges radiate electromagnetic energy to free space, described by the spectral radiance $I(K,\theta,T)$, which is emitted radiant power per unit frequency per unit surface per solid angle. The radiance is frequency ($f$) dependant, and is most usefully expressed as a function of spectroscopic wavevector $K = f / c$, temperature $T$, and azimuthal angle $\theta$, by Planck's law
\begin{equation}
	I\left(K, \theta, T\right) = 2 h c  \frac{K^3}{e^{h c K / k_{\mathrm{B}} T} - 1} \epsilon\left(K, \theta, T\right), \label{eq:Planck}
\end{equation}
in which $k_{\mathrm{B}}$ is the Boltzmann constant, $h$ is Planck's constant, $c$ is the speed of light and $\epsilon\left(K, \theta, T\right)$ is the emitter's emissivity. This relation can be derived considering the emitter to be in thermal equilibrium with the photon bath in the environment, as illustrated in Fig.~\ref{fig:Figthermal}a. When this photon bath consists of free photons, as shown in the upper panel, it's density of states monotonously grows with energy while it's occupancy falls. The spectral radiance in Eq.~\ref{eq:Planck}, which is the product of these factors, then has the shape of an ideal blackbody whose emissivity is equal to unity. In an ideal blackbody emission occurs over a broad wavelength range, such emitters are often utilised as broad-spectrum thermal sources in free-space optics, for example as sources in FTIR systems. These are typically glowbar type hot-filament sources, operating similarly to a traditional lightbulb. \\
For many applications conventional incoherent broadband sources are inappropriate, particularly those requiring directional emission or efficient light generation in a specific spectral window. Narrowband thermal emission can be achieved by tailoring the surface emissivity in Eq.~\ref{eq:Planck}, which relates through Kirchoff's law to the surface absorption $\mathcal{A}$ . This means thermal emittance can be controlled by altering the surface absorption of nanophotonic structures, the process is illustrated by the lower panel of Fig.~\ref{fig:Figthermal}a. By creating structures with a sharply peaked photonic density of states, it is possible to funnel thermal radiation into narrow spectral regions.\\
Large area surface patterning can be utilised to control $\epsilon$ \cite{Baranov2019}. Disordered patterns can enhance broadband emittance by increasing photon absorption across a wide spectral band \cite{Vorobyev2009}. Alternatively highly ordered patterns can be utilised to generate directional, narrowband emission. The simplest scheme consists of a planar 1D grating, fabricated on negative-dielectric substrate \cite{Greffet2002, Biener2008, Mason2011}. In these systems the grating periodicity folds the dispersion of the surface polariton supported by the substrate inside the light line, leading to resonant peaks in the emissivity. Furthermore the long coherence length of the emitting mode permits thermal radiation generated at different points to interfere, yielding lobes in the angular emission spectrum rather than the quasi-Lambertian expected for a blackbody. This concept can be further extended to control the angular spread of light utilising photonic crystals in two or more dimensions \cite{Pralle2002, Tsai2006, Celanovic2008, Lin2003}. The performance of photonic crystal emitters can be greatly improved on by utilising metamaterials, which have been proposed as perfect absorbers \cite{Landy2008, Liu2010}. In these systems the emissivity can approach the blackbody limit $\epsilon \approx 1$ over a wide, user-controlled spectral range \cite{Liu2011, Wu2012} and angular distribution \cite{Jiang2011}. A simpler approach is to exploit arrays of nanoscale cavities or antennas, whose well-defined mode spectrum contributes to the emissivity \cite{Ikeda2008, Puscasu2008}. This approach has even been shown to work down to the single antenna level \cite{Schuller2008}. Recently dynamic manipulation of thermal emission has been demonstrated by electrical modulation \cite{Brar2014,Wojszvzyk2021} or the use of photochromic materials \cite{Kats2013}.\\
Thermal emitters have excellent prospects in areas such as passive cooling \cite{Raman2014} but are ultimately limited by the spontaneous emission process through which they are powered. As emission is not stimulated, temporal coherence and lasing cannot be achieved, and radiance can only be increased by higher working temperature. This in turns leads to parasitic divergence in the short-wavelength spectral radiance, strongly limiting device efficiency in the target spectral window. Regardless, increasing the operating temperature is not always desirable without recourse to high-temperature materials \cite{Guler2014}.

\subsubsection{Semiconductor Emitters} 
Semiconductor emitters can surpass thermal emitters as they are not restricted by Planck's law. They can therefore emit with higher power per unit solid angle (brightness), making them both easier to use and in principle more efficient. However, the design of mid-infrared semiconductor emitters is challenging, particularly because of the narrow bandgaps required to facilitate mid-infrared transitions. Fig.~\ref{fig:Overview} outlines the spectral regions where different emission schemes operate across the mid-infrared region. At the short end of the mid-infrared spectrum narrow band-gap antimonide (III-V-Sb) structures are often employed in type-I structures where electrons and holes are confined within the same layer. The active region of such a device consists of multiple quantum wells sandwiched into a $p-n$ heterojunction. Electrons and holes are injected into the system, which recombine producing photons. For wavelengths greater than $2\mu$m the wall-plug efficiency of a diode laser drops dramatically while the threshold current density increases, primarily due to an increase in Auger scattering. A host of other issues compound to decrease device performances when increasing the wavelength, particularly a decrease in overlap between the optical mode and active region \cite{Rattunde2006}, difficulties in balancing strain in the structure with weakening confinement of holes \cite{Huang1995}, and more generally a correlation between smaller bandgaps and fragile materials. To the best of the authors knowledge, the longest wavelengths achieved by type-I diode lasers have been demonstrated at 3.8$\mu$m for pulsed operation using AlGaInAsSb barriers \cite{Nash2009}.\\
Rather than pushing against the limitations of natural materials it is possible to increase the operating wavelength of sources by designing a user-defined bandgap through quantum confinement of carriers. When electrons and holes are confined in a nanoscale layer the conduction and valence bands split into multiple quasi-parallel subbands with quantised momentum perpendicular to the layer. By confining electrons and holes on separate nanolayers in a type-II configuration energy transitions below the bandgap of the constituent materials can be engineered, extending the operating region of a given material system. Wall-plug efficiencies can be improved further by reducing the current density in the system at threshold, which in turn reduces parasitic Auger recombination. This can be achieved by wiring the multiple quantum wells comprising the device active region in series, so current passes through each in an interband cascade as proposed in 1995 by Yang \cite{Yang1995}. Compared to a diode laser this reduces the threshold current by a factor of the number of quantum wells, which can allow for operation beyond $6\mu$m \cite{Dallner2015}. Type-2 superlattices have been used to create interband cascade lasers in the mid infrared \cite{Vurgaftman2015}, but often suffer from low efficiency when operated at longer wavelengths. Type-2 superlattices also have the advantage that they can be operated to create LEDs with brightness higher than that of a blackbody \cite{Montealegre2021}. Generation of light in such an interband cascade system is illustrated in Fig.~\ref{fig:Overview}.
\\
Carrier recombination is eliminated entirely in a quantum cascade laser (QCL), first demonstrated by Faist et al. in 1995 \cite{Faist1995}. These are unipolar devices powered instead by intersubband electronic transitions, which occur between different subbands within the conduction band. A sketch of a single period of the active region of a QCL is shown in Fig.~\ref{fig:Overview}. In these devices each period consists of a transition region, where the lasing intersubband transition occurs, and an injection region in which electrons thermalize into the upper level of the subsequent transition region.\\
The lasing wavelength in a QCL is determined by the nanoscale structure of the transition region, which determines the energies of the electronic subbands. This can be engineered for emission across almost the entire mid-infrared spectral region, limited on the short-wavelength end by the conduction band offset and initially on the long-wavelength end in the neighbourhood of the optical phonon resonances of the material system. In this Reststrahlen region, between the longitudinal and transverse optical phonon frequencies (lower panel of Fig.~\ref{fig:Overview}), materials become highly reflective, preventing device operation. Devices can be operated below these energies in the THz spectral region, using appropriately modified designs  \cite{Williams2007,Tan2020b}.\\
 Room temperature, continuous wave outputs of order 1W with high wall-plug efficiencies has been demonstrated at $4.9\mathrm{\mu}$m \cite{Bai2011}. Longer wavelength THz  emission was first demonstrated in 2002 \cite{Kohler2002} but it still requires cryogenic operation \cite{Wade2009}. QCL are highly tuneable as the active region can be tailored to emit at multiple wavelengths, even in a single device by use of external cavities \cite{Hugi2009, Hugi2010}. \\

They do however have some important drawbacks.  Although Auger scattering is removed it is replaced by rapid LO phonon scattering. This gives intersubband optical transitions non-radiative lifetimes of the order of picoseconds, resulting in a larger threshold current and reducing wall-plug efficiency \cite{Faist2007}, although values reaching 50\% have been reported at low temperature \cite{Bai2010, Liu2010b}. Intersubband transitions are transverse magnetic (TM) polarised and couple only to TM polarised photons, meaning QCLs are intrinsically in-plane emitters. This precludes surface emission without fabrication of outcoupling structures \cite{Colombelli2003}. Additionally the wall-plug efficiency of QCLs rapid drops off at longer operating wavelengths as a result of an increase in the threshold current due to a reduction in gain cross-section \cite{Faist2007}.\\
The most serious drawback of QCLs is in our opinion their large cost. They are expensive to fabricate and complex to design. For good overlap with the cavity photon the device active region must be grown microns in thickness and will therefore contain many cascade periods, requiring the growth of hundreds to thousands of atomically precise layers. Notwithstanding technological improvements the cost of a QCL thus remains at least one order of magnitude too high for penetration in the consumer electronics market, especially at longer wavelengths.

\subsubsection{Thermal Detectors}
Thermal detectors are based on measuring changes in the optical properties of the material when it is heated by the absorption of infrared light. Common examples include thermopile detectors, pyroelectric detectors, photoaccoustic detectors, and bolomoetric detectors. The discussion of some of the most relevant infrared detectors has been adapted from Rogalski \cite{Rogalski2002}. Thermopile detectors are based on the Seebeck effect, where a voltage is generated across a metal heterojunction when there is a temperature difference. Whilst they typically are very slow (Hz), and have low specific detectivity $D^*\approx2\times10^{8} \mathrm{cm}^{0.5}\mathrm{Hz}^{0.5} \mathrm{W}^{-1}$, they are also relatively cheap, operate at room temperature and offer spectrally flat responsivity. This makes them useful for measuring the power of relatively strong continuous light sources, but limited in other applications. Pyroelectric detectors (such as those used commonly in infrared spectrometers) use the inherent electrical field inside certain materials to generate a voltage upon a change in temperature. Whilst they can operate at faster speeds than thermopile detectors (up to 1kHz), $D^*$ is still limited ($1\times10^{9} \mathrm{cm}^{0.5}\mathrm{Hz}^{0.5} \mathrm{W}^{-1}$) and they only produce a measurable voltage with modulated infrared light. Photo-acoustic infrared detectors are based on the expansion or contraction of gasses when they are heated  by infrared light, with the most well known being the Golay cell. Whilst these achieve much higher specific detectivity than thermopile or pyroelectric detectors ($D^*\approx 1\times10^{10} \mathrm{cm}^{0.5}\mathrm{Hz}^{0.5} \mathrm{W}^{-1}$), they are very sensitive to mechanical vibrations, and offer relatively low speeds, which make them excellent for accurate power and noise measurements, but of limited use beyond laboratory spaces.  The most advanced and sensitive type of thermal detectors are bolometers, which are based on the change in resistance in a device when it heats. Micro-bolometer arrays are the basis of uncooled thermal cameras \cite{Rogalski2012}, and cooled single element bolometers can offer extremely high values $D^*\approx 4\times10^{11} \mathrm{cm}^{0.5}\mathrm{Hz}^{0.5} \mathrm{W}^{-1}$ using conventional semiconductors. More modern superconducting bolometers achieve even higher sensitivities $D^*\approx 8\times10^{11} \mathrm{cm}^{0.5}\mathrm{Hz}^{0.5} \mathrm{W}^{-1}$, and even fast response speeds when based on the detection of hot electrons \cite{Klapwijk2017}. However, high sensitivity bolometers all require liquid helium cooling, which is extremely expensive and bulky. Thermal detectors have the major advantage in that they are highly linear and relatively simple in concept, However, room temperature versions tend to show low sensitivity and low speed \cite{Rogalski2002}.

\subsubsection{Semiconductor Detectors}
Semiconductor detectors can be operated identically to detectors in the near infrared or visible, in appropriate narrow band gap semiconductors \cite{Rogalski2005}.  Indium antimonide has a band gap as small as 0.3eV, and mercury cadmium telluride (MCT) based alloys have a band gap that can be tuned from 1.5 to -0.3eV (semimetallic behaviour). These detectors have been a material of choice for sensitive infrared detection for several decades, as they can be operated at relatively high speeds (greater than 1GHz), with high sensitivity \cite{Rogalski2005} and relatively high detectivities, with $D^*$ values ranging from $1\times10^{10} \mathrm{cm}^{0.5}\mathrm{Hz}^{0.5} \mathrm{W}^{-1}$ to $1\times10^{12} \mathrm{cm}^{0.5}\mathrm{Hz}^{0.5} \mathrm{W}^{-1}$, strongly dependent on cutoff wavelength.  They are routinely used in a wide range of applications in spectroscopy, typically operating at liquid nitrogen temperatures. One of the challenges of MCT detector technology is the relatively high defect density in materials, as well as a  drop in $D^*$ at longer wavelengths. Further, MCT detectors operated in photoconductive mode are renowned for showing nonlinear behaviours under even modest infrared illumination, strongly limiting the dynamic range of many measurements \cite{Theocharous2012}. Many of these properties are inherently linked to the properties of MCT alloys themselves, which has driven a push for other types of infrared detectors, especially at longer wavelengths. Whilst some 2D materials offer candidates here \cite{Koppens2014}, these will be addressed later in section 4 as relatively recent developments. The more established route has been through the development of quantum-engineered semiconductor heterostructures.

Quantum detectors based on III-V semiconductors exploit band structure engineering to control the energy levels in a material, based on mature III-V semiconductor growth. Two primary approaches have been studied: the quantum well infrared photodiode (QWIP) \cite{Schneider2007}, and the type-II superlattice (T2SL) \cite{Rogalski2017}. Both create an artificial band structure through the use of quantum wells, which allows the definition of a user-defined band gap to detect light. QWIPs are based on intersubband transitions in the conduction band, similarly to the operation of QCLs as discussed above. Whilst at low temperatures they are sensitive ($D^*\approx 1\times10^{10} \mathrm{cm}^{0.5}\mathrm{Hz}^{0.5} \mathrm{W}^{-1}$), QWIPs suffer from weak coupling to the out-of-plane dipole transitions in intersubbands without an extrinsic coupling mechanism such as a grating. Furthermore, QWIPS typically rely on multiple quantum wells operating in series, limiting maximum possible performance due to noise from each of the wells. Recently, there have been some exciting developments with the application of metallic antennas with QWIPS, which indicate the potential of nanophotonics to enhance infrared detection \cite{Palaferri2015,Palaferri2018,Vigneron2019,Miyazaki2020}, with $D^*$ values approaching $2 \times 10^{12} \mathrm{cm}^{0.5}\mathrm{Hz}^{0.5}\mathrm{W}^{-1}$. This enhancement comes from both the field enhancements inside the active layers, as well as the `antenna effect' where a small detector area can collect a large cross section of infrared light.

%The T2SLs structures are based on type-II band alignment, operating with both electron and holes. By forming a superlattice in semiconductors with a band offset (for instance InAsSb/GaSb). 

T2SLs have the advantage that they are much easier to couple to, with in-plane active transitions, similar to a narrow gap semiconductor, but they also offer significant control over the bandstructure. This makes it possible to design the bandstructure of T2SL structures with respect to multiple electronic and hole bands, reducing Auger recombination. It is this engineering that suggests that T2SLs can have a better efficiency than MCT detection technology, especially at longer wavelengths \cite{Rogalski2017}. However, recombination due to imperfections in the materials still limit this approach.

\section{SPhP Light Sources}
Having briefly outlined the state of the art in infrared optoelectronics, we will now turn to a detailed examination of how innovations in SPhP physics and technology can result in significant improvements for optoelectronic devices. While in the name of brevity we do not provide an in-depth review of SPhP physics here, we would invite the interested reader to examine recent reviews on the topic \cite{Foteinopoulou2019}. In general SPhPs offer a form of spectral control, through the choice of material and use of patterned surfaces, strong absorption, and strong field localization. Each of these properties offers specific advantages in terms of different concepts for optoelectronics, which we wlll break down individually in the subsequent sections.

\subsection{SPhP Based Thermal Emitters: Far Field}
SPhPs have great potential for the design of mid-infrared thermal emitters. Their narrow modal linewidths lead through Kirchoff's law to narrowband thermal emission, and their morphological dependence allows for both spectral and directional tuneablity. The highly refractory nature of typical polar dielectrics such as SiC makes them suitable for high-temperature operation.
Narrow linewidth SPhP modes can also act as narrowband thermal emitters \cite{Baranov2019}. The equality between absorption and thermal emission predicted by Kirchoff's law was first demonstrated for SPhPs in the work of Greffet {\it et al.} through measurements of both emissivity and reflectivity of a 1D 3C-SiC grating (Fig.~\ref{fig:Figthermal}c) \cite{Greffet1997, Greffet2002}. In this work the spatial coherence of the emitted light, a result of its emission from modes with well define wavevector, was also demonstrated by measuring the angular emission spectrum (Fig.~\ref{fig:Figthermal}d) which was seen to be highly directional. These results opened the way to design of narrowband, directional thermal sources.\\ Later studies have investigated thermal emission from 2D grating structures which allow for generation of collimated light, and for unpolarised emissivities approaching $100\%$ \cite{Arnold2012}. Periodic nanostructures have also been proposed as the basis of a thermal lens, whose thermal emission is focussed at a well-defined height above the grating structure \cite{Chalabi2016}. More sophisticated emission patterns can be realized by relaxing the periodicity of gratings, though the partial spatial coherence of the the SPhP mode needs to be accounted for in such designs \cite{Lu2021b}.\\
Similarly to these delocalised SPhPs in periodic structures, localised SPhPs on individual polar nanoresonators can also be utilised to enhance far-field thermal emission. In a nano-resonator bright modes with a net dipole moment couple naturally to free-space radiation and can modify the surface emissivity. The first demonstration of Kirchoff's law on the single-antenna scale was conducted in 2008 by Schuller {\it et al.} \cite{Schuller2008}, who demonstrated the equivalence between the extinction and emittance of long 3C-SiC nanowires. Subsequently the surface phonon polaritons of small polar nanospheres have been proposed as a route to radiative sky cooling \cite{Gentle2010}. While only the frequency-dependent emittance of SPhP has been exploited to this aim, directionality, possibly electrically tunable \cite{Dunkelberger2017}, could provide further fruitful venues to explore in the domain of smart thermal materials, for both civil and military applications.
Recent experimental studies have looked at the arrays of phonon polariton resonators on a same-material substrate typically utilised in optical experiments, particularly in a SiC bowtie nanoantennas configuration where tuneable mid-infrared emission was demonstrated with linewidth approaching $10$cm$^{-1}$ \cite{Wang2017}. Results from this work are shown in Fig.~\ref{fig:Figthermal}e, which compares the reflectivity and emissivity of such an array, showing correspondence between dips in the spectral reflectivity and peaks in the emissivity. The near field profiles of the contributing modes are shown in Fig.~\ref{fig:Figthermal}f.\\
Recent studies have looked at combining these elements in nano-resonator arrays to create hybrid thermal emitters, whose modes are mixtures of localised and propagating surface phonon polaritons \cite{Gubbin2016}. In these systems the discrete modes of individual nano-resonators interact with the propagative surface phonon polariton of the supporting substrate, which is leaky due to the periodic structure of the array. This allows for enhancement of the coherence length, increasing the directionality of emission, and for an increase in the modal quality factor \cite{Lu2021}.

\subsection{SPhP Based Thermal Emitters: Near Field}
Another advantage of utilising SPhPs as thermal sources is that they allow for super-Planckian energy transfer. Planck's law assumes that thermal radiation is transported from an emitter's near field to a far-field detector through propagative photons radiated from the emitter. This provides a lower limit to the true rate of thermal emission, when a source and sink of thermal radiation are placed in close proximity the rate of thermal exchange can increase dramatically through tunnelling of evanescent waves \cite{Francoeur2008, Pendry1999}.\\
This was first demonstrated in SPhP systems by Chen {\it et al.} in 2008 by measuring the heat transfer between parallel SiO$_2$ surfaces separated by a microscale gap \cite{Hu2008}, and reporting strong enhancement in the thermal heat transfer due to the propagative surface phonon polariton modes on each SiO$_2$ surface. An even stronger result was later demonstrated for heat-transfer between an SiO$_2$ microsphere and a flat surface \cite{Shen2009}, where for small separations a three order of magnitude enhancement in the radiative heat transfer was observed compared to the black body case. This is shown in Fig.~\ref{fig:Figthermal}g, where the calculated sphere-plate heat-transfer coefficient is shown as a function of separation compared to the black body case. The use of surface phonon polaritons for wide-area near-field radiative heat transfer has also been proposed \cite{Ghashami2018, Tang2020}. Strong near field radiative heat transfer is desirable for near-field thermophotovoltaics \cite{Fiorino2018, Narayanaswamy2003, Laroche2006}, for thermal management in integrated nanoscale devices \cite{Song2015} and for controlling heat flow in graphene \cite{Tielrooij2018}. Recently super-Planckian heat-transfer using surface phonon polaritons was demonstrated using nanoscopic SiO$_2$ films to dissipate heat injected from a metallic contact over 50\% more efficienctly than in bulk, as illustrated in Fig.~\ref{fig:Figthermal}h. These approaches have great potential in designing on-chip integrated cooling.

\subsection{SPhP Based Nonlinear Sources}
An alternative route to generate narrowband mid-infrared light is through nonlinear frequency conversion of emission from conventional light sources operating in the visible or near-infrared. Mid-infrared light could be generated either through difference frequency generation \cite{Constant2016, Gubbin2017}, or through four-wave mixing in an optical parametric oscillator \cite{Petrov2015, Gubbin2018}.\\
Although generation of SPhPs is yet to be demonstrated polar dielectrics are known to have strongly resonant second ($\chi^{(2)}$) and third ($\chi^{(3)}$) order nonlinear susceptibilities around their optical phonon frequencies \cite{Vanderbilt1986, Teitelbaum2018, Paarmann2015, Paarmann2016, Kitade2021}. Moreover, nonlinear processes have been investigated in the mid-infrared, when all photon frequencies lie in the neighbourhood of the optical phonons. Second harmonic generation has been demonstrated utilising propagative SPhPs on bulk 6H-SiC \cite{Passler2017}, in ultrathin AlN films supporting epsilon-near-zero modes \cite{Passler2019}, and in deep sub-diffraction polar nano-resonators \cite{Razdolski2016, Razdolski2018}.\\
Recently sub-diffraction imaging of mid-infrared resonances in polar nano-antennas was demonstrated utilising sum-frequency generation. As the sum-frequency beam is of significantly reduced wavelength compared to the mid-infrared this allows for a dramatic increase in resolution compared to mid-infrared imaging as shown in Fig.~\ref{fig:intersubbandEmitters}a  \cite{Kiessling2019}. This process is the inverse of SPhP difference frequency generation and it opens the way to all-visible resonant excitation of mid-infrared SPhP modes. Moreover strong third-order optical nonlinearities were also recently observed in SiC by measuring the self-phase modulation \cite{DeLeon2014} of a propagative surface phonon polariton as shown in Fig.~\ref{fig:intersubbandEmitters}b \cite{Kitade2021}. Here a three order of magnitude increase in the effective nonlinear index was observed compared to that expected for a non-resonant dielectric, confirming theoretical preditions, and presenting an promising result for excitation of SPhPs by four-wave mixing \cite{Gubbin2018}.

\subsection{SPhP Based Intersubband Sources}
\begin{figure*}
    \centering
    \includegraphics[width=\textwidth]{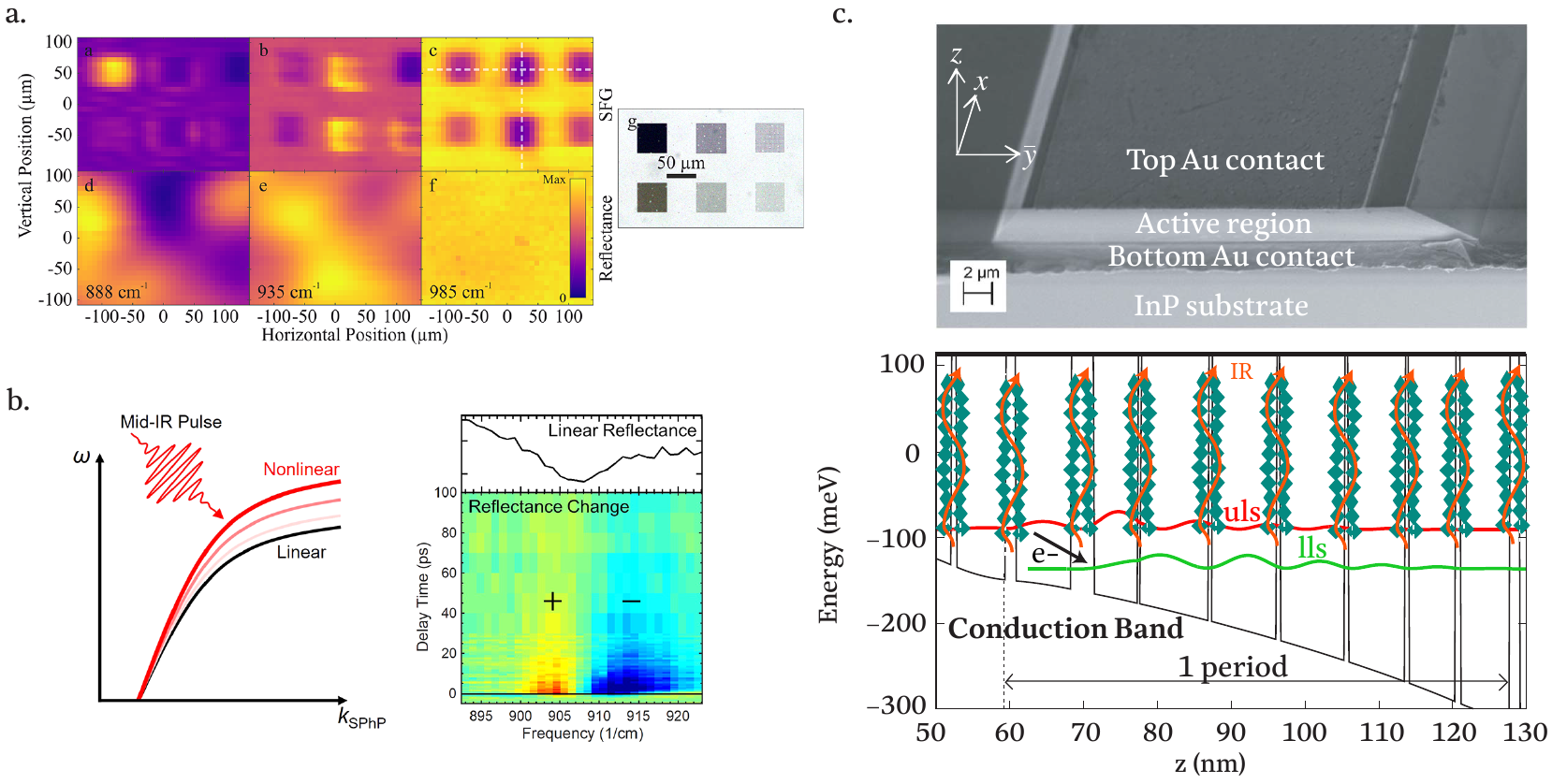}
    \caption{a. Surface phonon polariton imaging by frequency upconversion. Lower panels show mid-infrared images recorded from the polar resonator arrays shown on the right. Upper panels show the sum-frequency spectrum \cite{Kiessling2019} b. Measurement of the self-phase modulation of a surface phonon polariton, a static shift in frequency with varying illumination intensity as illustrated on the left. The right shows the change in linear reflectance in a two-pulse measurement as a function of delay time, a hallmark of the nonlinear interaction \cite{Kitade2021} c. Upper: SEM image of a phonon polariton emitting QCL. Lower: Conduction band structure of the QCL. Illustrated are the upper and lower lasing states (uls and lls), electronic transitions between which produce phonon-polaritions which are emitted from device end-facet. Photons are illustrated by orange arrows and the phonon oscillations by green dots. Adapted from Ref. \cite{Ohtani2019}}
    \label{fig:intersubbandEmitters}
\end{figure*}
In the introduction we discussed the QCL, the premier mid-infrared emitters due to their broad tuneability. There has been interest in utilising these systems to electrically excite phonon polariton modes in a planar heterostructure. In a remarkable work Ohtani {\it et al.} demonstrated phonon polaritons excitation by electrical injection in a quantum cascade structure \cite{Ohtani2019}. This study, illustrated in Fig.~\ref{fig:intersubbandEmitters}a., used phonon polaritons obtained hybridising TO phonons in the barrier layers as lasing mode.
The resulting device presents many interesting features, including the possibility to modulate gain by changing the light and matter components of the lasing mode, but it still required a standard quantum cascade structure for electron injection. 

One route to make QCLs more accessible is reducing the size of the active region. Due to the long photon wavelength in the mid-infrared, devices often require hundreds of repetitions of the cascade unit in order to achieve the good overlap between lasing mode and active region necessary for efficient operation. By utilising sub-diffraction SPhPs as the lasing mode of a QCL it could be possible to miniaturise the device, reducing the number of repetitions necessary. Moreover the low-loss, and narrow linewidths of SPhP could offset some of the limiting factors associated with plasmonic lasers \cite{Khurgin2014}. Phonon polariton powered QCLs could also have lower threshold voltages due to additional components opening up in the gain. They could also be highly tuneable, providing multi-wave mixing through the strong intrinsic nonlinearity of the optic phonons as discussed above.

\section{SPhP light Detectors}

\begin{figure}
    \centering
    \includegraphics[width=8cm]{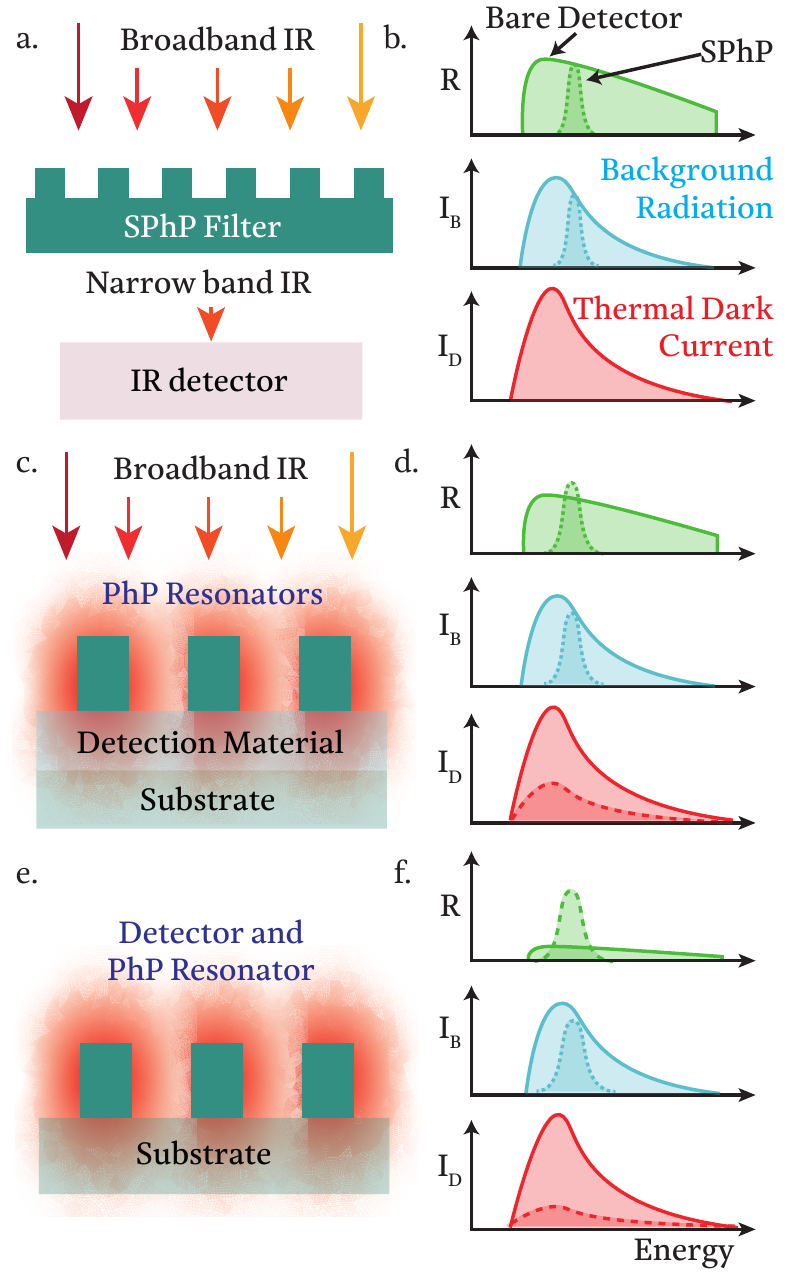}
    \caption{Illustration of different detector concepts when integrated with SPhPs.  a. \& b. a passive SPhP element is used as a spectral filter in otherwise challenging spectral ranges. c. \& d. by integrating the SPhP resonator onto the detector we can exploit field enhancements to improve responsitivity of the detector. e. \& f. Finally, by leveraging a detection mechansim inside the SPhP material itself we can maximize signal and minimize device volume}
    \label{fig:DetectorC}
\end{figure}

Similar to light emitters, phonon polaritons also offer a significant opportunity for light detection in the mid- to far-infrared where a narrowband response, commensurate with those of SPhPs is required. Three different general schemes for integration of SPhPs with detectors is illustrated in Fig. \ref{fig:DetectorC}. In the simplest case, light can simply be filtered by SPhP resonators, allowing the creation of band-pass filters in spectral ranges where conventional dielectrics are less well developed. This alters the absorbed wavelengths of the device and reduces any background-induced thermal noise, and hence results in a cleaner image. A second approach is the integration of phonon polariton resonators on top of an infrared detector. In this case we can exploit the field enhancement of the resonator to improve absorption responsivity, whilst also reducing the thickness of the detector, and thus both background and thermal induced noise. This could also improve the device response speed, due to any reductions in the thickness of the device active area. Finally, one could leverage optoelectronic transitions inside the SPhP resonator themselves, which allows us to further enhance field overlap, whist reducing the device area further, providing the strongest boosts to responsivity and decrease in background noise. We highlight that in all these approaches the bandwidth of the detector is reduced, making them appropriated for targeted applications where narrow spectral response is necessary (such as in spectroscopic sensors). Specific approaches to realizing these schemes will be discussed in the following subsections.

\subsection{Thermal detectors}
Thermal detectors are relatively simple in terms of operation, which makes them particularly appealing for SPhP enhancement. In a thermal detector it is only necessary to heat up the detection material with infrared light, without many considerations for quantum engineering required in semiconductor detectors. As such,  integrating a bolometric or pyroelectric material onto conventional SPhP resonators it should be possible to fabricate a narrowband SPhP detector, with possibility of on-site tuning. In such a detector, the infrared light at a specific frequency localized by the SPhP mode will be absorbed, creating a local rise in temperature and thus a detectable signal. A similar scheme has been used with pyroelectrics and plasmonics in the near infrared \cite{Stewart2020,Stewart2021}. An advantage of this approach is that the strong field localization of SPhPs mean that only a relatively small thermal detector will be required. A smaller amount of detecting material will decrease thermal mass, enabling rapid changes in temperature and a quicker response speed. Moreover, a secondary improvement will be an increase in sensitivity due to a reduction in background radiation flux from spectral filtering. Integration of mid-infrared SPhP resonators on thermal detectors seems a promising venue for early-stage commercial SPhP devices, and we expect to see multiple combinations of detectors and SPhP materials explored in the near future. 

A promising choice for such efforts would be the creation of a thermal detector based on vanadium oxide (VOx), which is already used in microbolometer array cameras \cite{Rogalski2017}. VOx can be conformally coated on in principle any material through atomic layer deposition \cite{Kozen2017}, and offers a strong bolometric effect. We envision that VOx could be coated on SiC nanopillars, similar to what achieved through coating of SiC resonators with AlOx \cite{Berte2018}. This would provide the perquisite localization of the infrared absorption inside the SPhP resonator with the thermal detection material. By measuring the resistance of the VOx film with and without infrared illumination, it should be possible to detect at the frequencies of the phonon polaritons. Although the proposed devices would leverages well established design rules for SPhP resonators, a couple of challenges would need to be overcome. First, the local heating must be closely localized at the location where the current travels through the material. For a nanopillar array current would flow through the VOx film at the base, which might limit the sensitivity, which might suggest different device geometry.  A second limitation will be the high thermal conductivity of SiC, which might result in low responsivity due to rapid thermal spreading, reducing the size of the electrical signature. If technical challenges could be overcome this might offer a number of advantages, notably improved response speed, and a controllable bandwidth. In terms of application performance, the improved response speed may offer advantages in terms of frame rate for thermal cameras targeting specific signatures. Current thermal cameras based on arrays of thermal detectors (often of VOx) are largely limited by the rate at which the material heats when subject to incident infrared. We highlight that state of the art modern superconducting bolometers might offer an alternative material system where some of the advantages of SPhPs can be exploited, however a serious examination of the interplay with these materials would require more significant design innovations.

\subsection{Semiconductor detectors}
Whilst nanophotonic enhancement can offer a significant improvement in thermal detector technology, further potential advantages present for semiconductor detectors. In particular III-V semiconductors have demonstrated their capacity to both support phonon polaritons \cite{Feng2015b}, and optoelectronic transitions \cite{Rogalski2017,Rogalski2012}. Indeed, it is already widely observed that close to the LO phonon of GaAs in terahertz QWIPs there is often a dramatic enhancement in the responsivity of devices, despite operation away from the strongest intersubband absorption \cite{Palaferri2015,Shao2019}. Fully leveraging and optimizing for coupling to the SPhPs should further enhance responsivity, as well as increase operational temperature. 

The rationale for operation of a semiconductor device with SPhPs can be seen clearly taking the specific example of QWIPs. Unless they are operated at extremely low temperatures (below liquid nitrogen), QWIPs typically exhibit performance limited by dark-current induced noise. Based on a simple electronic model, in thermal dark current limited operation one can show that the $D^*$ value is given by \cite{Schneider2007};
\begin{equation}
	D^* = \frac{\lambda \eta}{2hc}\sqrt{\frac{t_{c}}{N_{QW} N_{3D}L_p}}, \label{eq:QWIP}
\end{equation}
where $\lambda$ is the operational wavelength, $\eta$ is the internal absorption efficiency, $N_{QW}$ is the number of quantum wells, $t_c$ is the electron capture time, $N_{3D}$ is the temperature dependent 3D electron density, and $L_p$ is the quantum well period. By leveraging SPhPs it is possible to enhance the absorption of the detector, potentially up to an efficiency approaching 100\%. Further, by leveraging the wavelength compression associated with surface waves it is possible to reduce both the number of quantum wells, as well as potentially the periodicity between quantum wells. The result will be a strong drop in the noise of the device, which makes it possible to enhance the operational temperature of semiconductor detectors. Such advances have already been exploited using metal antennas in QWIPs to push operation to room temperature \cite{Palaferri2018}, and has been proposed for T2SL using related approaches \cite{Goldflam2016,Wang2018}.  Extending this approach to SPhPs seems to us a natural next step.

To give an example of specific materials where this idea can be applied we can examine conventional GaAs/AlGaAs based QWIPS, which have been most widely studied. The presence of a SPhP in a thin film of a material can induce extremely strong absorption close to the LO phonon, due to Berreman absorption \cite{Vassant2010}. Given that an GaAs/AlGaAs QWIP consists of a series of thin films of GaAs (the quantum wells), nested in AlGaAs barriers, it is feasible that the Berreman enhancement can be exploited for a detector operating close to the GaAs LO phonon. By enhancing the absorption from a single well dramatically, we anticipate that only a few quantum wells may be required for this type of detector, providing a high  $D^*$ when compared to conventional QWIPS. Indeed, similar concepts have already been applied to create an infrared modulator, where intersubband transitions and SPhP induced absorption were combined to create an infrared modulator \cite{Vassant2012} .  By exploiting alloys of AlGaAs it would subsequently be possible to tune the frequency of the mode within a narrow target band \cite{Kim1979}.  It is worth noting that T2SLs could also in principle be enhanced through a similar approach leveraging phonon polaritons. However, the optical phonons in GaSb/InAs based superlattices occur at significantly lower energies ($80\mu$m), which may limit applications. The integration of a second polar material may be more appropriate for efforts in this material system, and open up a significant wider application space, as will be discussed later in this perspectives.

\section{Longitudinal Transverse Coupled Devices}
\begin{figure*}
    \centering
    \includegraphics[width=17cm]{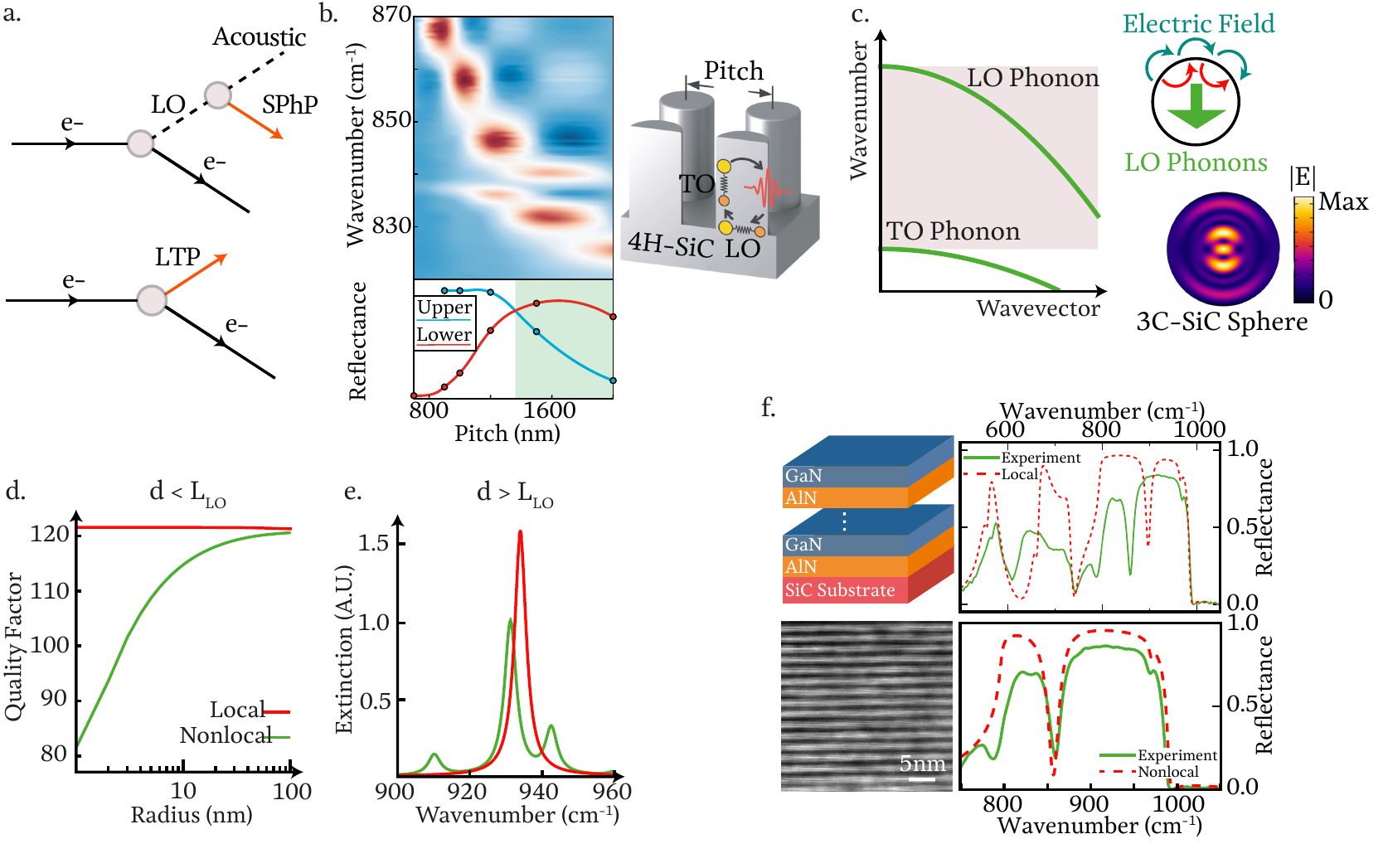}
    \caption{a. Illustrates generation of phonon polaritons through LO phonon decay and direct LTP emission b. Strong coupling between the resonances of a 4H-SiC nanopillar and a zone-folded optical phonon is illustrated by the reflectance map in the upper panel. The lower panel plots the reflectance dip magnitude, illustrating hybridisation between photon and phonon. Reproduced from Ref.  \cite{Gubbin2019} c. Illustrates the negative dispersion of phonon branches in a polar crystal and it's consequence. When SPhP energy sits in the lattice excitations it is transported by the phonon modes away from the supporting interface. d. The effect in thick layers is to increase loss, here we show the quality factor for a nanoscale 3C-SiC sphere in the local and nonlocal cases \cite{Gubbin2020}. e. In small particles the phonon spectrum is discrete, here we plot the extinction for a nanoscale 3C-SiC sphere, new modes in the nonlocal case correspond to localised phonon excitations. f. Left: Illustration of a crystal hybrid, formed from nano-layers of GaN/AlN on an SiC substrate, with SEM image of the fabricated hybrid Reproduced from Ref. \cite{Ratchford2018}. Right: Upper shows experimental reflectance from the hybrid, compared with that predicted from a local model (\cite{Ratchford2018}). Lower shows the same experimental reflectance compared with a nonlocal model (\cite{Gubbin2020})}
    \label{fig:LTPP}
\end{figure*}
This Section describes an alternative, conceptually simpler but still largely unexplored route to mid-infrared SPhP optoelectronics. Ohmic losses in polar dielectrics predominantly arise from emission of optical phonons via the Fr\"ohlich interaction \cite{Roldan1997}. These phonons are of LO polarisation and are usually decoupled from the transverse electromagnetic field which interacts instead with TO phonons to create standard phonon polaritons. Once generated LO phonons usually decay incoherently through the Ridley channel \cite{Chen2003} into an acoustic phonon plus a TO phonon.
As TO phonons can be part of hybrid SPhP excitations, the LO phonon decay can potentially also lead to the emission of SPhPs which can radiate to the far-field (top panel of Fig.~\ref{fig:LTPP}a). Such a process is nevertheless inefficient due to the large phase space available for LO-phonon disintegration.\\
Direct, resonant emission of SPhPs by electrical current would be possible by designing SPhP excitations with both LO and TO phonon components, the former coupling to the electrical currents and the latter to far-field electromagnetic radiation (bottom panel of Fig.~\ref{fig:LTPP}a). These excitations, termed longitudinal-transverse polaritons (LTP) were initially observed experimentally utilising the atomic-level structure of 4H-SiC to fold the LO phonon branch along the c-axis \cite{Gubbin2019}. In Fig.~\ref{fig:LTPP}b we reproduce figures from Ref. \cite{Gubbin2019} showing a clear anticrossing when the SPhP modes of a 4H-SiC nanopillar array is tuned across the folded LO phonon (also known as weak phonon mode \cite{Bluet1999}). Such anticrossing clearly signal that strong coupling between the contributing modes, indicating the system eigenmodes are a mixture of transverse SPhP and LO phonons \cite{Gubbin2016b}.

More recently the existence of LTPs has been demonstrated to be a general feature of polar materials in size regimes where the dispersion of optical phonons can no longer be safely neglected \cite{Gubbin2020,Gubbin2020b}. At large wavevectors optical phonon modes are dispersive as illustrated in Fig.~\ref{fig:LTPP}c, meaning they contribute to energy transport into the polar dielectric, an effect termed optical nonlocality by analogy with similar effects observed in other nanophotonic systems \cite{Ciraci2012,Rajabali2021}. In small polar dielectric particles this leads to increased losses as energy leaches away from the excitation spot, and transfer into short wavelength matter-like phonon modes as illustrated by the electric field plot for a nanoscale 3C-SiC sphere. The effect is illustrated in Fig.~\ref{fig:LTPP}d which shows the quality factor for a small 3C-SiC sphere as a function of radius in both the nonlocal and local case.\\

To account for the additional longitudinal degrees of freedom mechanical boundary conditions at the dielectric boundaries are needed to complement Maxwell boundary conditions, mixing longitudinal and transverse oscillation patterns with a coupling strength proportional to the phonon velocity and inversely proportional to the system size \cite{Gubbin2021}. For small systems this leads to the appearance of discrete LTP modes in the optical spectrum, which could be exploited for resonant electrical emission (Fig.~\ref{fig:LTPP}e). \\
Theory of nonlocality in dielectrics have also proved relevant to describe anomalous frequency shifts in crystalline hybrids, user defined materials created from many nanoscale layers of different polar media as illustrated in the left panel of Fig.~\ref{fig:LTPP}f \cite{Ratchford2018}. These materials are most promising for mid-infrared photonics and SPhP in particular but, as the layers comprising the crystalline hybrid are sufficiently small to exhibit nonlocal effects, the phonon modes observed in reflectance strongly deviate from what expected from a standard local treatment (upper right panel of Fig.~\ref{fig:LTPP}f. Correct results can instead be obtained utilising nonlocal response models \cite{Gubbin2020, Gubbin2020c}, illustrated in the lower right panel of Fig.~\ref{fig:LTPP}f, thus highlighting the relevant of LTP physics in this class of materials and thus the possibility to employ them in SPhP optoelectronics.\\
Many unsolved problems remain to the demonstration of LTP-based devices, as the appropriate nanostructuring required to match both momentum and energy conservation of the electron-to-SPhP conversion process, or the requirements to achieve efficient spontaneous emission or lasing for realistic current densities. 
Still, this mechanism promises the possibility to design mid-infrared optoelectronic devices powered directly through Ohmic losses, a process illustrated in Fig.~\ref{fig:LTPP}a. Achieving efficient electrical injection without requiring complex and expensive quantum cascade structures. 
The resulting devices, fabricated using optical lithography on commercially available wafers, would substantially cut the cost of mid-infrared capabilities, allowing them to be integrated in consumer electronics. 
We can note moreover that, should an LTP laser be achieved, the electron-electron scattering would become stimulated, thus potentially opening a novel way to ultrafast optical modulation of electrical resistivity at the picosecond scale.

\section{Materials challenges and considerations}
Whilst above we have detailed the opportunities to leveraging nanophonics for enhancing infrared optoelectronics, phonon polaritons offer some unique materials challenges. For plasmonics the only requirement is to have a population of free electrons, and can be supported in amorphous or crystalline materials. This means many plasmonic schemes can be realized simply by depositing metals on top of a material with optoelectronic properties. SPhPs, on the other hand, require crystalline materials, which will not always form using an arbritrary combination of growth material, method, and substrate. Further, the frequency and quality of the SPhP resonance is determined by its crystalline properties. Finally, even if an optoelectronic material supports SPhP modes itself, it may not be at a desirable frequency. This makes realizing SPhP optoelectronics a fundamental materials challenge, as many materials which support excellent PhP modes are challenging for optoelectronics, and vice versa. Therefore serious consideration needs to be given to the integration of SPhP modes with optoelectronic transitions. We have given a few specific examples, focused on SiC and III-V semiconductors, in the previous sections which naturally lend themselves to this integration. In this Section we aim to provide an outline of broader developments in materials science which could lead to new SPhP systems for optoelectronics.

\subsection{Wide band gap nitrides and oxides}

Wide band-gap materials such as GaN and AlN have seen significant development in the past decades, largely driven by efforts to create visible optoelectronics or power electronics. This means that a mature know-how of doping and quantum well formation have been developed for these materials. Further both GaN and AlN have been shown to support extremely high quality SPhPs in the infrared \cite{Passler2017,Feng2015,Khan2020,Ng2007}, with GaN even having a response controlled through carrier injection \cite{Dunkelberger2020}. Whilst studies into the wide bandgap materials have been limited by the challenging growth conditions, the increasing number of studies in this area suggest that this could be a promising area for SPhPs. Of particular relevance is that they possess some of the highest energy optical phonons for semiconducting materials, pushing into the LWIR transparency window. The nitrides are also interesting because in principle they could be used to create infrared optoelectronics \cite{Tsao2018}. In particular intersubband transitions have been demonstrated inside nitride materials, which suggests that an appropriately designed well could in principle support SPhP based emission or detection \cite{Lahnemann2017,Durmaz2016}. Indeed, the thinner active regions inherent to SPhPs would be a significant advantage for the growth of nitride electronics, where strain and defects often limit the thickness of epitaxial layers grown. Whilst undoubtedly extremely challenging systems in terms of materials.

Oxides present one of the oldest SPhP systems, with studies on quartz and sapphire dating back almost 50 years \cite{Falge1973}. Whilst the lack of optoelectronic functionality in the most common oxides has precluded significant applications, the family of oxides is extremely large. For instance, there has been significant interest in epitaxial oxide materials over the past several decades, including complex oxides \cite{Schlom2008} and semiconducting oxides \cite{Stepanov2016}. For the complex oxides (such as $SrTiO_3$, $LiNbO_3$) this has largely been due to the interest in their electro-optic properties, which are widely exploited in telecommunications to make modulators. Indeed, the piezoelectric properties of these materials have already been demonstrated as a way of controlling plasmonic resonators \cite{Beechem2018}. Another notable oxide is Vanadium Oxide, which supports a phase change in the infrared, making it particularly interesting for SPhPs \cite{Taboada-Gutierrez2020}, though large area single crystal growth is challenging. Many of these oxide materials possess strong Reststrahlen bands across the mid infrared. Further, emerging oxides, such as $\beta-Ga_2O_3$ \cite{Stepanov2016,Schubert2019}, are semiconductors which can be be doped. Whilst some groups have noted the relevance of the bulk polaritons modes in oxide materials such as these \cite{Kojima2018}, the surface modes have largely remained unexplored. This is because oxides often have a large number of optically active phonons, which makes studies more involved, although further exploration of the family of oxide materials could be beneficial for expanding the landscape of SPhPs applications. As a final note on the wide band gap materials and oxides, its worth noting that many of them have advantageous mechanical properties and thermal stability. This offers some major advantages versus more traditional MCT, III-V, and even silicon-based platforms in different environments. Further, this enables them to be effective thermal emitters even in ambient, or low vacuum environments, unlike many conventional materials.

\subsection{Heterogeneous integration of SPhP materials}

\begin{figure}
    \centering
    \includegraphics[width=8cm]{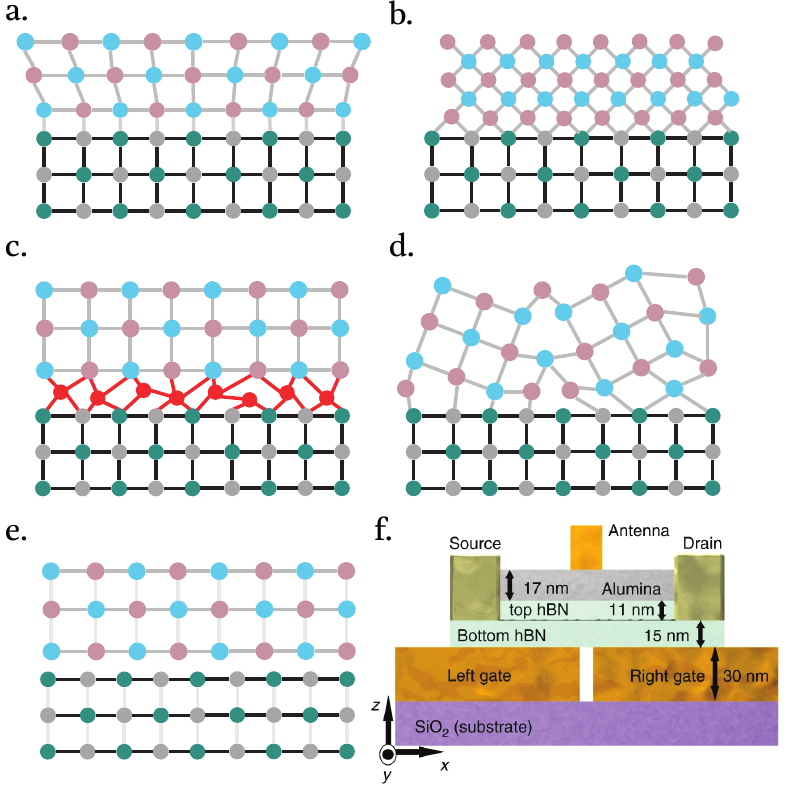}
    \caption{Illustration of different approaches for combining optoelectronic and SPhP materials a. illustrates the issue with lattice mismatch when a large lattice constant material (such as GaSb) is integrated on a small lattice constant material (such as 3C-SiC). b. Indirect lattice matching between materials with very different lattice constants for epitaxial integration c. Wafer bonding, using a thin film of amorphous material d. polycrystaline integration on a substrate e. integration based on 2D materials, wher ethe weak out of plane bonding allows different materials to couple without lattice matching f. recent demonstration of integration of 2D materials with optoelectronically active graphene, from reference \cite{Castilla2020}}
    \label{fig:Materials}
\end{figure}

Plasmonics has been very successful in realizing functionalities in different applications through heterogeneous integration of metal structures onto other materials and systems. The obvious approach to using SPhPs in optoelectronics would be to take this same approach and seek combinations of materials and systems where heteroepitaxial integration is possible. As noted above, such integration requires highly crystalline films, which can be challenging to realize when combining two different materials with different bonding, lattice constants, and potentially crystal structures. Perhaps the most obvious option is the heterogeneous expitaxial growth of a material with optoelectronic functionality on a substrate which supports SPhPs, or vice versa. An example might be the integration of a III-V active region onto SiC, as both are well demonstrated for optoelectronics and SPhP modes respectively. However, the different crystal structures between these materials will make it difficult, if not impossible, to grow these materials expitaxially due to the strain in lattice mismatched materials \ref{fig:Materials}a. A notable exception is the growth of cubic SiC on Si, which has a sufficiently similar lattice constant for growth up to several microns \cite{Ferro2015}, albeit with lower crystalline quality than that achievable in bulk single crystals of hexagonal SiC polytypes. Such material has already been exploited in the creation of SPhP devices in the infrared \cite{Gubbin2016,Howes2020}. Where direct lattice matching may not be possible, indirect lattice matching might also be an alternative for integration of SPhP materials (Fig.~\ref{fig:Materials}b). This process is applicable to the growth of complex oxides on semiconductors, as has been studied for electronics applications \cite{Kumah2020}. We suggest this system as one that could be studied in greater depth, where an indirect eptiaxial relationship is established.\\
A related approach to epitaxial growth is the use of wafer bonding in order to combine materials with different properties as depicted in Fig.~\ref{fig:Materials}c. Wafer bonding is a critical process in the semiconductor industry, and uses the application of temperature, pressure and radiation in order to physically join two wafers together. One of the most widely explored example of this has been the integration of III-Vs on silicon for the creation of photonic integrated circuits\cite{Roelkens2007}, and it has even been demonstrated that GaAs can be bonded onto silicon carbide \cite{Higurashi2015}. A related example is the integration of silicon on sapphire wafer technology, which could exploit SPhPs of the sapphire substrate, and bolometric or defect effects in silicon \cite{Baehr-Jones2010}. However, wafer bonding is prone to damaging the materials involved and  presents significant fabrication expense. Whilst it is foreseeable that there are other examples where integration via epitaxy or wafer bonding is possible, there is not for the moment a `one size fits all' approach applicable for the combination of two arbitrary materials to leverage SPhP modes.\\
An alternative to the integration of single crystal SPhP materials might the growth of polycrystaline SiOx or AlOx layers onto optoelectronic devices, which are widely deposited as dielectrics in semiconductor processing (Fig.~\ref{fig:Materials}d). Polycrystaline materials naturally have much higher crystal damping, and hence much higher losses than their monocrystaline counterparts. However, given that SPhP modes already have much lower losses than those seen in conventional metal or semiconductor plasmonics, this may still be a viable pathway to integration. For thin films techniques such as atomic layer deposition and pulsed laser deposition are highly suitable, and a wide range of oxides and nitrides have been studied using these techniques\cite{Berte2018,Beliaev2021}. Whilst the interaction between atomically thin  oxides and SPhPs have been explored in the context of enhanced sensing\cite{Berte2018}, more work is required in this area to assess if polycrystaline films are adequate to meet the requirements of relatively low absorption that would see amorphous integration practical.
\subsection{2D materials}

In recent years there has been growing interest in examining SPhP modes in the class of two dimensional (2D) materials. In general, 2D materials are characterized by a layered structure, which allows them to be exfoliated from bulk into thin films \cite{Novoselov2016}, supporting both polar optical phonons and optoelectronic transitions \cite{Basov2016}.  Further, as layers are only held together by Van-der-Walls forces, they can be stacked into arbitrary arrangements \cite{Novoselov2016}, including introducing a twist angle between layers to engineer new functionality \cite{Song2018,HerzigSheinfux2020}. Exploring SPhPs is a rapidly growing area for 2D materials research, due to the high energy of the polar phonons in hexagonal boron nitride \cite{Dai2014,Caldwell2014b} (with an LO phonon at $6.1 \mu m)$, as well as the anisotropic response of alpha molybdenum trioxide \cite{Ma2018,Zebo2018,Zheng2019}. They have been shown to exhibit some of the highest quality factors ($Q=400$) \cite{Tamagnone2020,Autore2021,Lee2020} due to the ability to isotopically enrich flakes of material. The ability to layer optoelectronic materials, such as monolayer \cite{Sun2014} or bilayer \cite{DeLiberato2015,Ju2017,Bandurin2018} graphene, with these polar material offers an exciting opportunities to create infrared detectors (Fig. \ref{fig:Materials}e). For example, recently it has been shown that by combining a metal antenna, phonon polaritons in hexagonal boron nitride, and graphene flakes it is possible to create an infrared detector \cite{Castilla2020}(Fig.~\ref{fig:Materials}f). It has also been possible to measure the photocurrent induced in twisted bilayer graphene, which highlights the new physics possible in this platform \cite{Sunku2021,Hesp2021}. Other materials, such as black phosphorous \cite{Rogalski2019} and platinum diselenide \cite{Yu2018} show potential for detection of infrared light.  A final option might be the combination of 2D materials with more conventional 3D semiconductor devices. One of the first demonstrations of this idea was the integration of graphene with terahertz lasers \cite{Chakraborty2016}, and has been achieved in the visible, with patterned gallium phosphide resonators used to control the emission from tungsten disulphide  \cite{Mey2019}. This requires careful co-location of the two materials, but may provide an interesting platform for novel devices. The main outstanding challenge for creating efficient infrared detectors from 2D materials is the  small size of the flakes created using exfoliation, the relatively technical fabrication processes involved \cite{Fan2020}, and lack of air stability for many of the materials \cite{Fan2020b}. However, given the ongoing development of 2D material growth and fabrication processes, these challenges may be overcome in the near future. This makes 2D materials a promising area to monitor as this material system continues to develop

\section{Outlook}

The last few years have seen a remarkable interest in the physics and technology of SPhP, with a number of proof-of-concepts demonstrated in top-tier research papers. While we are convinced such a line of investigation is still in its infancy, and many more exciting fundamental results will keep appearing in the next few years, we also think it is time  to put serious thought into maturing some of these proof-of-concepts and increase their technological readiness level into marketable devices.  We hope that this perspective provides a few different areas which are ripe for technological innovation, and spurs collaboration between photonics, optoelectronics and materials communities.
While the market landscape for SPhP devices remains yet unclear, their expected low cost, small dimension, and possibility to integrate them on existing platforms make us expect to see an intense effort to engineer SPhP-based commercially successful products.  For instance, we envision that SPhP thermals source can harvest waste heat generated by many consumer electronics \cite{Lu2020}, making them extremely efficient. Further, multi-spectral SPhP filters combined with conventional semiconductor technologies \cite{Dubrovkin2020} could offer a low technical barrier to entry. 
The low-cost optoelectronic devices which SPhPs promise would allow to integrate mid-infrared sensing capabilities in consumer electronics, gaining access to quantitative, real-time, localised information on our chemical environment. Smartphones capable to identify volatile organic compounds in the breath will revolutionize healthcare, while widespread real-time monitoring of sources of atmospheric pollution will provide authorities detailed information for policy enforcement. 
In the next few years we thus expect to see growing interest of deep-tech venture capital toward SPhP-based spin-offs, leading to the creation of multiple ventures exploring at least some of the pathways we sketch in this work.
We hope this perspective will help such a process, introducing new colleagues to the state-of-the-art in the field and serving as a white paper around which academic collaborations and commercial initiatives can coalesce.\\

\section{Funding}
S.D.L. is supported by a Royal Society Research fellowship and the  the Philip Leverhulme prize. S.D.L. and C.R.G. acknowledge support from the Royal Society Grant No. RGF\textbackslash EA\textbackslash181001 and the Leverhumlme Grant No. RPG-2019-174. T.G.F is supported through startup funds at the University of Iowa.

% If in two-column mode, this environment will change to single-column format so that long equations can be displayed. 
% Use only when necessary.eV
%\begin{widetext}eV, 
%$$\mbox{put long equation here}$$
%\end{widetext}

% Figures should be put into the text as floats. 
% Use the graphics or graphicx packages (distributed with LaTeX2e).
% Put the label that you will use with \ref{} command in the braces of the \label{} command.
%
% \begin{figure}
% \end{figure}

% \begin{table}
% \end{tabular}
% \end{table}

%\begin{acknowledgments}
% Put your acknowledgments here.
%\end{acknowledgments}

\bibliography{ARTICLE}

\providecommand{\noopsort}[1]{}\providecommand{\singleletter}[1]{#1}%
\begin{thebibliography}{200}%
\makeatletter
\providecommand \@ifxundefined [1]{%
 \@ifx{#1\undefined}
}%
\providecommand \@ifnum [1]{%
 \ifnum #1\expandafter \@firstoftwo
 \else \expandafter \@secondoftwo
 \fi
}%
\providecommand \@ifx [1]{%
 \ifx #1\expandafter \@firstoftwo
 \else \expandafter \@secondoftwo
 \fi
}%
\providecommand \natexlab [1]{#1}%
\providecommand \enquote  [1]{``#1''}%
\providecommand \bibnamefont  [1]{#1}%
\providecommand \bibfnamefont [1]{#1}%
\providecommand \citenamefont [1]{#1}%
\providecommand \href@noop [0]{\@secondoftwo}%
\providecommand \href [0]{\begingroup \@sanitize@url \@href}%
\providecommand \@href[1]{\@@startlink{#1}\@@href}%
\providecommand \@@href[1]{\endgroup#1\@@endlink}%
\providecommand \@sanitize@url [0]{\catcode `\\12\catcode `\$12\catcode
  `\&12\catcode `\#12\catcode `\^12\catcode `\_12\catcode `\%12\relax}%
\providecommand \@@startlink[1]{}%
\providecommand \@@endlink[0]{}%
\providecommand \url  [0]{\begingroup\@sanitize@url \@url }%
\providecommand \@url [1]{\endgroup\@href {#1}{\urlprefix }}%
\providecommand \urlprefix  [0]{URL }%
\providecommand \Eprint [0]{\href }%
\providecommand \doibase [0]{http://dx.doi.org/}%
\providecommand \selectlanguage [0]{\@gobble}%
\providecommand \bibinfo  [0]{\@secondoftwo}%
\providecommand \bibfield  [0]{\@secondoftwo}%
\providecommand \translation [1]{[#1]}%
\providecommand \BibitemOpen [0]{}%
\providecommand \bibitemStop [0]{}%
\providecommand \bibitemNoStop [0]{.\EOS\space}%
\providecommand \EOS [0]{\spacefactor3000\relax}%
\providecommand \BibitemShut  [1]{\csname bibitem#1\endcsname}%
\let\auto@bib@innerbib\@empty
%</preamble>
\bibitem [{\citenamefont {Purcell}(1946)}]{Purcell1946}%
  \BibitemOpen
  \bibfield  {author} {\bibinfo {author} {\bibfnamefont {E.~M.}\ \bibnamefont
  {Purcell}},\ }\href@noop {} {\bibfield  {journal} {\bibinfo  {journal}
  {Physical Review}\ }\textbf {\bibinfo {volume} {69}},\ \bibinfo {pages} {681}
  (\bibinfo {year} {1946})}\BibitemShut {NoStop}%
\bibitem [{\citenamefont {Ballarini}\ and\ \citenamefont {{De
  Liberato}}(2019)}]{Ballarini2019}%
  \BibitemOpen
  \bibfield  {author} {\bibinfo {author} {\bibfnamefont {D.}~\bibnamefont
  {Ballarini}}\ and\ \bibinfo {author} {\bibfnamefont {S.}~\bibnamefont {{De
  Liberato}}},\ }\href@noop {} {\bibfield  {journal} {\bibinfo  {journal}
  {Nanophotonics}\ }\textbf {\bibinfo {volume} {8}},\ \bibinfo {pages} {641}
  (\bibinfo {year} {2019})}\BibitemShut {NoStop}%
\bibitem [{\citenamefont {Tan}\ and\ \citenamefont {Mohseni}(2018)}]{Tan2018}%
  \BibitemOpen
  \bibfield  {author} {\bibinfo {author} {\bibfnamefont {C.~L.}\ \bibnamefont
  {Tan}}\ and\ \bibinfo {author} {\bibfnamefont {H.}~\bibnamefont {Mohseni}},\
  }\href {\doibase 10.1515/nanoph-2017-0061} {\bibfield  {journal} {\bibinfo
  {journal} {Nanophotonics}\ }\textbf {\bibinfo {volume} {7}},\ \bibinfo
  {pages} {169} (\bibinfo {year} {2018})}\BibitemShut {NoStop}%
\bibitem [{\citenamefont {Baranov}\ \emph {et~al.}(2019)\citenamefont
  {Baranov}, \citenamefont {Xiao}, \citenamefont {Nechepurenko}, \citenamefont
  {Krasnok}, \citenamefont {Al{\`{u}}},\ and\ \citenamefont
  {Kats}}]{Baranov2019}%
  \BibitemOpen
  \bibfield  {author} {\bibinfo {author} {\bibfnamefont {D.~G.}\ \bibnamefont
  {Baranov}}, \bibinfo {author} {\bibfnamefont {Y.}~\bibnamefont {Xiao}},
  \bibinfo {author} {\bibfnamefont {I.~A.}\ \bibnamefont {Nechepurenko}},
  \bibinfo {author} {\bibfnamefont {A.}~\bibnamefont {Krasnok}}, \bibinfo
  {author} {\bibfnamefont {A.}~\bibnamefont {Al{\`{u}}}}, \ and\ \bibinfo
  {author} {\bibfnamefont {M.~A.}\ \bibnamefont {Kats}},\ }\href {\doibase
  10.1038/s41563-019-0363-y} {\bibfield  {journal} {\bibinfo  {journal} {Nature
  Materials}\ }\textbf {\bibinfo {volume} {18}},\ \bibinfo {pages} {920}
  (\bibinfo {year} {2019})}\BibitemShut {NoStop}%
\bibitem [{\citenamefont {Kim}\ \emph {et~al.}(2021)\citenamefont {Kim},
  \citenamefont {Martins}, \citenamefont {Jang}, \citenamefont {Badloe},
  \citenamefont {Khadir}, \citenamefont {Jung}, \citenamefont {Kim},
  \citenamefont {Kim}, \citenamefont {Genevet},\ and\ \citenamefont
  {Rho}}]{Kim2021}%
  \BibitemOpen
  \bibfield  {author} {\bibinfo {author} {\bibfnamefont {I.}~\bibnamefont
  {Kim}}, \bibinfo {author} {\bibfnamefont {R.~J.}\ \bibnamefont {Martins}},
  \bibinfo {author} {\bibfnamefont {J.}~\bibnamefont {Jang}}, \bibinfo {author}
  {\bibfnamefont {T.}~\bibnamefont {Badloe}}, \bibinfo {author} {\bibfnamefont
  {S.}~\bibnamefont {Khadir}}, \bibinfo {author} {\bibfnamefont {H.-Y.}\
  \bibnamefont {Jung}}, \bibinfo {author} {\bibfnamefont {H.}~\bibnamefont
  {Kim}}, \bibinfo {author} {\bibfnamefont {J.}~\bibnamefont {Kim}}, \bibinfo
  {author} {\bibfnamefont {P.}~\bibnamefont {Genevet}}, \ and\ \bibinfo
  {author} {\bibfnamefont {J.}~\bibnamefont {Rho}},\ }\href {\doibase
  10.1038/s41565-021-00895-3} {\bibfield  {journal} {\bibinfo  {journal}
  {Nature Nanotechnology}\ }\textbf {\bibinfo {volume} {16}},\ \bibinfo {pages}
  {508} (\bibinfo {year} {2021})}\BibitemShut {NoStop}%
\bibitem [{\citenamefont {Stewart}\ \emph {et~al.}(2021)\citenamefont
  {Stewart}, \citenamefont {Wilson},\ and\ \citenamefont
  {Mikkelsen}}]{Stewart2021}%
  \BibitemOpen
  \bibfield  {author} {\bibinfo {author} {\bibfnamefont {J.~W.}\ \bibnamefont
  {Stewart}}, \bibinfo {author} {\bibfnamefont {N.~C.}\ \bibnamefont {Wilson}},
  \ and\ \bibinfo {author} {\bibfnamefont {M.~H.}\ \bibnamefont {Mikkelsen}},\
  }\href {\doibase 10.1021/acsphotonics.0c01068} {\bibfield  {journal}
  {\bibinfo  {journal} {ACS Photonics}\ }\textbf {\bibinfo {volume} {8}},\
  \bibinfo {pages} {71} (\bibinfo {year} {2021})}\BibitemShut {NoStop}%
\bibitem [{\citenamefont {Rogalski}(2012)}]{Rogalski2012}%
  \BibitemOpen
  \bibfield  {author} {\bibinfo {author} {\bibfnamefont {A.}~\bibnamefont
  {Rogalski}},\ }\href {\doibase 10.2478/s11772} {\bibfield  {journal}
  {\bibinfo  {journal} {Opto-Electronics Review}\ }\textbf {\bibinfo {volume}
  {20}},\ \bibinfo {pages} {279} (\bibinfo {year} {2012})}\BibitemShut
  {NoStop}%
\bibitem [{\citenamefont {Maier}(2007)}]{Maier2007}%
  \BibitemOpen
  \bibfield  {author} {\bibinfo {author} {\bibfnamefont {S.~A.}\ \bibnamefont
  {Maier}},\ }\href@noop {} {\emph {\bibinfo {title} {{Plasmonics: Fundamentals
  and Applications}}}}\ (\bibinfo  {publisher} {Springer},\ \bibinfo {address}
  {Berlin},\ \bibinfo {year} {2007})\BibitemShut {NoStop}%
\bibitem [{\citenamefont {Naik}\ \emph {et~al.}(2013)\citenamefont {Naik},
  \citenamefont {Shalaev},\ and\ \citenamefont {Boltasseva}}]{Naik2013}%
  \BibitemOpen
  \bibfield  {author} {\bibinfo {author} {\bibfnamefont {G.~V.}\ \bibnamefont
  {Naik}}, \bibinfo {author} {\bibfnamefont {V.~M.}\ \bibnamefont {Shalaev}}, \
  and\ \bibinfo {author} {\bibfnamefont {A.}~\bibnamefont {Boltasseva}},\
  }\href@noop {} {\bibfield  {journal} {\bibinfo  {journal} {Advanced
  Materials}\ }\textbf {\bibinfo {volume} {25}},\ \bibinfo {pages} {3264}
  (\bibinfo {year} {2013})}\BibitemShut {NoStop}%
\bibitem [{\citenamefont {Taliercio}\ and\ \citenamefont
  {Biagioni}(2019)}]{Taliercio2019}%
  \BibitemOpen
  \bibfield  {author} {\bibinfo {author} {\bibfnamefont {T.}~\bibnamefont
  {Taliercio}}\ and\ \bibinfo {author} {\bibfnamefont {P.}~\bibnamefont
  {Biagioni}},\ }\href {\doibase 10.1515/nanoph-2019-0077} {\bibfield
  {journal} {\bibinfo  {journal} {Nanophotonics}\ }\textbf {\bibinfo {volume}
  {8}},\ \bibinfo {pages} {949} (\bibinfo {year} {2019})}\BibitemShut {NoStop}%
\bibitem [{\citenamefont {Tredicucci}\ \emph {et~al.}(2000)\citenamefont
  {Tredicucci}, \citenamefont {Gmachl}, \citenamefont {Capasso}, \citenamefont
  {Hutchinson}, \citenamefont {Sivco},\ and\ \citenamefont
  {Cho}}]{Tredicucci2000}%
  \BibitemOpen
  \bibfield  {author} {\bibinfo {author} {\bibfnamefont {A.}~\bibnamefont
  {Tredicucci}}, \bibinfo {author} {\bibfnamefont {C.}~\bibnamefont {Gmachl}},
  \bibinfo {author} {\bibfnamefont {F.}~\bibnamefont {Capasso}}, \bibinfo
  {author} {\bibfnamefont {A.~L.}\ \bibnamefont {Hutchinson}}, \bibinfo
  {author} {\bibfnamefont {D.~L.}\ \bibnamefont {Sivco}}, \ and\ \bibinfo
  {author} {\bibfnamefont {A.~Y.}\ \bibnamefont {Cho}},\ }\href {\doibase
  10.1063/1.126183} {\bibfield  {journal} {\bibinfo  {journal} {Applied Physics
  Letters}\ }\textbf {\bibinfo {volume} {76}},\ \bibinfo {pages} {2164}
  (\bibinfo {year} {2000})}\BibitemShut {NoStop}%
\bibitem [{\citenamefont {Azzam}\ \emph {et~al.}(2020)\citenamefont {Azzam},
  \citenamefont {Kildishev}, \citenamefont {Ma}, \citenamefont {Ning},
  \citenamefont {Oulton}, \citenamefont {Shalaev}, \citenamefont {Stockman},
  \citenamefont {Xu},\ and\ \citenamefont {Zhang}}]{Azzam2020}%
  \BibitemOpen
  \bibfield  {author} {\bibinfo {author} {\bibfnamefont {S.~I.}\ \bibnamefont
  {Azzam}}, \bibinfo {author} {\bibfnamefont {A.~V.}\ \bibnamefont
  {Kildishev}}, \bibinfo {author} {\bibfnamefont {R.-M.}\ \bibnamefont {Ma}},
  \bibinfo {author} {\bibfnamefont {C.-Z.}\ \bibnamefont {Ning}}, \bibinfo
  {author} {\bibfnamefont {R.}~\bibnamefont {Oulton}}, \bibinfo {author}
  {\bibfnamefont {V.~M.}\ \bibnamefont {Shalaev}}, \bibinfo {author}
  {\bibfnamefont {M.~I.}\ \bibnamefont {Stockman}}, \bibinfo {author}
  {\bibfnamefont {J.-L.}\ \bibnamefont {Xu}}, \ and\ \bibinfo {author}
  {\bibfnamefont {X.}~\bibnamefont {Zhang}},\ }\href {\doibase
  10.1038/s41377-020-0319-7} {\bibfield  {journal} {\bibinfo  {journal} {Light:
  Science \& Applications}\ }\textbf {\bibinfo {volume} {9}},\ \bibinfo {pages}
  {90} (\bibinfo {year} {2020})}\BibitemShut {NoStop}%
\bibitem [{\citenamefont {K{\"o}hler}\ \emph {et~al.}(2002)\citenamefont
  {K{\"o}hler}, \citenamefont {Tredicucci}, \citenamefont {Beltram},
  \citenamefont {Beere}, \citenamefont {Linfield}, \citenamefont {Davies},
  \citenamefont {Ritchie}, \citenamefont {Iotti},\ and\ \citenamefont
  {F.}}]{Kohler2002}%
  \BibitemOpen
  \bibfield  {author} {\bibinfo {author} {\bibfnamefont {R.}~\bibnamefont
  {K{\"o}hler}}, \bibinfo {author} {\bibfnamefont {A.}~\bibnamefont
  {Tredicucci}}, \bibinfo {author} {\bibfnamefont {F.}~\bibnamefont {Beltram}},
  \bibinfo {author} {\bibfnamefont {H.~E.}\ \bibnamefont {Beere}}, \bibinfo
  {author} {\bibfnamefont {E.~H.}\ \bibnamefont {Linfield}}, \bibinfo {author}
  {\bibfnamefont {A.~G.}\ \bibnamefont {Davies}}, \bibinfo {author}
  {\bibfnamefont {D.~A.}\ \bibnamefont {Ritchie}}, \bibinfo {author}
  {\bibfnamefont {R.~C.}\ \bibnamefont {Iotti}}, \ and\ \bibinfo {author}
  {\bibfnamefont {R.}~\bibnamefont {F.}},\ }\href@noop {} {\bibfield  {journal}
  {\bibinfo  {journal} {Nature}\ }\textbf {\bibinfo {volume} {417}},\ \bibinfo
  {pages} {156} (\bibinfo {year} {2002})}\BibitemShut {NoStop}%
\bibitem [{\citenamefont {Sirtori}\ \emph {et~al.}(2013)\citenamefont
  {Sirtori}, \citenamefont {Barbieri},\ and\ \citenamefont
  {Colombelli}}]{Sirtori2013}%
  \BibitemOpen
  \bibfield  {author} {\bibinfo {author} {\bibfnamefont {C.}~\bibnamefont
  {Sirtori}}, \bibinfo {author} {\bibfnamefont {S.}~\bibnamefont {Barbieri}}, \
  and\ \bibinfo {author} {\bibfnamefont {R.}~\bibnamefont {Colombelli}},\
  }\href {\doibase 10.1038/nphoton.2013.208 LB - DFB;Photonics;QCL} {\bibfield
  {journal} {\bibinfo  {journal} {Nature Photonics}\ }\textbf {\bibinfo
  {volume} {7}},\ \bibinfo {pages} {691} (\bibinfo {year} {2013})}\BibitemShut
  {NoStop}%
\bibitem [{\citenamefont {Briggs}\ \emph {et~al.}(2020)\citenamefont {Briggs},
  \citenamefont {Nordin}, \citenamefont {Muhowski}, \citenamefont {Simmons},
  \citenamefont {Dhingra}, \citenamefont {Lee}, \citenamefont {Podolskiy},
  \citenamefont {Wasserman},\ and\ \citenamefont {Bank}}]{Briggs2020}%
  \BibitemOpen
  \bibfield  {author} {\bibinfo {author} {\bibfnamefont {A.~F.}\ \bibnamefont
  {Briggs}}, \bibinfo {author} {\bibfnamefont {L.}~\bibnamefont {Nordin}},
  \bibinfo {author} {\bibfnamefont {A.~J.}\ \bibnamefont {Muhowski}}, \bibinfo
  {author} {\bibfnamefont {E.}~\bibnamefont {Simmons}}, \bibinfo {author}
  {\bibfnamefont {P.}~\bibnamefont {Dhingra}}, \bibinfo {author} {\bibfnamefont
  {M.~L.}\ \bibnamefont {Lee}}, \bibinfo {author} {\bibfnamefont {V.~A.}\
  \bibnamefont {Podolskiy}}, \bibinfo {author} {\bibfnamefont {D.}~\bibnamefont
  {Wasserman}}, \ and\ \bibinfo {author} {\bibfnamefont {S.~R.}\ \bibnamefont
  {Bank}},\ }\href {\doibase 10.1364/OPTICA.402208} {\bibfield  {journal}
  {\bibinfo  {journal} {Optica}\ }\textbf {\bibinfo {volume} {7}},\ \bibinfo
  {pages} {1355} (\bibinfo {year} {2020})}\BibitemShut {NoStop}%
\bibitem [{\citenamefont {Nordin}\ \emph {et~al.}(2020)\citenamefont {Nordin},
  \citenamefont {Li}, \citenamefont {Briggs}, \citenamefont {Simmons},
  \citenamefont {Bank}, \citenamefont {Podolskiy},\ and\ \citenamefont
  {Wasserman}}]{Nordin2020}%
  \BibitemOpen
  \bibfield  {author} {\bibinfo {author} {\bibfnamefont {L.}~\bibnamefont
  {Nordin}}, \bibinfo {author} {\bibfnamefont {K.}~\bibnamefont {Li}}, \bibinfo
  {author} {\bibfnamefont {A.}~\bibnamefont {Briggs}}, \bibinfo {author}
  {\bibfnamefont {E.}~\bibnamefont {Simmons}}, \bibinfo {author} {\bibfnamefont
  {S.~R.}\ \bibnamefont {Bank}}, \bibinfo {author} {\bibfnamefont {V.~A.}\
  \bibnamefont {Podolskiy}}, \ and\ \bibinfo {author} {\bibfnamefont
  {D.}~\bibnamefont {Wasserman}},\ }\href {\doibase 10.1063/1.5132311}
  {\bibfield  {journal} {\bibinfo  {journal} {Applied Physics Letters}\
  }\textbf {\bibinfo {volume} {116}},\ \bibinfo {pages} {21102} (\bibinfo
  {year} {2020})}\BibitemShut {NoStop}%
\bibitem [{\citenamefont {Palaferri}\ \emph {et~al.}(2018)\citenamefont
  {Palaferri}, \citenamefont {Todorov}, \citenamefont {Bigioli}, \citenamefont
  {Mottaghizadeh}, \citenamefont {Gacemi}, \citenamefont {Calabrese},
  \citenamefont {Vasanelli}, \citenamefont {Li}, \citenamefont {Davies},
  \citenamefont {Linfield}, \citenamefont {Kapsalidis}, \citenamefont {Beck},
  \citenamefont {Faist},\ and\ \citenamefont {Sirtori}}]{Palaferri2018}%
  \BibitemOpen
  \bibfield  {author} {\bibinfo {author} {\bibfnamefont {D.}~\bibnamefont
  {Palaferri}}, \bibinfo {author} {\bibfnamefont {Y.}~\bibnamefont {Todorov}},
  \bibinfo {author} {\bibfnamefont {A.}~\bibnamefont {Bigioli}}, \bibinfo
  {author} {\bibfnamefont {A.}~\bibnamefont {Mottaghizadeh}}, \bibinfo {author}
  {\bibfnamefont {D.}~\bibnamefont {Gacemi}}, \bibinfo {author} {\bibfnamefont
  {A.}~\bibnamefont {Calabrese}}, \bibinfo {author} {\bibfnamefont
  {A.}~\bibnamefont {Vasanelli}}, \bibinfo {author} {\bibfnamefont
  {L.}~\bibnamefont {Li}}, \bibinfo {author} {\bibfnamefont {A.~G.}\
  \bibnamefont {Davies}}, \bibinfo {author} {\bibfnamefont {E.~H.}\
  \bibnamefont {Linfield}}, \bibinfo {author} {\bibfnamefont {F.}~\bibnamefont
  {Kapsalidis}}, \bibinfo {author} {\bibfnamefont {M.}~\bibnamefont {Beck}},
  \bibinfo {author} {\bibfnamefont {J.}~\bibnamefont {Faist}}, \ and\ \bibinfo
  {author} {\bibfnamefont {C.}~\bibnamefont {Sirtori}},\ }\href {\doibase
  10.1038/nature25790} {\bibfield  {journal} {\bibinfo  {journal} {Nature}\
  }\textbf {\bibinfo {volume} {556}},\ \bibinfo {pages} {85} (\bibinfo {year}
  {2018})}\BibitemShut {NoStop}%
\bibitem [{\citenamefont {Goldflam}\ \emph {et~al.}(2016)\citenamefont
  {Goldflam}, \citenamefont {Kadlec}, \citenamefont {Olson}, \citenamefont
  {Klem}, \citenamefont {Hawkins}, \citenamefont {Parameswaran}, \citenamefont
  {Coon}, \citenamefont {Keeler}, \citenamefont {Fortune}, \citenamefont
  {Tauke-Pedretti}, \citenamefont {Wendt}, \citenamefont {Shaner},
  \citenamefont {Davids}, \citenamefont {Kim},\ and\ \citenamefont
  {Peters}}]{Goldflam2016}%
  \BibitemOpen
  \bibfield  {author} {\bibinfo {author} {\bibfnamefont {M.~D.}\ \bibnamefont
  {Goldflam}}, \bibinfo {author} {\bibfnamefont {E.~A.}\ \bibnamefont
  {Kadlec}}, \bibinfo {author} {\bibfnamefont {B.~V.}\ \bibnamefont {Olson}},
  \bibinfo {author} {\bibfnamefont {J.~F.}\ \bibnamefont {Klem}}, \bibinfo
  {author} {\bibfnamefont {S.~D.}\ \bibnamefont {Hawkins}}, \bibinfo {author}
  {\bibfnamefont {S.}~\bibnamefont {Parameswaran}}, \bibinfo {author}
  {\bibfnamefont {W.~T.}\ \bibnamefont {Coon}}, \bibinfo {author}
  {\bibfnamefont {G.~A.}\ \bibnamefont {Keeler}}, \bibinfo {author}
  {\bibfnamefont {T.~R.}\ \bibnamefont {Fortune}}, \bibinfo {author}
  {\bibfnamefont {A.}~\bibnamefont {Tauke-Pedretti}}, \bibinfo {author}
  {\bibfnamefont {J.~R.}\ \bibnamefont {Wendt}}, \bibinfo {author}
  {\bibfnamefont {E.~A.}\ \bibnamefont {Shaner}}, \bibinfo {author}
  {\bibfnamefont {P.~S.}\ \bibnamefont {Davids}}, \bibinfo {author}
  {\bibfnamefont {J.~K.}\ \bibnamefont {Kim}}, \ and\ \bibinfo {author}
  {\bibfnamefont {D.~W.}\ \bibnamefont {Peters}},\ }\href {\doibase
  10.1063/1.4972844} {\bibfield  {journal} {\bibinfo  {journal} {Applied
  Physics Letters}\ }\textbf {\bibinfo {volume} {109}},\ \bibinfo {pages}
  {251103} (\bibinfo {year} {2016})}\BibitemShut {NoStop}%
\bibitem [{\citenamefont {Khurgin}(2015)}]{Khurgin2015}%
  \BibitemOpen
  \bibfield  {author} {\bibinfo {author} {\bibfnamefont {J.}~\bibnamefont
  {Khurgin}},\ }\href@noop {} {\bibfield  {journal} {\bibinfo  {journal}
  {Nature Nanotechnology}\ }\textbf {\bibinfo {volume} {10}},\ \bibinfo {pages}
  {2} (\bibinfo {year} {2015})}\BibitemShut {NoStop}%
\bibitem [{\citenamefont {Piotrowska}\ \emph {et~al.}(1983)\citenamefont
  {Piotrowska}, \citenamefont {Guivarc'h},\ and\ \citenamefont
  {Pelous}}]{Piotrowska1983}%
  \BibitemOpen
  \bibfield  {author} {\bibinfo {author} {\bibfnamefont {A.}~\bibnamefont
  {Piotrowska}}, \bibinfo {author} {\bibfnamefont {A.}~\bibnamefont
  {Guivarc'h}}, \ and\ \bibinfo {author} {\bibfnamefont {G.}~\bibnamefont
  {Pelous}},\ }\href {\doibase https://doi.org/10.1016/0038-1101(83)90083-7}
  {\bibfield  {journal} {\bibinfo  {journal} {Solid-State Electronics}\
  }\textbf {\bibinfo {volume} {26}},\ \bibinfo {pages} {179} (\bibinfo {year}
  {1983})}\BibitemShut {NoStop}%
\bibitem [{\citenamefont {Khurgin}\ and\ \citenamefont
  {Sun}(2013)}]{Khurgin2014}%
  \BibitemOpen
  \bibfield  {author} {\bibinfo {author} {\bibfnamefont {J.}~\bibnamefont
  {Khurgin}}\ and\ \bibinfo {author} {\bibfnamefont {G.}~\bibnamefont {Sun}},\
  }\href@noop {} {\bibfield  {journal} {\bibinfo  {journal} {Nature Photonics}\
  }\textbf {\bibinfo {volume} {8}} (\bibinfo {year} {2013})}\BibitemShut
  {NoStop}%
\bibitem [{\citenamefont {Breslin}\ \emph {et~al.}(2021)\citenamefont
  {Breslin}, \citenamefont {Ratchford}, \citenamefont {Giles}, \citenamefont
  {Dunkelberger},\ and\ \citenamefont {Owrutsky}}]{Breslin2021}%
  \BibitemOpen
  \bibfield  {author} {\bibinfo {author} {\bibfnamefont {V.~M.}\ \bibnamefont
  {Breslin}}, \bibinfo {author} {\bibfnamefont {D.~C.}\ \bibnamefont
  {Ratchford}}, \bibinfo {author} {\bibfnamefont {A.~J.}\ \bibnamefont
  {Giles}}, \bibinfo {author} {\bibfnamefont {A.~D.}\ \bibnamefont
  {Dunkelberger}}, \ and\ \bibinfo {author} {\bibfnamefont {J.~C.}\
  \bibnamefont {Owrutsky}},\ }\href {\doibase 10.1364/OE.417405} {\bibfield
  {journal} {\bibinfo  {journal} {Optics Express}\ }\textbf {\bibinfo {volume}
  {29}},\ \bibinfo {pages} {11760} (\bibinfo {year} {2021})}\BibitemShut
  {NoStop}%
\bibitem [{\citenamefont {Caldwell}\ \emph {et~al.}(2014)\citenamefont
  {Caldwell}, \citenamefont {Kretinin}, \citenamefont {Chen}, \citenamefont
  {Giannini}, \citenamefont {Fogler}, \citenamefont {Francescato},
  \citenamefont {Ellis}, \citenamefont {Tischler}, \citenamefont {Woods},
  \citenamefont {Giles}, \citenamefont {Hong}, \citenamefont {Watanabe},
  \citenamefont {Taniguchi}, \citenamefont {Maier},\ and\ \citenamefont
  {Novoselov}}]{Caldwell2014b}%
  \BibitemOpen
  \bibfield  {author} {\bibinfo {author} {\bibfnamefont {J.~D.}\ \bibnamefont
  {Caldwell}}, \bibinfo {author} {\bibfnamefont {A.~V.}\ \bibnamefont
  {Kretinin}}, \bibinfo {author} {\bibfnamefont {Y.}~\bibnamefont {Chen}},
  \bibinfo {author} {\bibfnamefont {V.}~\bibnamefont {Giannini}}, \bibinfo
  {author} {\bibfnamefont {M.~M.}\ \bibnamefont {Fogler}}, \bibinfo {author}
  {\bibfnamefont {Y.}~\bibnamefont {Francescato}}, \bibinfo {author}
  {\bibfnamefont {C.~T.}\ \bibnamefont {Ellis}}, \bibinfo {author}
  {\bibfnamefont {J.~G.}\ \bibnamefont {Tischler}}, \bibinfo {author}
  {\bibfnamefont {C.~R.}\ \bibnamefont {Woods}}, \bibinfo {author}
  {\bibfnamefont {A.~J.}\ \bibnamefont {Giles}}, \bibinfo {author}
  {\bibfnamefont {M.}~\bibnamefont {Hong}}, \bibinfo {author} {\bibfnamefont
  {K.}~\bibnamefont {Watanabe}}, \bibinfo {author} {\bibfnamefont
  {T.}~\bibnamefont {Taniguchi}}, \bibinfo {author} {\bibfnamefont {S.~A.}\
  \bibnamefont {Maier}}, \ and\ \bibinfo {author} {\bibfnamefont {K.~S.}\
  \bibnamefont {Novoselov}},\ }\href {\doibase 10.1038/ncomms6221} {\bibfield
  {journal} {\bibinfo  {journal} {Nature Communications}\ }\textbf {\bibinfo
  {volume} {5}},\ \bibinfo {pages} {5221} (\bibinfo {year} {2014})}\BibitemShut
  {NoStop}%
\bibitem [{\citenamefont {Dai}\ \emph {et~al.}(2014)\citenamefont {Dai},
  \citenamefont {Fei}, \citenamefont {Ma}, \citenamefont {Rodin}, \citenamefont
  {Wagner}, \citenamefont {McLeod}, \citenamefont {Liu}, \citenamefont
  {Gannett}, \citenamefont {Regan}, \citenamefont {Watanabe}, \citenamefont
  {Taniguchi}, \citenamefont {Thiemens}, \citenamefont {Dominguez},
  \citenamefont {Neto}, \citenamefont {Zettl}, \citenamefont {Keilmann},
  \citenamefont {Jarillo-Herrero}, \citenamefont {Fogler},\ and\ \citenamefont
  {Basov}}]{Dai2014}%
  \BibitemOpen
  \bibfield  {author} {\bibinfo {author} {\bibfnamefont {S.}~\bibnamefont
  {Dai}}, \bibinfo {author} {\bibfnamefont {Z.}~\bibnamefont {Fei}}, \bibinfo
  {author} {\bibfnamefont {Q.}~\bibnamefont {Ma}}, \bibinfo {author}
  {\bibfnamefont {A.~S.}\ \bibnamefont {Rodin}}, \bibinfo {author}
  {\bibfnamefont {M.}~\bibnamefont {Wagner}}, \bibinfo {author} {\bibfnamefont
  {A.~S.}\ \bibnamefont {McLeod}}, \bibinfo {author} {\bibfnamefont {M.~K.}\
  \bibnamefont {Liu}}, \bibinfo {author} {\bibfnamefont {W.}~\bibnamefont
  {Gannett}}, \bibinfo {author} {\bibfnamefont {W.}~\bibnamefont {Regan}},
  \bibinfo {author} {\bibfnamefont {K.}~\bibnamefont {Watanabe}}, \bibinfo
  {author} {\bibfnamefont {T.}~\bibnamefont {Taniguchi}}, \bibinfo {author}
  {\bibfnamefont {M.}~\bibnamefont {Thiemens}}, \bibinfo {author}
  {\bibfnamefont {G.}~\bibnamefont {Dominguez}}, \bibinfo {author}
  {\bibfnamefont {A.~H.~C.}\ \bibnamefont {Neto}}, \bibinfo {author}
  {\bibfnamefont {A.}~\bibnamefont {Zettl}}, \bibinfo {author} {\bibfnamefont
  {F.}~\bibnamefont {Keilmann}}, \bibinfo {author} {\bibfnamefont
  {P.}~\bibnamefont {Jarillo-Herrero}}, \bibinfo {author} {\bibfnamefont
  {M.~M.}\ \bibnamefont {Fogler}}, \ and\ \bibinfo {author} {\bibfnamefont
  {D.~N.}\ \bibnamefont {Basov}},\ }\href {\doibase 10.1126/science.1246833}
  {\bibfield  {journal} {\bibinfo  {journal} {Science}\ }\textbf {\bibinfo
  {volume} {343}},\ \bibinfo {pages} {1125} (\bibinfo {year}
  {2014})}\BibitemShut {NoStop}%
\bibitem [{\citenamefont {Dai}\ \emph {et~al.}(2015)\citenamefont {Dai},
  \citenamefont {Ma}, \citenamefont {Andersen}, \citenamefont {Mcleod},
  \citenamefont {Fei}, \citenamefont {Liu}, \citenamefont {Wagner},
  \citenamefont {Watanabe}, \citenamefont {Taniguchi}, \citenamefont
  {Thiemens}, \citenamefont {Keilmann}, \citenamefont {Jarillo-Herrero},
  \citenamefont {Fogler},\ and\ \citenamefont {Basov}}]{Dai2015}%
  \BibitemOpen
  \bibfield  {author} {\bibinfo {author} {\bibfnamefont {S.}~\bibnamefont
  {Dai}}, \bibinfo {author} {\bibfnamefont {Q.}~\bibnamefont {Ma}}, \bibinfo
  {author} {\bibfnamefont {T.}~\bibnamefont {Andersen}}, \bibinfo {author}
  {\bibfnamefont {A.~S.}\ \bibnamefont {Mcleod}}, \bibinfo {author}
  {\bibfnamefont {Z.}~\bibnamefont {Fei}}, \bibinfo {author} {\bibfnamefont
  {M.~K.}\ \bibnamefont {Liu}}, \bibinfo {author} {\bibfnamefont
  {M.}~\bibnamefont {Wagner}}, \bibinfo {author} {\bibfnamefont
  {K.}~\bibnamefont {Watanabe}}, \bibinfo {author} {\bibfnamefont
  {T.}~\bibnamefont {Taniguchi}}, \bibinfo {author} {\bibfnamefont
  {M.}~\bibnamefont {Thiemens}}, \bibinfo {author} {\bibfnamefont
  {F.}~\bibnamefont {Keilmann}}, \bibinfo {author} {\bibfnamefont
  {P.}~\bibnamefont {Jarillo-Herrero}}, \bibinfo {author} {\bibfnamefont
  {M.~M.}\ \bibnamefont {Fogler}}, \ and\ \bibinfo {author} {\bibfnamefont
  {D.~N.}\ \bibnamefont {Basov}},\ }\href {\doibase 10.1038/ncomms7963}
  {\bibfield  {journal} {\bibinfo  {journal} {Nature Communications}\ }\textbf
  {\bibinfo {volume} {6}},\ \bibinfo {pages} {6963} (\bibinfo {year}
  {2015})}\BibitemShut {NoStop}%
\bibitem [{\citenamefont {Li}\ \emph {et~al.}(2015)\citenamefont {Li},
  \citenamefont {Lewin}, \citenamefont {Kretinin}, \citenamefont {Caldwell},
  \citenamefont {Novoselov}, \citenamefont {Taniguchi}, \citenamefont
  {Watanabe}, \citenamefont {Gaussmann},\ and\ \citenamefont
  {Taubner}}]{Li2015}%
  \BibitemOpen
  \bibfield  {author} {\bibinfo {author} {\bibfnamefont {P.}~\bibnamefont
  {Li}}, \bibinfo {author} {\bibfnamefont {M.}~\bibnamefont {Lewin}}, \bibinfo
  {author} {\bibfnamefont {A.~V.}\ \bibnamefont {Kretinin}}, \bibinfo {author}
  {\bibfnamefont {J.~D.}\ \bibnamefont {Caldwell}}, \bibinfo {author}
  {\bibfnamefont {K.~S.}\ \bibnamefont {Novoselov}}, \bibinfo {author}
  {\bibfnamefont {T.}~\bibnamefont {Taniguchi}}, \bibinfo {author}
  {\bibfnamefont {K.}~\bibnamefont {Watanabe}}, \bibinfo {author}
  {\bibfnamefont {F.}~\bibnamefont {Gaussmann}}, \ and\ \bibinfo {author}
  {\bibfnamefont {T.}~\bibnamefont {Taubner}},\ }\href@noop {} {\bibfield
  {journal} {\bibinfo  {journal} {Nature Communications}\ }\textbf {\bibinfo
  {volume} {6}},\ \bibinfo {pages} {7507} (\bibinfo {year} {2015})}\BibitemShut
  {NoStop}%
\bibitem [{\citenamefont {Hartstein}\ \emph {et~al.}(1975)\citenamefont
  {Hartstein}, \citenamefont {Burstein}, \citenamefont {Palik}, \citenamefont
  {Gammon},\ and\ \citenamefont {Henvis}}]{Hartstein1975}%
  \BibitemOpen
  \bibfield  {author} {\bibinfo {author} {\bibfnamefont {A.}~\bibnamefont
  {Hartstein}}, \bibinfo {author} {\bibfnamefont {E.}~\bibnamefont {Burstein}},
  \bibinfo {author} {\bibfnamefont {E.~D.}\ \bibnamefont {Palik}}, \bibinfo
  {author} {\bibfnamefont {R.~W.}\ \bibnamefont {Gammon}}, \ and\ \bibinfo
  {author} {\bibfnamefont {B.~W.}\ \bibnamefont {Henvis}},\ }\href {\doibase
  10.1103/PhysRevB.12.3186} {\bibfield  {journal} {\bibinfo  {journal}
  {Physical Review B}\ }\textbf {\bibinfo {volume} {12}},\ \bibinfo {pages}
  {3186} (\bibinfo {year} {1975})}\BibitemShut {NoStop}%
\bibitem [{\citenamefont {Caldwell}\ \emph {et~al.}(2013)\citenamefont
  {Caldwell}, \citenamefont {Glembocki}, \citenamefont {Francescato},
  \citenamefont {Sharac}, \citenamefont {Giannini}, \citenamefont {Bezares},
  \citenamefont {Long}, \citenamefont {Owrutsky}, \citenamefont {Vurgaftman},
  \citenamefont {Tischler}, \citenamefont {Wheeler}, \citenamefont {Bassim},
  \citenamefont {Shirey}, \citenamefont {Kasica},\ and\ \citenamefont
  {Maier}}]{Caldwell2013}%
  \BibitemOpen
  \bibfield  {author} {\bibinfo {author} {\bibfnamefont {J.~D.}\ \bibnamefont
  {Caldwell}}, \bibinfo {author} {\bibfnamefont {O.~J.}\ \bibnamefont
  {Glembocki}}, \bibinfo {author} {\bibfnamefont {Y.}~\bibnamefont
  {Francescato}}, \bibinfo {author} {\bibfnamefont {N.}~\bibnamefont {Sharac}},
  \bibinfo {author} {\bibfnamefont {V.}~\bibnamefont {Giannini}}, \bibinfo
  {author} {\bibfnamefont {F.~J.}\ \bibnamefont {Bezares}}, \bibinfo {author}
  {\bibfnamefont {J.~P.}\ \bibnamefont {Long}}, \bibinfo {author}
  {\bibfnamefont {J.~C.}\ \bibnamefont {Owrutsky}}, \bibinfo {author}
  {\bibfnamefont {I.}~\bibnamefont {Vurgaftman}}, \bibinfo {author}
  {\bibfnamefont {J.~G.}\ \bibnamefont {Tischler}}, \bibinfo {author}
  {\bibfnamefont {V.~D.}\ \bibnamefont {Wheeler}}, \bibinfo {author}
  {\bibfnamefont {N.~D.}\ \bibnamefont {Bassim}}, \bibinfo {author}
  {\bibfnamefont {L.~M.}\ \bibnamefont {Shirey}}, \bibinfo {author}
  {\bibfnamefont {R.}~\bibnamefont {Kasica}}, \ and\ \bibinfo {author}
  {\bibfnamefont {S.~A.}\ \bibnamefont {Maier}},\ }\href {\doibase
  10.1021/nl401590g} {\bibfield  {journal} {\bibinfo  {journal} {Nano Letters}\
  }\textbf {\bibinfo {volume} {13}},\ \bibinfo {pages} {3690} (\bibinfo {year}
  {2013})},\ \bibinfo {note} {pMID: 23815389}\BibitemShut {NoStop}%
\bibitem [{\citenamefont {Chen}\ \emph {et~al.}(2014)\citenamefont {Chen},
  \citenamefont {Francescato}, \citenamefont {Caldwell}, \citenamefont
  {Giannini}, \citenamefont {Maß}, \citenamefont {Glembocki}, \citenamefont
  {Bezares}, \citenamefont {Taubner}, \citenamefont {Kasica}, \citenamefont
  {Hong},\ and\ \citenamefont {Maier}}]{Chen2014}%
  \BibitemOpen
  \bibfield  {author} {\bibinfo {author} {\bibfnamefont {Y.}~\bibnamefont
  {Chen}}, \bibinfo {author} {\bibfnamefont {Y.}~\bibnamefont {Francescato}},
  \bibinfo {author} {\bibfnamefont {J.~D.}\ \bibnamefont {Caldwell}}, \bibinfo
  {author} {\bibfnamefont {V.}~\bibnamefont {Giannini}}, \bibinfo {author}
  {\bibfnamefont {T.~W.~W.}\ \bibnamefont {Maß}}, \bibinfo {author}
  {\bibfnamefont {O.~J.}\ \bibnamefont {Glembocki}}, \bibinfo {author}
  {\bibfnamefont {F.~J.}\ \bibnamefont {Bezares}}, \bibinfo {author}
  {\bibfnamefont {T.}~\bibnamefont {Taubner}}, \bibinfo {author} {\bibfnamefont
  {R.}~\bibnamefont {Kasica}}, \bibinfo {author} {\bibfnamefont
  {M.}~\bibnamefont {Hong}}, \ and\ \bibinfo {author} {\bibfnamefont {S.~A.}\
  \bibnamefont {Maier}},\ }\href {\doibase 10.1021/ph500143u} {\bibfield
  {journal} {\bibinfo  {journal} {ACS Photonics}\ }\textbf {\bibinfo {volume}
  {1}},\ \bibinfo {pages} {718} (\bibinfo {year} {2014})}\BibitemShut {NoStop}%
\bibitem [{\citenamefont {Wang}\ \emph {et~al.}(2013)\citenamefont {Wang},
  \citenamefont {Li}, \citenamefont {Hauer}, \citenamefont {Chigrin},\ and\
  \citenamefont {Taubner}}]{Wang2013}%
  \BibitemOpen
  \bibfield  {author} {\bibinfo {author} {\bibfnamefont {T.}~\bibnamefont
  {Wang}}, \bibinfo {author} {\bibfnamefont {P.}~\bibnamefont {Li}}, \bibinfo
  {author} {\bibfnamefont {B.}~\bibnamefont {Hauer}}, \bibinfo {author}
  {\bibfnamefont {D.~N.}\ \bibnamefont {Chigrin}}, \ and\ \bibinfo {author}
  {\bibfnamefont {T.}~\bibnamefont {Taubner}},\ }\href {\doibase
  10.1021/nl4020342} {\bibfield  {journal} {\bibinfo  {journal} {Nano Letters}\
  }\textbf {\bibinfo {volume} {13}},\ \bibinfo {pages} {5051} (\bibinfo {year}
  {2013})}\BibitemShut {NoStop}%
\bibitem [{\citenamefont {Feng}\ \emph
  {et~al.}(2015{\natexlab{a}})\citenamefont {Feng}, \citenamefont {Streyer},
  \citenamefont {Islam}, \citenamefont {Verma}, \citenamefont {Jena},
  \citenamefont {Wasserman},\ and\ \citenamefont {Hoffman}}]{Feng2015}%
  \BibitemOpen
  \bibfield  {author} {\bibinfo {author} {\bibfnamefont {K.}~\bibnamefont
  {Feng}}, \bibinfo {author} {\bibfnamefont {W.}~\bibnamefont {Streyer}},
  \bibinfo {author} {\bibfnamefont {S.~M.}\ \bibnamefont {Islam}}, \bibinfo
  {author} {\bibfnamefont {J.}~\bibnamefont {Verma}}, \bibinfo {author}
  {\bibfnamefont {D.}~\bibnamefont {Jena}}, \bibinfo {author} {\bibfnamefont
  {D.}~\bibnamefont {Wasserman}}, \ and\ \bibinfo {author} {\bibfnamefont
  {A.~J.}\ \bibnamefont {Hoffman}},\ }\href {\doibase 10.1063/1.4929502}
  {\bibfield  {journal} {\bibinfo  {journal} {Applied Physics Letters}\
  }\textbf {\bibinfo {volume} {107}},\ \bibinfo {pages} {081108} (\bibinfo
  {year} {2015}{\natexlab{a}})}\BibitemShut {NoStop}%
\bibitem [{\citenamefont {Hafeli}\ \emph {et~al.}(2011)\citenamefont {Hafeli},
  \citenamefont {Rephaeli}, \citenamefont {Fan}, \citenamefont {Cahill},\ and\
  \citenamefont {Tiwald}}]{Hafeli2011}%
  \BibitemOpen
  \bibfield  {author} {\bibinfo {author} {\bibfnamefont {A.~K.}\ \bibnamefont
  {Hafeli}}, \bibinfo {author} {\bibfnamefont {E.}~\bibnamefont {Rephaeli}},
  \bibinfo {author} {\bibfnamefont {S.}~\bibnamefont {Fan}}, \bibinfo {author}
  {\bibfnamefont {D.~G.}\ \bibnamefont {Cahill}}, \ and\ \bibinfo {author}
  {\bibfnamefont {T.~E.}\ \bibnamefont {Tiwald}},\ }\href {\doibase
  10.1063/1.3624603} {\bibfield  {journal} {\bibinfo  {journal} {Journal of
  Applied Physics}\ }\textbf {\bibinfo {volume} {110}},\ \bibinfo {pages}
  {043517} (\bibinfo {year} {2011})}\BibitemShut {NoStop}%
\bibitem [{\citenamefont {Berte}\ \emph {et~al.}(2018)\citenamefont {Berte},
  \citenamefont {Gubbin}, \citenamefont {Wheeler}, \citenamefont {Giles},
  \citenamefont {Giannini}, \citenamefont {Maier}, \citenamefont
  {De~Liberato},\ and\ \citenamefont {Caldwell}}]{Berte2018}%
  \BibitemOpen
  \bibfield  {author} {\bibinfo {author} {\bibfnamefont {R.}~\bibnamefont
  {Berte}}, \bibinfo {author} {\bibfnamefont {C.~R.}\ \bibnamefont {Gubbin}},
  \bibinfo {author} {\bibfnamefont {V.~D.}\ \bibnamefont {Wheeler}}, \bibinfo
  {author} {\bibfnamefont {A.~J.}\ \bibnamefont {Giles}}, \bibinfo {author}
  {\bibfnamefont {V.}~\bibnamefont {Giannini}}, \bibinfo {author}
  {\bibfnamefont {S.~A.}\ \bibnamefont {Maier}}, \bibinfo {author}
  {\bibfnamefont {S.}~\bibnamefont {De~Liberato}}, \ and\ \bibinfo {author}
  {\bibfnamefont {J.~D.}\ \bibnamefont {Caldwell}},\ }\href {\doibase
  10.1021/acsphotonics.7b01482} {\bibfield  {journal} {\bibinfo  {journal} {ACS
  Photonics}\ }\textbf {\bibinfo {volume} {5}},\ \bibinfo {pages} {2807}
  (\bibinfo {year} {2018})}\BibitemShut {NoStop}%
\bibitem [{\citenamefont {Vassant}\ \emph {et~al.}(2010)\citenamefont
  {Vassant}, \citenamefont {Marquier}, \citenamefont {Greffet}, \citenamefont
  {Pardo},\ and\ \citenamefont {Pelouard}}]{Vassant2010}%
  \BibitemOpen
  \bibfield  {author} {\bibinfo {author} {\bibfnamefont {S.}~\bibnamefont
  {Vassant}}, \bibinfo {author} {\bibfnamefont {F.}~\bibnamefont {Marquier}},
  \bibinfo {author} {\bibfnamefont {J.~J.}\ \bibnamefont {Greffet}}, \bibinfo
  {author} {\bibfnamefont {F.}~\bibnamefont {Pardo}}, \ and\ \bibinfo {author}
  {\bibfnamefont {J.~L.}\ \bibnamefont {Pelouard}},\ }\href {\doibase
  10.1063/1.3497645} {\bibfield  {journal} {\bibinfo  {journal} {Applied
  Physics Letters}\ }\textbf {\bibinfo {volume} {97}},\ \bibinfo {pages}
  {161101} (\bibinfo {year} {2010})}\BibitemShut {NoStop}%
\bibitem [{\citenamefont {Caldwell}\ \emph {et~al.}(2015)\citenamefont
  {Caldwell}, \citenamefont {Lindsay}, \citenamefont {Giannini}, \citenamefont
  {Vurgaftman}, \citenamefont {Reinecke}, \citenamefont {Maier},\ and\
  \citenamefont {Glembocki}}]{Caldwell2015}%
  \BibitemOpen
  \bibfield  {author} {\bibinfo {author} {\bibfnamefont {J.~D.}\ \bibnamefont
  {Caldwell}}, \bibinfo {author} {\bibfnamefont {L.}~\bibnamefont {Lindsay}},
  \bibinfo {author} {\bibfnamefont {V.}~\bibnamefont {Giannini}}, \bibinfo
  {author} {\bibfnamefont {I.}~\bibnamefont {Vurgaftman}}, \bibinfo {author}
  {\bibfnamefont {T.~L.}\ \bibnamefont {Reinecke}}, \bibinfo {author}
  {\bibfnamefont {S.~A.}\ \bibnamefont {Maier}}, \ and\ \bibinfo {author}
  {\bibfnamefont {O.~J.}\ \bibnamefont {Glembocki}},\ }\href {\doibase
  doi:10.1515/nanoph-2014-0003} {\bibfield  {journal} {\bibinfo  {journal}
  {Nanophotonics}\ }\textbf {\bibinfo {volume} {4}},\ \bibinfo {pages} {44}
  (\bibinfo {year} {2015})}\BibitemShut {NoStop}%
\bibitem [{\citenamefont {Caldwell}\ \emph {et~al.}(2016)\citenamefont
  {Caldwell}, \citenamefont {Vurgaftman}, \citenamefont {Tischler},
  \citenamefont {Glembocki}, \citenamefont {Owrutsky},\ and\ \citenamefont
  {Reinecke}}]{Caldwell2016}%
  \BibitemOpen
  \bibfield  {author} {\bibinfo {author} {\bibfnamefont {J.~D.}\ \bibnamefont
  {Caldwell}}, \bibinfo {author} {\bibfnamefont {I.}~\bibnamefont
  {Vurgaftman}}, \bibinfo {author} {\bibfnamefont {J.~G.}\ \bibnamefont
  {Tischler}}, \bibinfo {author} {\bibfnamefont {O.~J.}\ \bibnamefont
  {Glembocki}}, \bibinfo {author} {\bibfnamefont {J.~C.}\ \bibnamefont
  {Owrutsky}}, \ and\ \bibinfo {author} {\bibfnamefont {T.~L.}\ \bibnamefont
  {Reinecke}},\ }\href {\doibase 10.1038/nnano.2015.305} {\bibfield  {journal}
  {\bibinfo  {journal} {Nature Nanotechnology}\ }\textbf {\bibinfo {volume}
  {11}},\ \bibinfo {pages} {9} (\bibinfo {year} {2016})}\BibitemShut {NoStop}%
\bibitem [{\citenamefont {Ratchford}\ \emph {et~al.}(2019)\citenamefont
  {Ratchford}, \citenamefont {Winta}, \citenamefont {Chatzakis}, \citenamefont
  {Ellis}, \citenamefont {Passler}, \citenamefont {Winterstein}, \citenamefont
  {Dev}, \citenamefont {Razdolski}, \citenamefont {Matson}, \citenamefont
  {Nolen}, \citenamefont {Tischler}, \citenamefont {Vurgaftman}, \citenamefont
  {Katz}, \citenamefont {Nepal}, \citenamefont {Hardy}, \citenamefont
  {Hachtel}, \citenamefont {Idrobo}, \citenamefont {Reinecke}, \citenamefont
  {Giles}, \citenamefont {Katzer}, \citenamefont {Bassim}, \citenamefont
  {Stroud}, \citenamefont {Wolf}, \citenamefont {Paarmann},\ and\ \citenamefont
  {Caldwell}}]{Ratchford2018}%
  \BibitemOpen
  \bibfield  {author} {\bibinfo {author} {\bibfnamefont {D.~C.}\ \bibnamefont
  {Ratchford}}, \bibinfo {author} {\bibfnamefont {C.~J.}\ \bibnamefont
  {Winta}}, \bibinfo {author} {\bibfnamefont {I.}~\bibnamefont {Chatzakis}},
  \bibinfo {author} {\bibfnamefont {C.~T.}\ \bibnamefont {Ellis}}, \bibinfo
  {author} {\bibfnamefont {N.~C.}\ \bibnamefont {Passler}}, \bibinfo {author}
  {\bibfnamefont {J.}~\bibnamefont {Winterstein}}, \bibinfo {author}
  {\bibfnamefont {P.}~\bibnamefont {Dev}}, \bibinfo {author} {\bibfnamefont
  {I.}~\bibnamefont {Razdolski}}, \bibinfo {author} {\bibfnamefont {J.~R.}\
  \bibnamefont {Matson}}, \bibinfo {author} {\bibfnamefont {J.~R.}\
  \bibnamefont {Nolen}}, \bibinfo {author} {\bibfnamefont {J.~G.}\ \bibnamefont
  {Tischler}}, \bibinfo {author} {\bibfnamefont {I.}~\bibnamefont
  {Vurgaftman}}, \bibinfo {author} {\bibfnamefont {M.~B.}\ \bibnamefont
  {Katz}}, \bibinfo {author} {\bibfnamefont {N.}~\bibnamefont {Nepal}},
  \bibinfo {author} {\bibfnamefont {M.~T.}\ \bibnamefont {Hardy}}, \bibinfo
  {author} {\bibfnamefont {J.~A.}\ \bibnamefont {Hachtel}}, \bibinfo {author}
  {\bibfnamefont {J.-C.}\ \bibnamefont {Idrobo}}, \bibinfo {author}
  {\bibfnamefont {T.~L.}\ \bibnamefont {Reinecke}}, \bibinfo {author}
  {\bibfnamefont {A.~J.}\ \bibnamefont {Giles}}, \bibinfo {author}
  {\bibfnamefont {D.~S.}\ \bibnamefont {Katzer}}, \bibinfo {author}
  {\bibfnamefont {N.~D.}\ \bibnamefont {Bassim}}, \bibinfo {author}
  {\bibfnamefont {R.~M.}\ \bibnamefont {Stroud}}, \bibinfo {author}
  {\bibfnamefont {M.}~\bibnamefont {Wolf}}, \bibinfo {author} {\bibfnamefont
  {A.}~\bibnamefont {Paarmann}}, \ and\ \bibinfo {author} {\bibfnamefont
  {J.~D.}\ \bibnamefont {Caldwell}},\ }\href {\doibase 10.1021/acsnano.9b01275}
  {\bibfield  {journal} {\bibinfo  {journal} {ACS Nano}\ }\textbf {\bibinfo
  {volume} {13}},\ \bibinfo {pages} {6730} (\bibinfo {year} {2019})},\ \bibinfo
  {note} {pMID: 31184132}\BibitemShut {NoStop}%
\bibitem [{\citenamefont {Dunkelberger}\ \emph {et~al.}(2020)\citenamefont
  {Dunkelberger}, \citenamefont {Ratchford}, \citenamefont {Grafton},
  \citenamefont {Breslin}, \citenamefont {Ryland}, \citenamefont {Katzer},
  \citenamefont {Fears}, \citenamefont {Weiblen}, \citenamefont {Vurgaftman},
  \citenamefont {Giles}, \citenamefont {Ellis}, \citenamefont {Tischler},
  \citenamefont {Caldwell},\ and\ \citenamefont {Owrutsky}}]{Dunkelberger2020}%
  \BibitemOpen
  \bibfield  {author} {\bibinfo {author} {\bibfnamefont {A.~D.}\ \bibnamefont
  {Dunkelberger}}, \bibinfo {author} {\bibfnamefont {D.~C.}\ \bibnamefont
  {Ratchford}}, \bibinfo {author} {\bibfnamefont {A.~B.}\ \bibnamefont
  {Grafton}}, \bibinfo {author} {\bibfnamefont {V.~M.}\ \bibnamefont
  {Breslin}}, \bibinfo {author} {\bibfnamefont {E.~S.}\ \bibnamefont {Ryland}},
  \bibinfo {author} {\bibfnamefont {D.~S.}\ \bibnamefont {Katzer}}, \bibinfo
  {author} {\bibfnamefont {K.~P.}\ \bibnamefont {Fears}}, \bibinfo {author}
  {\bibfnamefont {R.~J.}\ \bibnamefont {Weiblen}}, \bibinfo {author}
  {\bibfnamefont {I.}~\bibnamefont {Vurgaftman}}, \bibinfo {author}
  {\bibfnamefont {A.~J.}\ \bibnamefont {Giles}}, \bibinfo {author}
  {\bibfnamefont {C.~T.}\ \bibnamefont {Ellis}}, \bibinfo {author}
  {\bibfnamefont {J.~G.}\ \bibnamefont {Tischler}}, \bibinfo {author}
  {\bibfnamefont {J.~D.}\ \bibnamefont {Caldwell}}, \ and\ \bibinfo {author}
  {\bibfnamefont {J.~C.}\ \bibnamefont {Owrutsky}},\ }\href {\doibase
  10.1021/acsphotonics.9b01578} {\bibfield  {journal} {\bibinfo  {journal} {ACS
  Photonics}\ }\textbf {\bibinfo {volume} {7}},\ \bibinfo {pages} {279}
  (\bibinfo {year} {2020})}\BibitemShut {NoStop}%
\bibitem [{\citenamefont {Vassant}\ \emph {et~al.}(2012)\citenamefont
  {Vassant}, \citenamefont {Archambault}, \citenamefont {Marquier},
  \citenamefont {Pardo}, \citenamefont {Gennser}, \citenamefont {Cavanna},
  \citenamefont {Pelouard},\ and\ \citenamefont {Greffet}}]{Vassant2012}%
  \BibitemOpen
  \bibfield  {author} {\bibinfo {author} {\bibfnamefont {S.}~\bibnamefont
  {Vassant}}, \bibinfo {author} {\bibfnamefont {A.}~\bibnamefont
  {Archambault}}, \bibinfo {author} {\bibfnamefont {F.}~\bibnamefont
  {Marquier}}, \bibinfo {author} {\bibfnamefont {F.}~\bibnamefont {Pardo}},
  \bibinfo {author} {\bibfnamefont {U.}~\bibnamefont {Gennser}}, \bibinfo
  {author} {\bibfnamefont {A.}~\bibnamefont {Cavanna}}, \bibinfo {author}
  {\bibfnamefont {J.~L.}\ \bibnamefont {Pelouard}}, \ and\ \bibinfo {author}
  {\bibfnamefont {J.~J.}\ \bibnamefont {Greffet}},\ }\href {\doibase
  10.1103/PhysRevLett.109.237401} {\bibfield  {journal} {\bibinfo  {journal}
  {Physical Review Letters}\ }\textbf {\bibinfo {volume} {109}},\ \bibinfo
  {pages} {237401} (\bibinfo {year} {2012})}\BibitemShut {NoStop}%
\bibitem [{\citenamefont {Li}\ \emph {et~al.}(2016)\citenamefont {Li},
  \citenamefont {Yang}, \citenamefont {Ma{\ss}}, \citenamefont {Hanss},
  \citenamefont {Lewin}, \citenamefont {Michel}, \citenamefont {Wuttig},\ and\
  \citenamefont {Taubner}}]{Li2016}%
  \BibitemOpen
  \bibfield  {author} {\bibinfo {author} {\bibfnamefont {P.}~\bibnamefont
  {Li}}, \bibinfo {author} {\bibfnamefont {X.}~\bibnamefont {Yang}}, \bibinfo
  {author} {\bibfnamefont {T.~W.~W.}\ \bibnamefont {Ma{\ss}}}, \bibinfo
  {author} {\bibfnamefont {J.}~\bibnamefont {Hanss}}, \bibinfo {author}
  {\bibfnamefont {M.}~\bibnamefont {Lewin}}, \bibinfo {author} {\bibfnamefont
  {A.-K.~U.}\ \bibnamefont {Michel}}, \bibinfo {author} {\bibfnamefont
  {M.}~\bibnamefont {Wuttig}}, \ and\ \bibinfo {author} {\bibfnamefont
  {T.}~\bibnamefont {Taubner}},\ }\href {\doibase 10.1038/nmat4649} {\bibfield
  {journal} {\bibinfo  {journal} {Nature Materials}\ }\textbf {\bibinfo
  {volume} {15}},\ \bibinfo {pages} {870} (\bibinfo {year} {2016})}\BibitemShut
  {NoStop}%
\bibitem [{\citenamefont {Folland}\ \emph {et~al.}(2018)\citenamefont
  {Folland}, \citenamefont {Fali}, \citenamefont {White}, \citenamefont
  {Matson}, \citenamefont {Liu}, \citenamefont {Aghamiri}, \citenamefont
  {Edgar}, \citenamefont {Haglund}, \citenamefont {Abate},\ and\ \citenamefont
  {Caldwell}}]{Folland2018}%
  \BibitemOpen
  \bibfield  {author} {\bibinfo {author} {\bibfnamefont {T.~G.}\ \bibnamefont
  {Folland}}, \bibinfo {author} {\bibfnamefont {A.}~\bibnamefont {Fali}},
  \bibinfo {author} {\bibfnamefont {S.~T.}\ \bibnamefont {White}}, \bibinfo
  {author} {\bibfnamefont {J.~R.}\ \bibnamefont {Matson}}, \bibinfo {author}
  {\bibfnamefont {S.}~\bibnamefont {Liu}}, \bibinfo {author} {\bibfnamefont
  {N.~A.}\ \bibnamefont {Aghamiri}}, \bibinfo {author} {\bibfnamefont {J.~H.}\
  \bibnamefont {Edgar}}, \bibinfo {author} {\bibfnamefont {R.~F.}\ \bibnamefont
  {Haglund}}, \bibinfo {author} {\bibfnamefont {Y.}~\bibnamefont {Abate}}, \
  and\ \bibinfo {author} {\bibfnamefont {J.~D.}\ \bibnamefont {Caldwell}},\
  }\href {\doibase 10.1038/s41467-018-06858-y} {\bibfield  {journal} {\bibinfo
  {journal} {Nature Communications}\ }\textbf {\bibinfo {volume} {9}},\
  \bibinfo {pages} {4371} (\bibinfo {year} {2018})}\BibitemShut {NoStop}%
\bibitem [{\citenamefont {Chaudhary}\ \emph {et~al.}(2019)\citenamefont
  {Chaudhary}, \citenamefont {Tamagnone}, \citenamefont {Yin}, \citenamefont
  {Sp{\"{a}}gele}, \citenamefont {Oscurato}, \citenamefont {Li}, \citenamefont
  {Persch}, \citenamefont {Li}, \citenamefont {Rubin}, \citenamefont
  {Jauregui}, \citenamefont {Watanabe}, \citenamefont {Taniguchi},
  \citenamefont {Kim}, \citenamefont {Wuttig}, \citenamefont {Edgar},
  \citenamefont {Ambrosio},\ and\ \citenamefont {Capasso}}]{Chaudhary2019}%
  \BibitemOpen
  \bibfield  {author} {\bibinfo {author} {\bibfnamefont {K.}~\bibnamefont
  {Chaudhary}}, \bibinfo {author} {\bibfnamefont {M.}~\bibnamefont
  {Tamagnone}}, \bibinfo {author} {\bibfnamefont {X.}~\bibnamefont {Yin}},
  \bibinfo {author} {\bibfnamefont {C.~M.}\ \bibnamefont {Sp{\"{a}}gele}},
  \bibinfo {author} {\bibfnamefont {S.~L.}\ \bibnamefont {Oscurato}}, \bibinfo
  {author} {\bibfnamefont {J.}~\bibnamefont {Li}}, \bibinfo {author}
  {\bibfnamefont {C.}~\bibnamefont {Persch}}, \bibinfo {author} {\bibfnamefont
  {R.}~\bibnamefont {Li}}, \bibinfo {author} {\bibfnamefont {N.~A.}\
  \bibnamefont {Rubin}}, \bibinfo {author} {\bibfnamefont {L.~A.}\ \bibnamefont
  {Jauregui}}, \bibinfo {author} {\bibfnamefont {K.}~\bibnamefont {Watanabe}},
  \bibinfo {author} {\bibfnamefont {T.}~\bibnamefont {Taniguchi}}, \bibinfo
  {author} {\bibfnamefont {P.}~\bibnamefont {Kim}}, \bibinfo {author}
  {\bibfnamefont {M.}~\bibnamefont {Wuttig}}, \bibinfo {author} {\bibfnamefont
  {J.~H.}\ \bibnamefont {Edgar}}, \bibinfo {author} {\bibfnamefont
  {A.}~\bibnamefont {Ambrosio}}, \ and\ \bibinfo {author} {\bibfnamefont
  {F.}~\bibnamefont {Capasso}},\ }\href {\doibase 10.1038/s41467-019-12439-4}
  {\bibfield  {journal} {\bibinfo  {journal} {Nature Communications}\ }\textbf
  {\bibinfo {volume} {10}},\ \bibinfo {pages} {4487} (\bibinfo {year}
  {2019})}\BibitemShut {NoStop}%
\bibitem [{\citenamefont {Dubrovkin}\ \emph {et~al.}(2020)\citenamefont
  {Dubrovkin}, \citenamefont {Qiang}, \citenamefont {Salim}, \citenamefont
  {Nam}, \citenamefont {Zheludev},\ and\ \citenamefont {Wang}}]{Dubrovkin2020}%
  \BibitemOpen
  \bibfield  {author} {\bibinfo {author} {\bibfnamefont {A.~M.}\ \bibnamefont
  {Dubrovkin}}, \bibinfo {author} {\bibfnamefont {B.}~\bibnamefont {Qiang}},
  \bibinfo {author} {\bibfnamefont {T.}~\bibnamefont {Salim}}, \bibinfo
  {author} {\bibfnamefont {D.}~\bibnamefont {Nam}}, \bibinfo {author}
  {\bibfnamefont {N.~I.}\ \bibnamefont {Zheludev}}, \ and\ \bibinfo {author}
  {\bibfnamefont {Q.~J.}\ \bibnamefont {Wang}},\ }\href {\doibase
  10.1038/s41467-020-15767-y} {\bibfield  {journal} {\bibinfo  {journal}
  {Nature Communications}\ }\textbf {\bibinfo {volume} {11}},\ \bibinfo {pages}
  {1863} (\bibinfo {year} {2020})}\BibitemShut {NoStop}%
\bibitem [{\citenamefont {Rattunde}\ \emph {et~al.}(2006)\citenamefont
  {Rattunde}, \citenamefont {Schmitz}, \citenamefont {Mermelstein},
  \citenamefont {Kiefer},\ and\ \citenamefont {Wagner}}]{Rattunde2006}%
  \BibitemOpen
  \bibfield  {author} {\bibinfo {author} {\bibfnamefont {M.}~\bibnamefont
  {Rattunde}}, \bibinfo {author} {\bibfnamefont {J.}~\bibnamefont {Schmitz}},
  \bibinfo {author} {\bibfnamefont {C.}~\bibnamefont {Mermelstein}}, \bibinfo
  {author} {\bibfnamefont {R.}~\bibnamefont {Kiefer}}, \ and\ \bibinfo {author}
  {\bibfnamefont {J.}~\bibnamefont {Wagner}},\ }\enquote {\bibinfo {title}
  {Iii-sb-based type-i qw diode lasers},}\ in\ \href {\doibase
  10.1007/1-84628-209-8_3} {\emph {\bibinfo {booktitle} {Mid-infrared
  Semiconductor Optoelectronics}}},\ \bibinfo {editor} {edited by\ \bibinfo
  {editor} {\bibfnamefont {A.}~\bibnamefont {Krier}}}\ (\bibinfo  {publisher}
  {Springer London},\ \bibinfo {address} {London},\ \bibinfo {year} {2006})\
  pp.\ \bibinfo {pages} {131--157}\BibitemShut {NoStop}%
\bibitem [{\citenamefont {De~Liberato}\ and\ \citenamefont
  {Ciuti}(2012)}]{DeLiberato2012}%
  \BibitemOpen
  \bibfield  {author} {\bibinfo {author} {\bibfnamefont {S.}~\bibnamefont
  {De~Liberato}}\ and\ \bibinfo {author} {\bibfnamefont {C.}~\bibnamefont
  {Ciuti}},\ }\href {\doibase 10.1103/PhysRevB.85.125302} {\bibfield  {journal}
  {\bibinfo  {journal} {Phys. Rev. B}\ }\textbf {\bibinfo {volume} {85}},\
  \bibinfo {pages} {125302} (\bibinfo {year} {2012})}\BibitemShut {NoStop}%
\bibitem [{\citenamefont {Manceau}\ \emph {et~al.}(2018)\citenamefont
  {Manceau}, \citenamefont {Tran}, \citenamefont {Biasiol}, \citenamefont
  {Laurent}, \citenamefont {Sagnes}, \citenamefont {Beaudoin}, \citenamefont
  {De~Liberato}, \citenamefont {Carusotto},\ and\ \citenamefont
  {Colombelli}}]{Manceau2018}%
  \BibitemOpen
  \bibfield  {author} {\bibinfo {author} {\bibfnamefont {J.-M.}\ \bibnamefont
  {Manceau}}, \bibinfo {author} {\bibfnamefont {N.-L.}\ \bibnamefont {Tran}},
  \bibinfo {author} {\bibfnamefont {G.}~\bibnamefont {Biasiol}}, \bibinfo
  {author} {\bibfnamefont {T.}~\bibnamefont {Laurent}}, \bibinfo {author}
  {\bibfnamefont {I.}~\bibnamefont {Sagnes}}, \bibinfo {author} {\bibfnamefont
  {G.}~\bibnamefont {Beaudoin}}, \bibinfo {author} {\bibfnamefont
  {S.}~\bibnamefont {De~Liberato}}, \bibinfo {author} {\bibfnamefont
  {I.}~\bibnamefont {Carusotto}}, \ and\ \bibinfo {author} {\bibfnamefont
  {R.}~\bibnamefont {Colombelli}},\ }\href {\doibase 10.1063/1.5029893}
  {\bibfield  {journal} {\bibinfo  {journal} {Applied Physics Letters}\
  }\textbf {\bibinfo {volume} {112}},\ \bibinfo {pages} {191106} (\bibinfo
  {year} {2018})}\BibitemShut {NoStop}%
\bibitem [{\citenamefont {Tamagnone}\ \emph {et~al.}(2020)\citenamefont
  {Tamagnone}, \citenamefont {Meretska}, \citenamefont {Chaudhary},
  \citenamefont {Spagele}, \citenamefont {Zhu}, \citenamefont {Li},
  \citenamefont {Edgar}, \citenamefont {Ambrosio},\ and\ \citenamefont
  {Capasso}}]{Tamagnone2020}%
  \BibitemOpen
  \bibfield  {author} {\bibinfo {author} {\bibfnamefont {M.}~\bibnamefont
  {Tamagnone}}, \bibinfo {author} {\bibfnamefont {M.}~\bibnamefont {Meretska}},
  \bibinfo {author} {\bibfnamefont {K.}~\bibnamefont {Chaudhary}}, \bibinfo
  {author} {\bibfnamefont {C.~M.}\ \bibnamefont {Spagele}}, \bibinfo {author}
  {\bibfnamefont {A.}~\bibnamefont {Zhu}}, \bibinfo {author} {\bibfnamefont
  {J.}~\bibnamefont {Li}}, \bibinfo {author} {\bibfnamefont {J.~H.}\
  \bibnamefont {Edgar}}, \bibinfo {author} {\bibfnamefont {A.}~\bibnamefont
  {Ambrosio}}, \ and\ \bibinfo {author} {\bibfnamefont {F.}~\bibnamefont
  {Capasso}},\ }in\ \href {\doibase 10.1364/CLEO_QELS.2020.FTh4N.2} {\emph
  {\bibinfo {booktitle} {Conference on Lasers and Electro-Optics}}},\ \bibinfo
  {series and number} {OSA Technical Digest}\ (\bibinfo  {publisher} {Optical
  Society of America},\ \bibinfo {address} {Washington, DC},\ \bibinfo {year}
  {2020})\ p.\ \bibinfo {pages} {FTh4N.2}\BibitemShut {NoStop}%
\bibitem [{\citenamefont {Gubbin}\ \emph {et~al.}(2019)\citenamefont {Gubbin},
  \citenamefont {Berte}, \citenamefont {Meeker}, \citenamefont {Giles},
  \citenamefont {Ellis}, \citenamefont {Tischler}, \citenamefont {Wheeler},
  \citenamefont {Maier}, \citenamefont {Caldwell},\ and\ \citenamefont
  {De~Liberato}}]{Gubbin2019}%
  \BibitemOpen
  \bibfield  {author} {\bibinfo {author} {\bibfnamefont {C.~R.}\ \bibnamefont
  {Gubbin}}, \bibinfo {author} {\bibfnamefont {R.}~\bibnamefont {Berte}},
  \bibinfo {author} {\bibfnamefont {M.~A.}\ \bibnamefont {Meeker}}, \bibinfo
  {author} {\bibfnamefont {A.~J.}\ \bibnamefont {Giles}}, \bibinfo {author}
  {\bibfnamefont {C.~T.}\ \bibnamefont {Ellis}}, \bibinfo {author}
  {\bibfnamefont {J.~G.}\ \bibnamefont {Tischler}}, \bibinfo {author}
  {\bibfnamefont {V.~D.}\ \bibnamefont {Wheeler}}, \bibinfo {author}
  {\bibfnamefont {S.~A.}\ \bibnamefont {Maier}}, \bibinfo {author}
  {\bibfnamefont {J.~D.}\ \bibnamefont {Caldwell}}, \ and\ \bibinfo {author}
  {\bibfnamefont {S.}~\bibnamefont {De~Liberato}},\ }\href@noop {} {\bibfield
  {journal} {\bibinfo  {journal} {Nature Communications}\ }\textbf {\bibinfo
  {volume} {10}},\ \bibinfo {pages} {1682} (\bibinfo {year}
  {2019})}\BibitemShut {NoStop}%
\bibitem [{\citenamefont {Ma}\ \emph {et~al.}(2018)\citenamefont {Ma},
  \citenamefont {Alonso-Gonz{\'{a}}lez}, \citenamefont {Li}, \citenamefont
  {Nikitin}, \citenamefont {Yuan}, \citenamefont {Mart{\'{i}}n-S{\'{a}}nchez},
  \citenamefont {Taboada-Guti{\'{e}}rrez}, \citenamefont {Amenabar},
  \citenamefont {Li}, \citenamefont {V{\'{e}}lez}, \citenamefont {Tollan},
  \citenamefont {Dai}, \citenamefont {Zhang}, \citenamefont {Sriram},
  \citenamefont {Kalantar-Zadeh}, \citenamefont {Lee}, \citenamefont
  {Hillenbrand},\ and\ \citenamefont {Bao}}]{Ma2018}%
  \BibitemOpen
  \bibfield  {author} {\bibinfo {author} {\bibfnamefont {W.}~\bibnamefont
  {Ma}}, \bibinfo {author} {\bibfnamefont {P.}~\bibnamefont
  {Alonso-Gonz{\'{a}}lez}}, \bibinfo {author} {\bibfnamefont {S.}~\bibnamefont
  {Li}}, \bibinfo {author} {\bibfnamefont {A.~Y.}\ \bibnamefont {Nikitin}},
  \bibinfo {author} {\bibfnamefont {J.}~\bibnamefont {Yuan}}, \bibinfo {author}
  {\bibfnamefont {J.}~\bibnamefont {Mart{\'{i}}n-S{\'{a}}nchez}}, \bibinfo
  {author} {\bibfnamefont {J.}~\bibnamefont {Taboada-Guti{\'{e}}rrez}},
  \bibinfo {author} {\bibfnamefont {I.}~\bibnamefont {Amenabar}}, \bibinfo
  {author} {\bibfnamefont {P.}~\bibnamefont {Li}}, \bibinfo {author}
  {\bibfnamefont {S.}~\bibnamefont {V{\'{e}}lez}}, \bibinfo {author}
  {\bibfnamefont {C.}~\bibnamefont {Tollan}}, \bibinfo {author} {\bibfnamefont
  {Z.}~\bibnamefont {Dai}}, \bibinfo {author} {\bibfnamefont {Y.}~\bibnamefont
  {Zhang}}, \bibinfo {author} {\bibfnamefont {S.}~\bibnamefont {Sriram}},
  \bibinfo {author} {\bibfnamefont {K.}~\bibnamefont {Kalantar-Zadeh}},
  \bibinfo {author} {\bibfnamefont {S.-T.}\ \bibnamefont {Lee}}, \bibinfo
  {author} {\bibfnamefont {R.}~\bibnamefont {Hillenbrand}}, \ and\ \bibinfo
  {author} {\bibfnamefont {Q.}~\bibnamefont {Bao}},\ }\href {\doibase
  10.1038/s41586-018-0618-9} {\bibfield  {journal} {\bibinfo  {journal}
  {Nature}\ }\textbf {\bibinfo {volume} {562}},\ \bibinfo {pages} {557}
  (\bibinfo {year} {2018})}\BibitemShut {NoStop}%
\bibitem [{\citenamefont {Hu}\ \emph {et~al.}(2020)\citenamefont {Hu},
  \citenamefont {Ou}, \citenamefont {Si}, \citenamefont {Wu}, \citenamefont
  {Wu}, \citenamefont {Dai}, \citenamefont {Krasnok}, \citenamefont {Mazor},
  \citenamefont {Zhang}, \citenamefont {Bao}, \citenamefont {Qiu},\ and\
  \citenamefont {Al{\`{u}}}}]{Hu2020}%
  \BibitemOpen
  \bibfield  {author} {\bibinfo {author} {\bibfnamefont {G.}~\bibnamefont
  {Hu}}, \bibinfo {author} {\bibfnamefont {Q.}~\bibnamefont {Ou}}, \bibinfo
  {author} {\bibfnamefont {G.}~\bibnamefont {Si}}, \bibinfo {author}
  {\bibfnamefont {Y.}~\bibnamefont {Wu}}, \bibinfo {author} {\bibfnamefont
  {J.}~\bibnamefont {Wu}}, \bibinfo {author} {\bibfnamefont {Z.}~\bibnamefont
  {Dai}}, \bibinfo {author} {\bibfnamefont {A.}~\bibnamefont {Krasnok}},
  \bibinfo {author} {\bibfnamefont {Y.}~\bibnamefont {Mazor}}, \bibinfo
  {author} {\bibfnamefont {Q.}~\bibnamefont {Zhang}}, \bibinfo {author}
  {\bibfnamefont {Q.}~\bibnamefont {Bao}}, \bibinfo {author} {\bibfnamefont
  {C.-W.}\ \bibnamefont {Qiu}}, \ and\ \bibinfo {author} {\bibfnamefont
  {A.}~\bibnamefont {Al{\`{u}}}},\ }\href {\doibase 10.1038/s41586-020-2359-9}
  {\bibfield  {journal} {\bibinfo  {journal} {Nature}\ }\textbf {\bibinfo
  {volume} {582}},\ \bibinfo {pages} {209} (\bibinfo {year}
  {2020})}\BibitemShut {NoStop}%
\bibitem [{\citenamefont {Wu}\ \emph {et~al.}(2020)\citenamefont {Wu},
  \citenamefont {Ou}, \citenamefont {Yin}, \citenamefont {Li}, \citenamefont
  {Ma}, \citenamefont {Yu}, \citenamefont {Liu}, \citenamefont {Cui},
  \citenamefont {Bao}, \citenamefont {Duan}, \citenamefont
  {{\'{A}}lvarez-P{\'{e}}rez}, \citenamefont {Dai}, \citenamefont {Shabbir},
  \citenamefont {Medhekar}, \citenamefont {Li}, \citenamefont {Li},
  \citenamefont {Alonso-Gonz{\'{a}}lez},\ and\ \citenamefont {Bao}}]{Wu2020}%
  \BibitemOpen
  \bibfield  {author} {\bibinfo {author} {\bibfnamefont {Y.}~\bibnamefont
  {Wu}}, \bibinfo {author} {\bibfnamefont {Q.}~\bibnamefont {Ou}}, \bibinfo
  {author} {\bibfnamefont {Y.}~\bibnamefont {Yin}}, \bibinfo {author}
  {\bibfnamefont {Y.}~\bibnamefont {Li}}, \bibinfo {author} {\bibfnamefont
  {W.}~\bibnamefont {Ma}}, \bibinfo {author} {\bibfnamefont {W.}~\bibnamefont
  {Yu}}, \bibinfo {author} {\bibfnamefont {G.}~\bibnamefont {Liu}}, \bibinfo
  {author} {\bibfnamefont {X.}~\bibnamefont {Cui}}, \bibinfo {author}
  {\bibfnamefont {X.}~\bibnamefont {Bao}}, \bibinfo {author} {\bibfnamefont
  {J.}~\bibnamefont {Duan}}, \bibinfo {author} {\bibfnamefont {G.}~\bibnamefont
  {{\'{A}}lvarez-P{\'{e}}rez}}, \bibinfo {author} {\bibfnamefont
  {Z.}~\bibnamefont {Dai}}, \bibinfo {author} {\bibfnamefont {B.}~\bibnamefont
  {Shabbir}}, \bibinfo {author} {\bibfnamefont {N.}~\bibnamefont {Medhekar}},
  \bibinfo {author} {\bibfnamefont {X.}~\bibnamefont {Li}}, \bibinfo {author}
  {\bibfnamefont {C.-M.}\ \bibnamefont {Li}}, \bibinfo {author} {\bibfnamefont
  {P.}~\bibnamefont {Alonso-Gonz{\'{a}}lez}}, \ and\ \bibinfo {author}
  {\bibfnamefont {Q.}~\bibnamefont {Bao}},\ }\href {\doibase
  10.1038/s41467-020-16459-3} {\bibfield  {journal} {\bibinfo  {journal}
  {Nature Communications}\ }\textbf {\bibinfo {volume} {11}},\ \bibinfo {pages}
  {2646} (\bibinfo {year} {2020})}\BibitemShut {NoStop}%
\bibitem [{\citenamefont {Greffet}\ \emph {et~al.}(2002)\citenamefont
  {Greffet}, \citenamefont {Carminati}, \citenamefont {Joulain}, \citenamefont
  {Mulet}, \citenamefont {Mainguy},\ and\ \citenamefont {Chen}}]{Greffet2002}%
  \BibitemOpen
  \bibfield  {author} {\bibinfo {author} {\bibfnamefont {J.-J.}\ \bibnamefont
  {Greffet}}, \bibinfo {author} {\bibfnamefont {R.}~\bibnamefont {Carminati}},
  \bibinfo {author} {\bibfnamefont {K.}~\bibnamefont {Joulain}}, \bibinfo
  {author} {\bibfnamefont {J.-P.}\ \bibnamefont {Mulet}}, \bibinfo {author}
  {\bibfnamefont {S.}~\bibnamefont {Mainguy}}, \ and\ \bibinfo {author}
  {\bibfnamefont {Y.}~\bibnamefont {Chen}},\ }\href {\doibase 10.1038/416061a}
  {\bibfield  {journal} {\bibinfo  {journal} {Nature}\ }\textbf {\bibinfo
  {volume} {416}},\ \bibinfo {pages} {61} (\bibinfo {year} {2002})}\BibitemShut
  {NoStop}%
\bibitem [{\citenamefont {Hillenbrand}\ \emph {et~al.}(2002)\citenamefont
  {Hillenbrand}, \citenamefont {Taubner},\ and\ \citenamefont
  {Keilmann}}]{Hillenbrand2002}%
  \BibitemOpen
  \bibfield  {author} {\bibinfo {author} {\bibfnamefont {R.}~\bibnamefont
  {Hillenbrand}}, \bibinfo {author} {\bibfnamefont {T.}~\bibnamefont
  {Taubner}}, \ and\ \bibinfo {author} {\bibfnamefont {F.}~\bibnamefont
  {Keilmann}},\ }\href@noop {} {\bibfield  {journal} {\bibinfo  {journal}
  {Nature}\ ,\ \bibinfo {pages} {159}} (\bibinfo {year} {2002})}\BibitemShut
  {NoStop}%
\bibitem [{\citenamefont {Feng}\ \emph
  {et~al.}(2015{\natexlab{b}})\citenamefont {Feng}, \citenamefont {Streyer},
  \citenamefont {Zhong}, \citenamefont {Hoffman},\ and\ \citenamefont
  {Wasserman}}]{Feng2015b}%
  \BibitemOpen
  \bibfield  {author} {\bibinfo {author} {\bibfnamefont {K.}~\bibnamefont
  {Feng}}, \bibinfo {author} {\bibfnamefont {W.}~\bibnamefont {Streyer}},
  \bibinfo {author} {\bibfnamefont {Y.}~\bibnamefont {Zhong}}, \bibinfo
  {author} {\bibfnamefont {A.~J.}\ \bibnamefont {Hoffman}}, \ and\ \bibinfo
  {author} {\bibfnamefont {D.}~\bibnamefont {Wasserman}},\ }\href {\doibase
  10.1364/OE.23.0A1418} {\bibfield  {journal} {\bibinfo  {journal} {Optics
  Express}\ }\textbf {\bibinfo {volume} {23}},\ \bibinfo {pages} {A1418}
  (\bibinfo {year} {2015}{\natexlab{b}})}\BibitemShut {NoStop}%
\bibitem [{\citenamefont {Chen}\ \emph {et~al.}(2020)\citenamefont {Chen},
  \citenamefont {Hsieh}, \citenamefont {Chao}, \citenamefont {Yang},
  \citenamefont {Cheng}, \citenamefont {Chan}, \citenamefont {Lu},
  \citenamefont {Meng},\ and\ \citenamefont {Zan}}]{Chen2020}%
  \BibitemOpen
  \bibfield  {author} {\bibinfo {author} {\bibfnamefont {C.-C.}\ \bibnamefont
  {Chen}}, \bibinfo {author} {\bibfnamefont {J.-C.}\ \bibnamefont {Hsieh}},
  \bibinfo {author} {\bibfnamefont {C.-H.}\ \bibnamefont {Chao}}, \bibinfo
  {author} {\bibfnamefont {W.-S.}\ \bibnamefont {Yang}}, \bibinfo {author}
  {\bibfnamefont {H.-T.}\ \bibnamefont {Cheng}}, \bibinfo {author}
  {\bibfnamefont {C.-K.}\ \bibnamefont {Chan}}, \bibinfo {author}
  {\bibfnamefont {C.-J.}\ \bibnamefont {Lu}}, \bibinfo {author} {\bibfnamefont
  {H.-F.}\ \bibnamefont {Meng}}, \ and\ \bibinfo {author} {\bibfnamefont
  {H.-W.}\ \bibnamefont {Zan}},\ }\href {\doibase 10.1088/1752-7163/ab728b}
  {\bibfield  {journal} {\bibinfo  {journal} {Journal of Breath Research}\
  }\textbf {\bibinfo {volume} {14}},\ \bibinfo {pages} {036002} (\bibinfo
  {year} {2020})}\BibitemShut {NoStop}%
\bibitem [{\citenamefont {Lu}\ \emph {et~al.}(2016)\citenamefont {Lu},
  \citenamefont {Li}, \citenamefont {Mizaikoff}, \citenamefont {Katzir},
  \citenamefont {Raichlin}, \citenamefont {Sheng},\ and\ \citenamefont
  {Yu}}]{Lu2016}%
  \BibitemOpen
  \bibfield  {author} {\bibinfo {author} {\bibfnamefont {R.}~\bibnamefont
  {Lu}}, \bibinfo {author} {\bibfnamefont {W.-W.}\ \bibnamefont {Li}}, \bibinfo
  {author} {\bibfnamefont {B.}~\bibnamefont {Mizaikoff}}, \bibinfo {author}
  {\bibfnamefont {A.}~\bibnamefont {Katzir}}, \bibinfo {author} {\bibfnamefont
  {Y.}~\bibnamefont {Raichlin}}, \bibinfo {author} {\bibfnamefont {G.-P.}\
  \bibnamefont {Sheng}}, \ and\ \bibinfo {author} {\bibfnamefont {H.-Q.}\
  \bibnamefont {Yu}},\ }\href {\doibase 10.1038/nprot.2016.013} {\bibfield
  {journal} {\bibinfo  {journal} {Nature Protocols}\ }\textbf {\bibinfo
  {volume} {11}},\ \bibinfo {pages} {377} (\bibinfo {year} {2016})}\BibitemShut
  {NoStop}%
\bibitem [{\citenamefont {Sobrino}\ \emph {et~al.}(2016)\citenamefont
  {Sobrino}, \citenamefont {Frate}, \citenamefont {Drusch}, \citenamefont
  {Jim{\'{e}}nez-Mu{\~{n}}oz}, \citenamefont {Manunta},\ and\ \citenamefont
  {Regan}}]{Sobrino2016}%
  \BibitemOpen
  \bibfield  {author} {\bibinfo {author} {\bibfnamefont {J.~A.}\ \bibnamefont
  {Sobrino}}, \bibinfo {author} {\bibfnamefont {F.~D.}\ \bibnamefont {Frate}},
  \bibinfo {author} {\bibfnamefont {M.}~\bibnamefont {Drusch}}, \bibinfo
  {author} {\bibfnamefont {J.~C.}\ \bibnamefont {Jim{\'{e}}nez-Mu{\~{n}}oz}},
  \bibinfo {author} {\bibfnamefont {P.}~\bibnamefont {Manunta}}, \ and\
  \bibinfo {author} {\bibfnamefont {A.}~\bibnamefont {Regan}},\ }\href
  {\doibase 10.1109/TGRS.2015.2509179} {\bibfield  {journal} {\bibinfo
  {journal} {IEEE Transactions on Geoscience and Remote Sensing}\ }\textbf
  {\bibinfo {volume} {54}},\ \bibinfo {pages} {2963} (\bibinfo {year}
  {2016})}\BibitemShut {NoStop}%
\bibitem [{\citenamefont {Folland}\ \emph {et~al.}(2019)\citenamefont
  {Folland}, \citenamefont {Nordin}, \citenamefont {Wasserman},\ and\
  \citenamefont {Caldwell}}]{Folland2019}%
  \BibitemOpen
  \bibfield  {author} {\bibinfo {author} {\bibfnamefont {T.}~\bibnamefont
  {Folland}}, \bibinfo {author} {\bibfnamefont {L.}~\bibnamefont {Nordin}},
  \bibinfo {author} {\bibfnamefont {D.}~\bibnamefont {Wasserman}}, \ and\
  \bibinfo {author} {\bibfnamefont {J.~D.}\ \bibnamefont {Caldwell}},\ }\href
  {\doibase 10.1063/1.5090777} {\bibfield  {journal} {\bibinfo  {journal}
  {Journal of Applied Physics}\ }\textbf {\bibinfo {volume} {125}},\ \bibinfo
  {pages} {191102} (\bibinfo {year} {2019})}\BibitemShut {NoStop}%
\bibitem [{\citenamefont {Day}\ \emph {et~al.}(2013)\citenamefont {Day},
  \citenamefont {Pushkarsky}, \citenamefont {Caffey}, \citenamefont
  {Cecchetti}, \citenamefont {Arp}, \citenamefont {Whitmore}, \citenamefont
  {Henson},\ and\ \citenamefont {Takeuchi}}]{Day2013}%
  \BibitemOpen
  \bibfield  {author} {\bibinfo {author} {\bibfnamefont {T.}~\bibnamefont
  {Day}}, \bibinfo {author} {\bibfnamefont {M.}~\bibnamefont {Pushkarsky}},
  \bibinfo {author} {\bibfnamefont {D.}~\bibnamefont {Caffey}}, \bibinfo
  {author} {\bibfnamefont {K.}~\bibnamefont {Cecchetti}}, \bibinfo {author}
  {\bibfnamefont {R.}~\bibnamefont {Arp}}, \bibinfo {author} {\bibfnamefont
  {A.}~\bibnamefont {Whitmore}}, \bibinfo {author} {\bibfnamefont
  {M.}~\bibnamefont {Henson}}, \ and\ \bibinfo {author} {\bibfnamefont {E.~B.}\
  \bibnamefont {Takeuchi}},\ }in\ \href {\doibase 10.1117/12.2031536} {\emph
  {\bibinfo {booktitle} {Proc.SPIE}}},\ Vol.\ \bibinfo {volume} {8898}\
  (\bibinfo {year} {2013})\BibitemShut {NoStop}%
\bibitem [{\citenamefont {Wang}\ \emph {et~al.}(2017)\citenamefont {Wang},
  \citenamefont {Li}, \citenamefont {Chigrin}, \citenamefont {Giles},
  \citenamefont {Bezares}, \citenamefont {Glembocki}, \citenamefont
  {Caldwell},\ and\ \citenamefont {Taubner}}]{Wang2017}%
  \BibitemOpen
  \bibfield  {author} {\bibinfo {author} {\bibfnamefont {T.}~\bibnamefont
  {Wang}}, \bibinfo {author} {\bibfnamefont {P.}~\bibnamefont {Li}}, \bibinfo
  {author} {\bibfnamefont {D.~N.}\ \bibnamefont {Chigrin}}, \bibinfo {author}
  {\bibfnamefont {A.~J.}\ \bibnamefont {Giles}}, \bibinfo {author}
  {\bibfnamefont {F.~J.}\ \bibnamefont {Bezares}}, \bibinfo {author}
  {\bibfnamefont {O.~J.}\ \bibnamefont {Glembocki}}, \bibinfo {author}
  {\bibfnamefont {J.~D.}\ \bibnamefont {Caldwell}}, \ and\ \bibinfo {author}
  {\bibfnamefont {T.}~\bibnamefont {Taubner}},\ }\href {\doibase
  10.1021/acsphotonics.7b00321} {\bibfield  {journal} {\bibinfo  {journal} {ACS
  Photonics}\ }\textbf {\bibinfo {volume} {4}},\ \bibinfo {pages} {1753}
  (\bibinfo {year} {2017})}\BibitemShut {NoStop}%
\bibitem [{\citenamefont {Shen}\ \emph {et~al.}(2009)\citenamefont {Shen},
  \citenamefont {Narayanaswamy},\ and\ \citenamefont {Chen}}]{Shen2009}%
  \BibitemOpen
  \bibfield  {author} {\bibinfo {author} {\bibfnamefont {S.}~\bibnamefont
  {Shen}}, \bibinfo {author} {\bibfnamefont {A.}~\bibnamefont {Narayanaswamy}},
  \ and\ \bibinfo {author} {\bibfnamefont {G.}~\bibnamefont {Chen}},\ }\href
  {\doibase 10.1021/nl901208v} {\bibfield  {journal} {\bibinfo  {journal} {Nano
  Letters}\ }\textbf {\bibinfo {volume} {9}},\ \bibinfo {pages} {2909}
  (\bibinfo {year} {2009})},\ \bibinfo {note} {pMID: 19719110}\BibitemShut
  {NoStop}%
\bibitem [{\citenamefont {Tranchant}\ \emph {et~al.}(2019)\citenamefont
  {Tranchant}, \citenamefont {Hamamura}, \citenamefont {Ordonez-Miranda},
  \citenamefont {Yabuki}, \citenamefont {Vega-Flick}, \citenamefont
  {Cervantes-Alvarez}, \citenamefont {Alvarado-Gil}, \citenamefont {Volz},\
  and\ \citenamefont {Miyazaki}}]{Tranchant2019}%
  \BibitemOpen
  \bibfield  {author} {\bibinfo {author} {\bibfnamefont {L.}~\bibnamefont
  {Tranchant}}, \bibinfo {author} {\bibfnamefont {S.}~\bibnamefont {Hamamura}},
  \bibinfo {author} {\bibfnamefont {J.}~\bibnamefont {Ordonez-Miranda}},
  \bibinfo {author} {\bibfnamefont {T.}~\bibnamefont {Yabuki}}, \bibinfo
  {author} {\bibfnamefont {A.}~\bibnamefont {Vega-Flick}}, \bibinfo {author}
  {\bibfnamefont {F.}~\bibnamefont {Cervantes-Alvarez}}, \bibinfo {author}
  {\bibfnamefont {J.~J.}\ \bibnamefont {Alvarado-Gil}}, \bibinfo {author}
  {\bibfnamefont {S.}~\bibnamefont {Volz}}, \ and\ \bibinfo {author}
  {\bibfnamefont {K.}~\bibnamefont {Miyazaki}},\ }\href {\doibase
  10.1021/acs.nanolett.9b02214} {\bibfield  {journal} {\bibinfo  {journal}
  {Nano Letters}\ }\textbf {\bibinfo {volume} {19}},\ \bibinfo {pages} {6924}
  (\bibinfo {year} {2019})},\ \bibinfo {note} {pMID: 31525061}\BibitemShut
  {NoStop}%
\bibitem [{\citenamefont {Vorobyev}\ \emph {et~al.}(2009)\citenamefont
  {Vorobyev}, \citenamefont {Makin},\ and\ \citenamefont {Guo}}]{Vorobyev2009}%
  \BibitemOpen
  \bibfield  {author} {\bibinfo {author} {\bibfnamefont {A.~Y.}\ \bibnamefont
  {Vorobyev}}, \bibinfo {author} {\bibfnamefont {V.~S.}\ \bibnamefont {Makin}},
  \ and\ \bibinfo {author} {\bibfnamefont {C.}~\bibnamefont {Guo}},\
  }\href@noop {} {\bibfield  {journal} {\bibinfo  {journal} {Phys. Rev. Lett.}\
  }\textbf {\bibinfo {volume} {102}},\ \bibinfo {pages} {234301} (\bibinfo
  {year} {2009})}\BibitemShut {NoStop}%
\bibitem [{\citenamefont {Biener}\ \emph {et~al.}(2008)\citenamefont {Biener},
  \citenamefont {Dahan}, \citenamefont {Niv}, \citenamefont {Kleiner},\ and\
  \citenamefont {Hasman}}]{Biener2008}%
  \BibitemOpen
  \bibfield  {author} {\bibinfo {author} {\bibfnamefont {G.}~\bibnamefont
  {Biener}}, \bibinfo {author} {\bibfnamefont {N.}~\bibnamefont {Dahan}},
  \bibinfo {author} {\bibfnamefont {A.}~\bibnamefont {Niv}}, \bibinfo {author}
  {\bibfnamefont {V.}~\bibnamefont {Kleiner}}, \ and\ \bibinfo {author}
  {\bibfnamefont {E.}~\bibnamefont {Hasman}},\ }\href {\doibase
  10.1063/1.2883948} {\bibfield  {journal} {\bibinfo  {journal} {Applied
  Physics Letters}\ }\textbf {\bibinfo {volume} {92}},\ \bibinfo {pages}
  {081913} (\bibinfo {year} {2008})}\BibitemShut {NoStop}%
\bibitem [{\citenamefont {Mason}\ \emph {et~al.}(2011)\citenamefont {Mason},
  \citenamefont {Smith},\ and\ \citenamefont {Wasserman}}]{Mason2011}%
  \BibitemOpen
  \bibfield  {author} {\bibinfo {author} {\bibfnamefont {J.~A.}\ \bibnamefont
  {Mason}}, \bibinfo {author} {\bibfnamefont {S.}~\bibnamefont {Smith}}, \ and\
  \bibinfo {author} {\bibfnamefont {D.}~\bibnamefont {Wasserman}},\ }\href
  {\doibase 10.1063/1.3600779} {\bibfield  {journal} {\bibinfo  {journal}
  {Applied Physics Letters}\ }\textbf {\bibinfo {volume} {98}},\ \bibinfo
  {pages} {241105} (\bibinfo {year} {2011})}\BibitemShut {NoStop}%
\bibitem [{\citenamefont {Pralle}\ \emph {et~al.}(2002)\citenamefont {Pralle},
  \citenamefont {Moelders}, \citenamefont {McNeal}, \citenamefont {Puscasu},
  \citenamefont {Greenwald}, \citenamefont {Daly}, \citenamefont {Johnson},
  \citenamefont {George}, \citenamefont {Choi}, \citenamefont {El-Kady},\ and\
  \citenamefont {Biswas}}]{Pralle2002}%
  \BibitemOpen
  \bibfield  {author} {\bibinfo {author} {\bibfnamefont {M.~U.}\ \bibnamefont
  {Pralle}}, \bibinfo {author} {\bibfnamefont {N.}~\bibnamefont {Moelders}},
  \bibinfo {author} {\bibfnamefont {M.~P.}\ \bibnamefont {McNeal}}, \bibinfo
  {author} {\bibfnamefont {I.}~\bibnamefont {Puscasu}}, \bibinfo {author}
  {\bibfnamefont {A.~C.}\ \bibnamefont {Greenwald}}, \bibinfo {author}
  {\bibfnamefont {J.~T.}\ \bibnamefont {Daly}}, \bibinfo {author}
  {\bibfnamefont {E.~A.}\ \bibnamefont {Johnson}}, \bibinfo {author}
  {\bibfnamefont {T.}~\bibnamefont {George}}, \bibinfo {author} {\bibfnamefont
  {D.~S.}\ \bibnamefont {Choi}}, \bibinfo {author} {\bibfnamefont
  {I.}~\bibnamefont {El-Kady}}, \ and\ \bibinfo {author} {\bibfnamefont
  {R.}~\bibnamefont {Biswas}},\ }\href {\doibase 10.1063/1.1526919} {\bibfield
  {journal} {\bibinfo  {journal} {Applied Physics Letters}\ }\textbf {\bibinfo
  {volume} {81}},\ \bibinfo {pages} {4685} (\bibinfo {year}
  {2002})}\BibitemShut {NoStop}%
\bibitem [{\citenamefont {Tsai}\ \emph {et~al.}(2006)\citenamefont {Tsai},
  \citenamefont {Chuang}, \citenamefont {Meng}, \citenamefont {Chang},\ and\
  \citenamefont {Lee}}]{Tsai2006}%
  \BibitemOpen
  \bibfield  {author} {\bibinfo {author} {\bibfnamefont {M.-W.}\ \bibnamefont
  {Tsai}}, \bibinfo {author} {\bibfnamefont {T.-H.}\ \bibnamefont {Chuang}},
  \bibinfo {author} {\bibfnamefont {C.-Y.}\ \bibnamefont {Meng}}, \bibinfo
  {author} {\bibfnamefont {Y.-T.}\ \bibnamefont {Chang}}, \ and\ \bibinfo
  {author} {\bibfnamefont {S.-C.}\ \bibnamefont {Lee}},\ }\href {\doibase
  10.1063/1.2364860} {\bibfield  {journal} {\bibinfo  {journal} {Applied
  Physics Letters}\ }\textbf {\bibinfo {volume} {89}},\ \bibinfo {pages}
  {173116} (\bibinfo {year} {2006})}\BibitemShut {NoStop}%
\bibitem [{\citenamefont {Celanovic}\ \emph {et~al.}(2008)\citenamefont
  {Celanovic}, \citenamefont {Jovanovic},\ and\ \citenamefont
  {Kassakian}}]{Celanovic2008}%
  \BibitemOpen
  \bibfield  {author} {\bibinfo {author} {\bibfnamefont {I.}~\bibnamefont
  {Celanovic}}, \bibinfo {author} {\bibfnamefont {N.}~\bibnamefont
  {Jovanovic}}, \ and\ \bibinfo {author} {\bibfnamefont {J.}~\bibnamefont
  {Kassakian}},\ }\href {\doibase 10.1063/1.2927484} {\bibfield  {journal}
  {\bibinfo  {journal} {Applied Physics Letters}\ }\textbf {\bibinfo {volume}
  {92}},\ \bibinfo {pages} {193101} (\bibinfo {year} {2008})}\BibitemShut
  {NoStop}%
\bibitem [{\citenamefont {Lin}\ \emph {et~al.}(2003)\citenamefont {Lin},
  \citenamefont {Moreno},\ and\ \citenamefont {Fleming}}]{Lin2003}%
  \BibitemOpen
  \bibfield  {author} {\bibinfo {author} {\bibfnamefont {S.~Y.}\ \bibnamefont
  {Lin}}, \bibinfo {author} {\bibfnamefont {J.}~\bibnamefont {Moreno}}, \ and\
  \bibinfo {author} {\bibfnamefont {J.~G.}\ \bibnamefont {Fleming}},\ }\href
  {\doibase 10.1063/1.1592614} {\bibfield  {journal} {\bibinfo  {journal}
  {Applied Physics Letters}\ }\textbf {\bibinfo {volume} {83}},\ \bibinfo
  {pages} {380} (\bibinfo {year} {2003})}\BibitemShut {NoStop}%
\bibitem [{\citenamefont {Landy}\ \emph {et~al.}(2008)\citenamefont {Landy},
  \citenamefont {Sajuyigbe}, \citenamefont {Mock}, \citenamefont {Smith},\ and\
  \citenamefont {Padilla}}]{Landy2008}%
  \BibitemOpen
  \bibfield  {author} {\bibinfo {author} {\bibfnamefont {N.~I.}\ \bibnamefont
  {Landy}}, \bibinfo {author} {\bibfnamefont {S.}~\bibnamefont {Sajuyigbe}},
  \bibinfo {author} {\bibfnamefont {J.~J.}\ \bibnamefont {Mock}}, \bibinfo
  {author} {\bibfnamefont {D.~R.}\ \bibnamefont {Smith}}, \ and\ \bibinfo
  {author} {\bibfnamefont {W.~J.}\ \bibnamefont {Padilla}},\ }\href {\doibase
  10.1103/PhysRevLett.100.207402} {\bibfield  {journal} {\bibinfo  {journal}
  {Phys. Rev. Lett.}\ }\textbf {\bibinfo {volume} {100}},\ \bibinfo {pages}
  {207402} (\bibinfo {year} {2008})}\BibitemShut {NoStop}%
\bibitem [{\citenamefont {Liu}\ \emph {et~al.}(2010{\natexlab{a}})\citenamefont
  {Liu}, \citenamefont {Starr}, \citenamefont {Starr},\ and\ \citenamefont
  {Padilla}}]{Liu2010}%
  \BibitemOpen
  \bibfield  {author} {\bibinfo {author} {\bibfnamefont {X.}~\bibnamefont
  {Liu}}, \bibinfo {author} {\bibfnamefont {T.}~\bibnamefont {Starr}}, \bibinfo
  {author} {\bibfnamefont {A.~F.}\ \bibnamefont {Starr}}, \ and\ \bibinfo
  {author} {\bibfnamefont {W.~J.}\ \bibnamefont {Padilla}},\ }\href {\doibase
  10.1103/PhysRevLett.104.207403} {\bibfield  {journal} {\bibinfo  {journal}
  {Phys. Rev. Lett.}\ }\textbf {\bibinfo {volume} {104}},\ \bibinfo {pages}
  {207403} (\bibinfo {year} {2010}{\natexlab{a}})}\BibitemShut {NoStop}%
\bibitem [{\citenamefont {Liu}\ \emph {et~al.}(2011)\citenamefont {Liu},
  \citenamefont {Tyler}, \citenamefont {Starr}, \citenamefont {Starr},
  \citenamefont {Jokerst},\ and\ \citenamefont {Padilla}}]{Liu2011}%
  \BibitemOpen
  \bibfield  {author} {\bibinfo {author} {\bibfnamefont {X.}~\bibnamefont
  {Liu}}, \bibinfo {author} {\bibfnamefont {T.}~\bibnamefont {Tyler}}, \bibinfo
  {author} {\bibfnamefont {T.}~\bibnamefont {Starr}}, \bibinfo {author}
  {\bibfnamefont {A.~F.}\ \bibnamefont {Starr}}, \bibinfo {author}
  {\bibfnamefont {N.~M.}\ \bibnamefont {Jokerst}}, \ and\ \bibinfo {author}
  {\bibfnamefont {W.~J.}\ \bibnamefont {Padilla}},\ }\href {\doibase
  10.1103/PhysRevLett.107.045901} {\bibfield  {journal} {\bibinfo  {journal}
  {Phys. Rev. Lett.}\ }\textbf {\bibinfo {volume} {107}},\ \bibinfo {pages}
  {045901} (\bibinfo {year} {2011})}\BibitemShut {NoStop}%
\bibitem [{\citenamefont {Wu}\ \emph {et~al.}(2012)\citenamefont {Wu},
  \citenamefont {III}, \citenamefont {John}, \citenamefont {Milder},
  \citenamefont {Zollars}, \citenamefont {Savoy},\ and\ \citenamefont
  {Shvets}}]{Wu2012}%
  \BibitemOpen
  \bibfield  {author} {\bibinfo {author} {\bibfnamefont {C.}~\bibnamefont
  {Wu}}, \bibinfo {author} {\bibfnamefont {B.~N.}\ \bibnamefont {III}},
  \bibinfo {author} {\bibfnamefont {J.}~\bibnamefont {John}}, \bibinfo {author}
  {\bibfnamefont {A.}~\bibnamefont {Milder}}, \bibinfo {author} {\bibfnamefont
  {B.}~\bibnamefont {Zollars}}, \bibinfo {author} {\bibfnamefont
  {S.}~\bibnamefont {Savoy}}, \ and\ \bibinfo {author} {\bibfnamefont
  {G.}~\bibnamefont {Shvets}},\ }\href {\doibase 10.1088/2040-8978/14/2/024005}
  {\bibfield  {journal} {\bibinfo  {journal} {Journal of Optics}\ }\textbf
  {\bibinfo {volume} {14}},\ \bibinfo {pages} {024005} (\bibinfo {year}
  {2012})}\BibitemShut {NoStop}%
\bibitem [{\citenamefont {Jiang}\ \emph {et~al.}(2011)\citenamefont {Jiang},
  \citenamefont {Yun}, \citenamefont {Toor}, \citenamefont {Werner},\ and\
  \citenamefont {Mayer}}]{Jiang2011}%
  \BibitemOpen
  \bibfield  {author} {\bibinfo {author} {\bibfnamefont {Z.~H.}\ \bibnamefont
  {Jiang}}, \bibinfo {author} {\bibfnamefont {S.}~\bibnamefont {Yun}}, \bibinfo
  {author} {\bibfnamefont {F.}~\bibnamefont {Toor}}, \bibinfo {author}
  {\bibfnamefont {D.~H.}\ \bibnamefont {Werner}}, \ and\ \bibinfo {author}
  {\bibfnamefont {T.~S.}\ \bibnamefont {Mayer}},\ }\href {\doibase
  10.1021/nn2004603} {\bibfield  {journal} {\bibinfo  {journal} {ACS Nano}\
  }\textbf {\bibinfo {volume} {5}},\ \bibinfo {pages} {4641} (\bibinfo {year}
  {2011})},\ \bibinfo {note} {pMID: 21456579}\BibitemShut {NoStop}%
\bibitem [{\citenamefont {Ikeda}\ \emph {et~al.}(2008)\citenamefont {Ikeda},
  \citenamefont {Miyazaki}, \citenamefont {Kasaya}, \citenamefont {Yamamoto},
  \citenamefont {Inoue}, \citenamefont {Fujimura}, \citenamefont {Kanakugi},
  \citenamefont {Okada}, \citenamefont {Hatade},\ and\ \citenamefont
  {Kitagawa}}]{Ikeda2008}%
  \BibitemOpen
  \bibfield  {author} {\bibinfo {author} {\bibfnamefont {K.}~\bibnamefont
  {Ikeda}}, \bibinfo {author} {\bibfnamefont {H.~T.}\ \bibnamefont {Miyazaki}},
  \bibinfo {author} {\bibfnamefont {T.}~\bibnamefont {Kasaya}}, \bibinfo
  {author} {\bibfnamefont {K.}~\bibnamefont {Yamamoto}}, \bibinfo {author}
  {\bibfnamefont {Y.}~\bibnamefont {Inoue}}, \bibinfo {author} {\bibfnamefont
  {K.}~\bibnamefont {Fujimura}}, \bibinfo {author} {\bibfnamefont
  {T.}~\bibnamefont {Kanakugi}}, \bibinfo {author} {\bibfnamefont
  {M.}~\bibnamefont {Okada}}, \bibinfo {author} {\bibfnamefont
  {K.}~\bibnamefont {Hatade}}, \ and\ \bibinfo {author} {\bibfnamefont
  {S.}~\bibnamefont {Kitagawa}},\ }\href {\doibase 10.1063/1.2834903}
  {\bibfield  {journal} {\bibinfo  {journal} {Applied Physics Letters}\
  }\textbf {\bibinfo {volume} {92}},\ \bibinfo {pages} {021117} (\bibinfo
  {year} {2008})}\BibitemShut {NoStop}%
\bibitem [{\citenamefont {Puscasu}\ and\ \citenamefont
  {Schaich}(2008)}]{Puscasu2008}%
  \BibitemOpen
  \bibfield  {author} {\bibinfo {author} {\bibfnamefont {I.}~\bibnamefont
  {Puscasu}}\ and\ \bibinfo {author} {\bibfnamefont {W.~L.}\ \bibnamefont
  {Schaich}},\ }\href {\doibase 10.1063/1.2938716} {\bibfield  {journal}
  {\bibinfo  {journal} {Applied Physics Letters}\ }\textbf {\bibinfo {volume}
  {92}},\ \bibinfo {pages} {233102} (\bibinfo {year} {2008})}\BibitemShut
  {NoStop}%
\bibitem [{\citenamefont {Schuller}\ \emph {et~al.}(2009)\citenamefont
  {Schuller}, \citenamefont {Taubner},\ and\ \citenamefont
  {Brongersma}}]{Schuller2008}%
  \BibitemOpen
  \bibfield  {author} {\bibinfo {author} {\bibfnamefont {J.}~\bibnamefont
  {Schuller}}, \bibinfo {author} {\bibfnamefont {T.}~\bibnamefont {Taubner}}, \
  and\ \bibinfo {author} {\bibfnamefont {M.}~\bibnamefont {Brongersma}},\
  }\href {\doibase 10.1038/nphoton.2009.188} {\bibfield  {journal} {\bibinfo
  {journal} {Nature Photonics}\ }\textbf {\bibinfo {volume} {3}},\ \bibinfo
  {pages} {658} (\bibinfo {year} {2009})}\BibitemShut {NoStop}%
\bibitem [{\citenamefont {Brar}\ \emph {et~al.}(2014)\citenamefont {Brar},
  \citenamefont {Sherrott}, \citenamefont {Jang}, \citenamefont {Kim},
  \citenamefont {Kim}, \citenamefont {Choi}, \citenamefont {Sweatlock},\ and\
  \citenamefont {Atwater}}]{Brar2014}%
  \BibitemOpen
  \bibfield  {author} {\bibinfo {author} {\bibfnamefont {V.~W.}\ \bibnamefont
  {Brar}}, \bibinfo {author} {\bibfnamefont {M.~C.}\ \bibnamefont {Sherrott}},
  \bibinfo {author} {\bibfnamefont {M.~S.}\ \bibnamefont {Jang}}, \bibinfo
  {author} {\bibfnamefont {S.}~\bibnamefont {Kim}}, \bibinfo {author}
  {\bibfnamefont {L.}~\bibnamefont {Kim}}, \bibinfo {author} {\bibfnamefont
  {M.}~\bibnamefont {Choi}}, \bibinfo {author} {\bibfnamefont {L.~A.}\
  \bibnamefont {Sweatlock}}, \ and\ \bibinfo {author} {\bibfnamefont {H.~A.}\
  \bibnamefont {Atwater}},\ }\href {\doibase 10.1038/ncomms8032} {\bibfield
  {journal} {\bibinfo  {journal} {Nature Communications}\ }\textbf {\bibinfo
  {volume} {6}} (\bibinfo {year} {2014}),\ 10.1038/ncomms8032}\BibitemShut
  {NoStop}%
\bibitem [{\citenamefont {Wojszvzyk}\ \emph {et~al.}(2021)\citenamefont
  {Wojszvzyk}, \citenamefont {Nguyen}, \citenamefont {Coutrot}, \citenamefont
  {Zhang}, \citenamefont {Vest},\ and\ \citenamefont
  {Greffet}}]{Wojszvzyk2021}%
  \BibitemOpen
  \bibfield  {author} {\bibinfo {author} {\bibfnamefont {L.}~\bibnamefont
  {Wojszvzyk}}, \bibinfo {author} {\bibfnamefont {A.}~\bibnamefont {Nguyen}},
  \bibinfo {author} {\bibfnamefont {L.-A.}\ \bibnamefont {Coutrot}}, \bibinfo
  {author} {\bibfnamefont {C.}~\bibnamefont {Zhang}}, \bibinfo {author}
  {\bibfnamefont {B.}~\bibnamefont {Vest}}, \ and\ \bibinfo {author}
  {\bibfnamefont {J.-J.}\ \bibnamefont {Greffet}},\ }\href@noop {} {\bibfield
  {journal} {\bibinfo  {journal} {Nature Communications}\ }\textbf {\bibinfo
  {volume} {12}},\ \bibinfo {pages} {1492} (\bibinfo {year}
  {2021})}\BibitemShut {NoStop}%
\bibitem [{\citenamefont {Kats}\ \emph {et~al.}(2013)\citenamefont {Kats},
  \citenamefont {Blanchard}, \citenamefont {Zhang}, \citenamefont {Genevet},
  \citenamefont {Ko}, \citenamefont {Ramanathan},\ and\ \citenamefont
  {Capasso}}]{Kats2013}%
  \BibitemOpen
  \bibfield  {author} {\bibinfo {author} {\bibfnamefont {M.~A.}\ \bibnamefont
  {Kats}}, \bibinfo {author} {\bibfnamefont {R.}~\bibnamefont {Blanchard}},
  \bibinfo {author} {\bibfnamefont {S.}~\bibnamefont {Zhang}}, \bibinfo
  {author} {\bibfnamefont {P.}~\bibnamefont {Genevet}}, \bibinfo {author}
  {\bibfnamefont {C.}~\bibnamefont {Ko}}, \bibinfo {author} {\bibfnamefont
  {S.}~\bibnamefont {Ramanathan}}, \ and\ \bibinfo {author} {\bibfnamefont
  {F.}~\bibnamefont {Capasso}},\ }\href {\doibase 10.1103/PhysRevX.3.041004}
  {\bibfield  {journal} {\bibinfo  {journal} {Phys. Rev. X}\ }\textbf {\bibinfo
  {volume} {3}},\ \bibinfo {pages} {041004} (\bibinfo {year}
  {2013})}\BibitemShut {NoStop}%
\bibitem [{\citenamefont {Raman}\ \emph {et~al.}(2014)\citenamefont {Raman},
  \citenamefont {Anoma}, \citenamefont {Zhu}, \citenamefont {Rephaeli},\ and\
  \citenamefont {Fan}}]{Raman2014}%
  \BibitemOpen
  \bibfield  {author} {\bibinfo {author} {\bibfnamefont {A.~P.}\ \bibnamefont
  {Raman}}, \bibinfo {author} {\bibfnamefont {M.~A.}\ \bibnamefont {Anoma}},
  \bibinfo {author} {\bibfnamefont {L.}~\bibnamefont {Zhu}}, \bibinfo {author}
  {\bibfnamefont {E.}~\bibnamefont {Rephaeli}}, \ and\ \bibinfo {author}
  {\bibfnamefont {S.}~\bibnamefont {Fan}},\ }\href@noop {} {\bibfield
  {journal} {\bibinfo  {journal} {Nature}\ }\textbf {\bibinfo {volume} {515}},\
  \bibinfo {pages} {540} (\bibinfo {year} {2014})}\BibitemShut {NoStop}%
\bibitem [{\citenamefont {Guler}\ \emph {et~al.}(2014)\citenamefont {Guler},
  \citenamefont {Boltasseva},\ and\ \citenamefont {Shalaev}}]{Guler2014}%
  \BibitemOpen
  \bibfield  {author} {\bibinfo {author} {\bibfnamefont {U.}~\bibnamefont
  {Guler}}, \bibinfo {author} {\bibfnamefont {A.}~\bibnamefont {Boltasseva}}, \
  and\ \bibinfo {author} {\bibfnamefont {V.~M.}\ \bibnamefont {Shalaev}},\
  }\href {\doibase 10.1126/science.1252722} {\bibfield  {journal} {\bibinfo
  {journal} {Science}\ }\textbf {\bibinfo {volume} {344}},\ \bibinfo {pages}
  {263} (\bibinfo {year} {2014})}\BibitemShut {NoStop}%
\bibitem [{\citenamefont {Huang}\ \emph {et~al.}(1995)\citenamefont {Huang},
  \citenamefont {Gumbs},\ and\ \citenamefont {Manasreh}}]{Huang1995}%
  \BibitemOpen
  \bibfield  {author} {\bibinfo {author} {\bibfnamefont {D.}~\bibnamefont
  {Huang}}, \bibinfo {author} {\bibfnamefont {G.}~\bibnamefont {Gumbs}}, \ and\
  \bibinfo {author} {\bibfnamefont {M.~O.}\ \bibnamefont {Manasreh}},\ }\href
  {\doibase 10.1103/PhysRevB.52.14126} {\bibfield  {journal} {\bibinfo
  {journal} {Phys. Rev. B}\ }\textbf {\bibinfo {volume} {52}},\ \bibinfo
  {pages} {14126} (\bibinfo {year} {1995})}\BibitemShut {NoStop}%
\bibitem [{\citenamefont {{Nash}}\ \emph {et~al.}(2009)\citenamefont {{Nash}},
  \citenamefont {{Forman}}, \citenamefont {{Smith}}, \citenamefont
  {{Robinson}}, \citenamefont {{Buckle}}, \citenamefont {{Coomber}},
  \citenamefont {{Emeny}}, \citenamefont {{Gordon}},\ and\ \citenamefont
  {{Ashley}}}]{Nash2009}%
  \BibitemOpen
  \bibfield  {author} {\bibinfo {author} {\bibfnamefont {G.~R.}\ \bibnamefont
  {{Nash}}}, \bibinfo {author} {\bibfnamefont {H.~L.}\ \bibnamefont
  {{Forman}}}, \bibinfo {author} {\bibfnamefont {S.~J.}\ \bibnamefont
  {{Smith}}}, \bibinfo {author} {\bibfnamefont {P.~B.}\ \bibnamefont
  {{Robinson}}}, \bibinfo {author} {\bibfnamefont {L.}~\bibnamefont
  {{Buckle}}}, \bibinfo {author} {\bibfnamefont {S.~D.}\ \bibnamefont
  {{Coomber}}}, \bibinfo {author} {\bibfnamefont {M.~T.}\ \bibnamefont
  {{Emeny}}}, \bibinfo {author} {\bibfnamefont {N.~T.}\ \bibnamefont
  {{Gordon}}}, \ and\ \bibinfo {author} {\bibfnamefont {T.}~\bibnamefont
  {{Ashley}}},\ }\href {\doibase 10.1109/JSEN.2009.2029815} {\bibfield
  {journal} {\bibinfo  {journal} {IEEE Sensors Journal}\ }\textbf {\bibinfo
  {volume} {9}},\ \bibinfo {pages} {1240} (\bibinfo {year} {2009})}\BibitemShut
  {NoStop}%
\bibitem [{\citenamefont {Yang}(1995)}]{Yang1995}%
  \BibitemOpen
  \bibfield  {author} {\bibinfo {author} {\bibfnamefont {R.~Q.}\ \bibnamefont
  {Yang}},\ }\href {\doibase https://doi.org/10.1006/spmi.1995.1017} {\bibfield
   {journal} {\bibinfo  {journal} {Superlattices and Microstructures}\ }\textbf
  {\bibinfo {volume} {17}},\ \bibinfo {pages} {77} (\bibinfo {year}
  {1995})}\BibitemShut {NoStop}%
\bibitem [{\citenamefont {Dallner}\ \emph {et~al.}(2015)\citenamefont
  {Dallner}, \citenamefont {Hau}, \citenamefont {Höfling},\ and\ \citenamefont
  {Kamp}}]{Dallner2015}%
  \BibitemOpen
  \bibfield  {author} {\bibinfo {author} {\bibfnamefont {M.}~\bibnamefont
  {Dallner}}, \bibinfo {author} {\bibfnamefont {F.}~\bibnamefont {Hau}},
  \bibinfo {author} {\bibfnamefont {S.}~\bibnamefont {Höfling}}, \ and\
  \bibinfo {author} {\bibfnamefont {M.}~\bibnamefont {Kamp}},\ }\href {\doibase
  10.1063/1.4907002} {\bibfield  {journal} {\bibinfo  {journal} {Applied
  Physics Letters}\ }\textbf {\bibinfo {volume} {106}},\ \bibinfo {pages}
  {041108} (\bibinfo {year} {2015})}\BibitemShut {NoStop}%
\bibitem [{\citenamefont {Vurgaftman}\ \emph {et~al.}(2015)\citenamefont
  {Vurgaftman}, \citenamefont {Weih}, \citenamefont {Kamp}, \citenamefont
  {Meyer}, \citenamefont {Canedy}, \citenamefont {Kim}, \citenamefont {Kim},
  \citenamefont {Bewley}, \citenamefont {Merritt}, \citenamefont {Abell},\ and\
  \citenamefont {H{\"{o}}fling}}]{Vurgaftman2015}%
  \BibitemOpen
  \bibfield  {author} {\bibinfo {author} {\bibfnamefont {I.}~\bibnamefont
  {Vurgaftman}}, \bibinfo {author} {\bibfnamefont {R.}~\bibnamefont {Weih}},
  \bibinfo {author} {\bibfnamefont {M.}~\bibnamefont {Kamp}}, \bibinfo {author}
  {\bibfnamefont {J.~R.}\ \bibnamefont {Meyer}}, \bibinfo {author}
  {\bibfnamefont {C.~L.}\ \bibnamefont {Canedy}}, \bibinfo {author}
  {\bibfnamefont {C.~S.}\ \bibnamefont {Kim}}, \bibinfo {author} {\bibfnamefont
  {M.}~\bibnamefont {Kim}}, \bibinfo {author} {\bibfnamefont {W.~W.}\
  \bibnamefont {Bewley}}, \bibinfo {author} {\bibfnamefont {C.~D.}\
  \bibnamefont {Merritt}}, \bibinfo {author} {\bibfnamefont {J.}~\bibnamefont
  {Abell}}, \ and\ \bibinfo {author} {\bibfnamefont {S.}~\bibnamefont
  {H{\"{o}}fling}},\ }\href {\doibase 10.1088/0022-3727/48/12/123001}
  {\bibfield  {journal} {\bibinfo  {journal} {Journal of Physics D: Applied
  Physics}\ }\textbf {\bibinfo {volume} {48}} (\bibinfo {year} {2015}),\
  10.1088/0022-3727/48/12/123001}\BibitemShut {NoStop}%
\bibitem [{\citenamefont {Montealegre}\ \emph {et~al.}(2021)\citenamefont
  {Montealegre}, \citenamefont {Schrock}, \citenamefont {Walhof}, \citenamefont
  {Muellerleile},\ and\ \citenamefont {Prineas}}]{Montealegre2021}%
  \BibitemOpen
  \bibfield  {author} {\bibinfo {author} {\bibfnamefont {D.~A.}\ \bibnamefont
  {Montealegre}}, \bibinfo {author} {\bibfnamefont {K.~N.}\ \bibnamefont
  {Schrock}}, \bibinfo {author} {\bibfnamefont {A.~C.}\ \bibnamefont {Walhof}},
  \bibinfo {author} {\bibfnamefont {A.~M.}\ \bibnamefont {Muellerleile}}, \
  and\ \bibinfo {author} {\bibfnamefont {J.~P.}\ \bibnamefont {Prineas}},\
  }\href {\doibase 10.1063/5.0039269} {\bibfield  {journal} {\bibinfo
  {journal} {Applied Physics Letters}\ }\textbf {\bibinfo {volume} {118}},\
  \bibinfo {pages} {71105} (\bibinfo {year} {2021})}\BibitemShut {NoStop}%
\bibitem [{\citenamefont {Faist}\ \emph {et~al.}(1994)\citenamefont {Faist},
  \citenamefont {Capasso}, \citenamefont {Sivco}, \citenamefont {Sirtori},
  \citenamefont {Hutchinson},\ and\ \citenamefont {Cho}}]{Faist1995}%
  \BibitemOpen
  \bibfield  {author} {\bibinfo {author} {\bibfnamefont {J.}~\bibnamefont
  {Faist}}, \bibinfo {author} {\bibfnamefont {F.}~\bibnamefont {Capasso}},
  \bibinfo {author} {\bibfnamefont {D.~L.}\ \bibnamefont {Sivco}}, \bibinfo
  {author} {\bibfnamefont {C.}~\bibnamefont {Sirtori}}, \bibinfo {author}
  {\bibfnamefont {A.~L.}\ \bibnamefont {Hutchinson}}, \ and\ \bibinfo {author}
  {\bibfnamefont {A.~Y.}\ \bibnamefont {Cho}},\ }\href {\doibase
  10.1126/science.264.5158.553} {\bibfield  {journal} {\bibinfo  {journal}
  {Science}\ }\textbf {\bibinfo {volume} {264}},\ \bibinfo {pages} {553}
  (\bibinfo {year} {1994})}\BibitemShut {NoStop}%
\bibitem [{\citenamefont {Williams}(2007)}]{Williams2007}%
  \BibitemOpen
  \bibfield  {author} {\bibinfo {author} {\bibfnamefont {B.}~\bibnamefont
  {Williams}},\ }\href@noop {} {\bibfield  {journal} {\bibinfo  {journal}
  {Nature Photonics}\ }\textbf {\bibinfo {volume} {1}},\ \bibinfo {pages} {517}
  (\bibinfo {year} {2007})}\BibitemShut {NoStop}%
\bibitem [{\citenamefont {Tan}\ \emph {et~al.}(2020)\citenamefont {Tan},
  \citenamefont {Wan},\ and\ \citenamefont {Cao}}]{Tan2020b}%
  \BibitemOpen
  \bibfield  {author} {\bibinfo {author} {\bibfnamefont {Z.-Y.}\ \bibnamefont
  {Tan}}, \bibinfo {author} {\bibfnamefont {W.-J.}\ \bibnamefont {Wan}}, \ and\
  \bibinfo {author} {\bibfnamefont {J.-C.}\ \bibnamefont {Cao}},\ }\href
  {\doibase 10.1088/1674-1056/aba945} {\bibfield  {journal} {\bibinfo
  {journal} {Chinese Physics B}\ }\textbf {\bibinfo {volume} {29}},\ \bibinfo
  {pages} {84212} (\bibinfo {year} {2020})}\BibitemShut {NoStop}%
\bibitem [{\citenamefont {Bai}\ \emph {et~al.}(2011)\citenamefont {Bai},
  \citenamefont {Bandyopadhyay}, \citenamefont {Tsao}, \citenamefont
  {Slivken},\ and\ \citenamefont {Razeghi}}]{Bai2011}%
  \BibitemOpen
  \bibfield  {author} {\bibinfo {author} {\bibfnamefont {Y.}~\bibnamefont
  {Bai}}, \bibinfo {author} {\bibfnamefont {N.}~\bibnamefont {Bandyopadhyay}},
  \bibinfo {author} {\bibfnamefont {S.}~\bibnamefont {Tsao}}, \bibinfo {author}
  {\bibfnamefont {S.}~\bibnamefont {Slivken}}, \ and\ \bibinfo {author}
  {\bibfnamefont {M.}~\bibnamefont {Razeghi}},\ }\href@noop {} {\bibfield
  {journal} {\bibinfo  {journal} {Applied Physics Letters}\ }\textbf {\bibinfo
  {volume} {98}},\ \bibinfo {pages} {181102} (\bibinfo {year}
  {2011})}\BibitemShut {NoStop}%
\bibitem [{\citenamefont {Wade}\ \emph {et~al.}(2009)\citenamefont {Wade},
  \citenamefont {Fedorov}, \citenamefont {Smirnov}, \citenamefont {Kumar},
  \citenamefont {Williams}, \citenamefont {Hu},\ and\ \citenamefont
  {Reno}}]{Wade2009}%
  \BibitemOpen
  \bibfield  {author} {\bibinfo {author} {\bibfnamefont {A.}~\bibnamefont
  {Wade}}, \bibinfo {author} {\bibfnamefont {G.}~\bibnamefont {Fedorov}},
  \bibinfo {author} {\bibfnamefont {D.}~\bibnamefont {Smirnov}}, \bibinfo
  {author} {\bibfnamefont {S.}~\bibnamefont {Kumar}}, \bibinfo {author}
  {\bibfnamefont {B.~S.}\ \bibnamefont {Williams}}, \bibinfo {author}
  {\bibfnamefont {Q.}~\bibnamefont {Hu}}, \ and\ \bibinfo {author}
  {\bibfnamefont {J.~L.}\ \bibnamefont {Reno}},\ }\href@noop {} {\bibfield
  {journal} {\bibinfo  {journal} {Nature Photonics}\ }\textbf {\bibinfo
  {volume} {3}},\ \bibinfo {pages} {41} (\bibinfo {year} {2009})}\BibitemShut
  {NoStop}%
\bibitem [{\citenamefont {Hugi}\ \emph {et~al.}(2009)\citenamefont {Hugi},
  \citenamefont {Terazzi}, \citenamefont {Bonetti}, \citenamefont {Wittmann},
  \citenamefont {Fischer}, \citenamefont {Beck}, \citenamefont {Faist},\ and\
  \citenamefont {Gini}}]{Hugi2009}%
  \BibitemOpen
  \bibfield  {author} {\bibinfo {author} {\bibfnamefont {A.}~\bibnamefont
  {Hugi}}, \bibinfo {author} {\bibfnamefont {R.}~\bibnamefont {Terazzi}},
  \bibinfo {author} {\bibfnamefont {Y.}~\bibnamefont {Bonetti}}, \bibinfo
  {author} {\bibfnamefont {A.}~\bibnamefont {Wittmann}}, \bibinfo {author}
  {\bibfnamefont {M.}~\bibnamefont {Fischer}}, \bibinfo {author} {\bibfnamefont
  {M.}~\bibnamefont {Beck}}, \bibinfo {author} {\bibfnamefont {J.}~\bibnamefont
  {Faist}}, \ and\ \bibinfo {author} {\bibfnamefont {E.}~\bibnamefont {Gini}},\
  }\href@noop {} {\bibfield  {journal} {\bibinfo  {journal} {Applied Physics
  Letters}\ }\textbf {\bibinfo {volume} {95}},\ \bibinfo {pages} {061103}
  (\bibinfo {year} {2009})}\BibitemShut {NoStop}%
\bibitem [{\citenamefont {Hugi}\ \emph {et~al.}(2010)\citenamefont {Hugi},
  \citenamefont {Maulini},\ and\ \citenamefont {Faist}}]{Hugi2010}%
  \BibitemOpen
  \bibfield  {author} {\bibinfo {author} {\bibfnamefont {A.}~\bibnamefont
  {Hugi}}, \bibinfo {author} {\bibfnamefont {R.}~\bibnamefont {Maulini}}, \
  and\ \bibinfo {author} {\bibfnamefont {J.}~\bibnamefont {Faist}},\ }\href
  {\doibase 10.1088/0268-1242/25/8/083001} {\bibfield  {journal} {\bibinfo
  {journal} {Semiconductor Science and Technology}\ }\textbf {\bibinfo {volume}
  {25}},\ \bibinfo {pages} {083001} (\bibinfo {year} {2010})}\BibitemShut
  {NoStop}%
\bibitem [{\citenamefont {Faist}(2007)}]{Faist2007}%
  \BibitemOpen
  \bibfield  {author} {\bibinfo {author} {\bibfnamefont {J.}~\bibnamefont
  {Faist}},\ }\href@noop {} {\bibfield  {journal} {\bibinfo  {journal} {Applied
  Physics Letters}\ }\textbf {\bibinfo {volume} {90}},\ \bibinfo {pages}
  {253512} (\bibinfo {year} {2007})}\BibitemShut {NoStop}%
\bibitem [{\citenamefont {Bai}\ \emph {et~al.}(2010)\citenamefont {Bai},
  \citenamefont {Slivken}, \citenamefont {Kuboya}, \citenamefont {Darvish},\
  and\ \citenamefont {Razeghi}}]{Bai2010}%
  \BibitemOpen
  \bibfield  {author} {\bibinfo {author} {\bibfnamefont {Y.}~\bibnamefont
  {Bai}}, \bibinfo {author} {\bibfnamefont {S.}~\bibnamefont {Slivken}},
  \bibinfo {author} {\bibfnamefont {S.}~\bibnamefont {Kuboya}}, \bibinfo
  {author} {\bibfnamefont {S.~R.}\ \bibnamefont {Darvish}}, \ and\ \bibinfo
  {author} {\bibfnamefont {M.}~\bibnamefont {Razeghi}},\ }\href {\doibase
  10.1038/nphoton.2009.263} {\bibfield  {journal} {\bibinfo  {journal} {Nature
  Photonics}\ }\textbf {\bibinfo {volume} {4}},\ \bibinfo {pages} {99}
  (\bibinfo {year} {2010})}\BibitemShut {NoStop}%
\bibitem [{\citenamefont {Liu}\ \emph {et~al.}(2010{\natexlab{b}})\citenamefont
  {Liu}, \citenamefont {Hoffman}, \citenamefont {Escarra}, \citenamefont
  {Franz}, \citenamefont {Khurgin}, \citenamefont {Dikmelik}, \citenamefont
  {Wang}, \citenamefont {Fan},\ and\ \citenamefont {Gmachl}}]{Liu2010b}%
  \BibitemOpen
  \bibfield  {author} {\bibinfo {author} {\bibfnamefont {P.~Q.}\ \bibnamefont
  {Liu}}, \bibinfo {author} {\bibfnamefont {A.~J.}\ \bibnamefont {Hoffman}},
  \bibinfo {author} {\bibfnamefont {M.~D.}\ \bibnamefont {Escarra}}, \bibinfo
  {author} {\bibfnamefont {K.~J.}\ \bibnamefont {Franz}}, \bibinfo {author}
  {\bibfnamefont {J.~B.}\ \bibnamefont {Khurgin}}, \bibinfo {author}
  {\bibfnamefont {Y.}~\bibnamefont {Dikmelik}}, \bibinfo {author}
  {\bibfnamefont {X.}~\bibnamefont {Wang}}, \bibinfo {author} {\bibfnamefont
  {J.-Y.}\ \bibnamefont {Fan}}, \ and\ \bibinfo {author} {\bibfnamefont
  {C.~F.}\ \bibnamefont {Gmachl}},\ }\href {\doibase 10.1038/nphoton.2009.262}
  {\bibfield  {journal} {\bibinfo  {journal} {Nature Photonics}\ }\textbf
  {\bibinfo {volume} {4}},\ \bibinfo {pages} {95} (\bibinfo {year}
  {2010}{\natexlab{b}})}\BibitemShut {NoStop}%
\bibitem [{\citenamefont {Colombelli}\ \emph {et~al.}(2003)\citenamefont
  {Colombelli}, \citenamefont {Srinivasan}, \citenamefont {Troccoli},
  \citenamefont {Painter}, \citenamefont {Gmachl}, \citenamefont {Tennant},
  \citenamefont {Sergent}, \citenamefont {Sivco}, \citenamefont {Cho},\ and\
  \citenamefont {Capasso}}]{Colombelli2003}%
  \BibitemOpen
  \bibfield  {author} {\bibinfo {author} {\bibfnamefont {R.}~\bibnamefont
  {Colombelli}}, \bibinfo {author} {\bibfnamefont {K.}~\bibnamefont
  {Srinivasan}}, \bibinfo {author} {\bibfnamefont {M.}~\bibnamefont
  {Troccoli}}, \bibinfo {author} {\bibfnamefont {O.}~\bibnamefont {Painter}},
  \bibinfo {author} {\bibfnamefont {C.~F.}\ \bibnamefont {Gmachl}}, \bibinfo
  {author} {\bibfnamefont {D.~M.}\ \bibnamefont {Tennant}}, \bibinfo {author}
  {\bibfnamefont {A.~M.}\ \bibnamefont {Sergent}}, \bibinfo {author}
  {\bibfnamefont {D.~L.}\ \bibnamefont {Sivco}}, \bibinfo {author}
  {\bibfnamefont {A.~Y.}\ \bibnamefont {Cho}}, \ and\ \bibinfo {author}
  {\bibfnamefont {F.}~\bibnamefont {Capasso}},\ }\href {\doibase
  10.1126/science.1090561} {\bibfield  {journal} {\bibinfo  {journal}
  {Science}\ }\textbf {\bibinfo {volume} {302}},\ \bibinfo {pages} {1374}
  (\bibinfo {year} {2003})}\BibitemShut {NoStop}%
\bibitem [{\citenamefont {Rogalski}(2002)}]{Rogalski2002}%
  \BibitemOpen
  \bibfield  {author} {\bibinfo {author} {\bibfnamefont {A.}~\bibnamefont
  {Rogalski}},\ }\href {\doibase https://doi.org/10.1016/S1350-4495(02)00140-8}
  {\bibfield  {journal} {\bibinfo  {journal} {Infrared Physics \& Technology}\
  }\textbf {\bibinfo {volume} {43}},\ \bibinfo {pages} {187} (\bibinfo {year}
  {2002})}\BibitemShut {NoStop}%
\bibitem [{\citenamefont {Klapwijk}\ and\ \citenamefont
  {Semenov}(2017)}]{Klapwijk2017}%
  \BibitemOpen
  \bibfield  {author} {\bibinfo {author} {\bibfnamefont {T.~M.}\ \bibnamefont
  {Klapwijk}}\ and\ \bibinfo {author} {\bibfnamefont {A.~V.}\ \bibnamefont
  {Semenov}},\ }\href {\doibase 10.1109/TTHZ.2017.2758267} {\bibfield
  {journal} {\bibinfo  {journal} {IEEE Transactions on Terahertz Science and
  Technology}\ }\textbf {\bibinfo {volume} {7}},\ \bibinfo {pages} {627}
  (\bibinfo {year} {2017})}\BibitemShut {NoStop}%
\bibitem [{\citenamefont {Rogalski}(2005)}]{Rogalski2005}%
  \BibitemOpen
  \bibfield  {author} {\bibinfo {author} {\bibfnamefont {A.}~\bibnamefont
  {Rogalski}},\ }\href {\doibase 10.1088/0034-4885/68/10/R01} {\bibfield
  {journal} {\bibinfo  {journal} {Reports on Progress in Physics}\ }\textbf
  {\bibinfo {volume} {68}},\ \bibinfo {pages} {2267} (\bibinfo {year}
  {2005})}\BibitemShut {NoStop}%
\bibitem [{\citenamefont {Theocharous}(2012)}]{Theocharous2012}%
  \BibitemOpen
  \bibfield  {author} {\bibinfo {author} {\bibfnamefont {E.}~\bibnamefont
  {Theocharous}},\ }\href {\doibase 10.1088/0026-1394/49/2/s99} {\bibfield
  {journal} {\bibinfo  {journal} {Metrologia}\ }\textbf {\bibinfo {volume}
  {49}},\ \bibinfo {pages} {S99} (\bibinfo {year} {2012})}\BibitemShut
  {NoStop}%
\bibitem [{\citenamefont {Koppens}\ \emph {et~al.}(2014)\citenamefont
  {Koppens}, \citenamefont {Mueller}, \citenamefont {Avouris}, \citenamefont
  {Ferrari}, \citenamefont {Vitiello},\ and\ \citenamefont
  {Polini}}]{Koppens2014}%
  \BibitemOpen
  \bibfield  {author} {\bibinfo {author} {\bibfnamefont {F.~H.~L.}\
  \bibnamefont {Koppens}}, \bibinfo {author} {\bibfnamefont {T.}~\bibnamefont
  {Mueller}}, \bibinfo {author} {\bibfnamefont {P.}~\bibnamefont {Avouris}},
  \bibinfo {author} {\bibfnamefont {A.~C.}\ \bibnamefont {Ferrari}}, \bibinfo
  {author} {\bibfnamefont {M.~S.}\ \bibnamefont {Vitiello}}, \ and\ \bibinfo
  {author} {\bibfnamefont {M.}~\bibnamefont {Polini}},\ }\href {\doibase
  10.1038/nnano.2014.215} {\bibfield  {journal} {\bibinfo  {journal} {Nature
  Nanotechnology}\ }\textbf {\bibinfo {volume} {9}},\ \bibinfo {pages} {780}
  (\bibinfo {year} {2014})}\BibitemShut {NoStop}%
\bibitem [{\citenamefont {Schneider}\ and\ \citenamefont
  {Liu}(2007)}]{Schneider2007}%
  \BibitemOpen
  \bibfield  {author} {\bibinfo {author} {\bibfnamefont {H.}~\bibnamefont
  {Schneider}}\ and\ \bibinfo {author} {\bibfnamefont {H.~C.}\ \bibnamefont
  {Liu}},\ }in\ \href {\doibase 10.1007/978-3-540-36324-8_4} {\emph {\bibinfo
  {booktitle} {Quantum Well Infrared Photodetectors}}}\ (\bibinfo  {publisher}
  {Springer Berlin Heidelberg},\ \bibinfo {year} {2007})\ pp.\ \bibinfo {pages}
  {45--81}\BibitemShut {NoStop}%
\bibitem [{\citenamefont {Rogalski}\ \emph {et~al.}(2017)\citenamefont
  {Rogalski}, \citenamefont {Martyniuk},\ and\ \citenamefont
  {Kopytko}}]{Rogalski2017}%
  \BibitemOpen
  \bibfield  {author} {\bibinfo {author} {\bibfnamefont {A.}~\bibnamefont
  {Rogalski}}, \bibinfo {author} {\bibfnamefont {P.}~\bibnamefont {Martyniuk}},
  \ and\ \bibinfo {author} {\bibfnamefont {M.}~\bibnamefont {Kopytko}},\ }\href
  {\doibase 10.1063/1.4999077} {\bibfield  {journal} {\bibinfo  {journal}
  {Applied Physics Reviews}\ }\textbf {\bibinfo {volume} {4}},\ \bibinfo
  {pages} {031304} (\bibinfo {year} {2017})}\BibitemShut {NoStop}%
\bibitem [{\citenamefont {Palaferri}\ \emph {et~al.}(2015)\citenamefont
  {Palaferri}, \citenamefont {Todorov}, \citenamefont {Chen}, \citenamefont
  {Madeo}, \citenamefont {Vasanelli}, \citenamefont {Li}, \citenamefont
  {Davies}, \citenamefont {Linfield},\ and\ \citenamefont
  {Sirtori}}]{Palaferri2015}%
  \BibitemOpen
  \bibfield  {author} {\bibinfo {author} {\bibfnamefont {D.}~\bibnamefont
  {Palaferri}}, \bibinfo {author} {\bibfnamefont {Y.}~\bibnamefont {Todorov}},
  \bibinfo {author} {\bibfnamefont {Y.~N.}\ \bibnamefont {Chen}}, \bibinfo
  {author} {\bibfnamefont {J.}~\bibnamefont {Madeo}}, \bibinfo {author}
  {\bibfnamefont {A.}~\bibnamefont {Vasanelli}}, \bibinfo {author}
  {\bibfnamefont {L.~H.}\ \bibnamefont {Li}}, \bibinfo {author} {\bibfnamefont
  {A.~G.}\ \bibnamefont {Davies}}, \bibinfo {author} {\bibfnamefont {E.~H.}\
  \bibnamefont {Linfield}}, \ and\ \bibinfo {author} {\bibfnamefont
  {C.}~\bibnamefont {Sirtori}},\ }\href {\doibase 10.1063/1.4918983} {\bibfield
   {journal} {\bibinfo  {journal} {Applied Physics Letters}\ }\textbf {\bibinfo
  {volume} {106}},\ \bibinfo {pages} {1} (\bibinfo {year} {2015})}\BibitemShut
  {NoStop}%
\bibitem [{\citenamefont {Vigneron}\ \emph {et~al.}(2019)\citenamefont
  {Vigneron}, \citenamefont {Pirotta}, \citenamefont {Carusotto}, \citenamefont
  {Tran}, \citenamefont {Biasiol}, \citenamefont {Manceau}, \citenamefont
  {Bousseksou},\ and\ \citenamefont {Colombelli}}]{Vigneron2019}%
  \BibitemOpen
  \bibfield  {author} {\bibinfo {author} {\bibfnamefont {P.-B.}\ \bibnamefont
  {Vigneron}}, \bibinfo {author} {\bibfnamefont {S.}~\bibnamefont {Pirotta}},
  \bibinfo {author} {\bibfnamefont {I.}~\bibnamefont {Carusotto}}, \bibinfo
  {author} {\bibfnamefont {N.-L.}\ \bibnamefont {Tran}}, \bibinfo {author}
  {\bibfnamefont {G.}~\bibnamefont {Biasiol}}, \bibinfo {author} {\bibfnamefont
  {J.-M.}\ \bibnamefont {Manceau}}, \bibinfo {author} {\bibfnamefont
  {A.}~\bibnamefont {Bousseksou}}, \ and\ \bibinfo {author} {\bibfnamefont
  {R.}~\bibnamefont {Colombelli}},\ }\href {\doibase 10.1063/1.5084112}
  {\bibfield  {journal} {\bibinfo  {journal} {Applied Physics Letters}\
  }\textbf {\bibinfo {volume} {114}},\ \bibinfo {pages} {131104} (\bibinfo
  {year} {2019})}\BibitemShut {NoStop}%
\bibitem [{\citenamefont {Miyazaki}\ \emph {et~al.}(2020)\citenamefont
  {Miyazaki}, \citenamefont {Mano}, \citenamefont {Kasaya}, \citenamefont
  {Osato}, \citenamefont {Watanabe}, \citenamefont {Sugimoto}, \citenamefont
  {Kawazu}, \citenamefont {Arai}, \citenamefont {Shigetou}, \citenamefont
  {Ochiai}, \citenamefont {Jimba},\ and\ \citenamefont
  {Miyazaki}}]{Miyazaki2020}%
  \BibitemOpen
  \bibfield  {author} {\bibinfo {author} {\bibfnamefont {H.~T.}\ \bibnamefont
  {Miyazaki}}, \bibinfo {author} {\bibfnamefont {T.}~\bibnamefont {Mano}},
  \bibinfo {author} {\bibfnamefont {T.}~\bibnamefont {Kasaya}}, \bibinfo
  {author} {\bibfnamefont {H.}~\bibnamefont {Osato}}, \bibinfo {author}
  {\bibfnamefont {K.}~\bibnamefont {Watanabe}}, \bibinfo {author}
  {\bibfnamefont {Y.}~\bibnamefont {Sugimoto}}, \bibinfo {author}
  {\bibfnamefont {T.}~\bibnamefont {Kawazu}}, \bibinfo {author} {\bibfnamefont
  {Y.}~\bibnamefont {Arai}}, \bibinfo {author} {\bibfnamefont {A.}~\bibnamefont
  {Shigetou}}, \bibinfo {author} {\bibfnamefont {T.}~\bibnamefont {Ochiai}},
  \bibinfo {author} {\bibfnamefont {Y.}~\bibnamefont {Jimba}}, \ and\ \bibinfo
  {author} {\bibfnamefont {H.}~\bibnamefont {Miyazaki}},\ }\href {\doibase
  10.1038/s41467-020-14426-6} {\bibfield  {journal} {\bibinfo  {journal}
  {Nature Communications}\ }\textbf {\bibinfo {volume} {11}},\ \bibinfo {pages}
  {565} (\bibinfo {year} {2020})}\BibitemShut {NoStop}%
\bibitem [{\citenamefont {Foteinopoulou}\ \emph {et~al.}(2019)\citenamefont
  {Foteinopoulou}, \citenamefont {{Devarapu Ganga Chinna}}, \citenamefont
  {{Subramania Ganapathi}}, \citenamefont {Krishna},\ and\ \citenamefont
  {Wasserman}}]{Foteinopoulou2019}%
  \BibitemOpen
  \bibfield  {author} {\bibinfo {author} {\bibfnamefont {S.}~\bibnamefont
  {Foteinopoulou}}, \bibinfo {author} {\bibfnamefont {R.}~\bibnamefont
  {{Devarapu Ganga Chinna}}}, \bibinfo {author} {\bibfnamefont
  {S.}~\bibnamefont {{Subramania Ganapathi}}}, \bibinfo {author} {\bibfnamefont
  {S.}~\bibnamefont {Krishna}}, \ and\ \bibinfo {author} {\bibfnamefont
  {D.}~\bibnamefont {Wasserman}},\ }\href {\doibase 10.1515/nanoph-2019-0232}
  {\enquote {\bibinfo {title} {{Phonon-polaritonics: enabling powerful
  capabilities for infrared photonics}},}\ } (\bibinfo {year}
  {2019})\BibitemShut {NoStop}%
\bibitem [{\citenamefont {Le~Gall}\ \emph {et~al.}(1997)\citenamefont
  {Le~Gall}, \citenamefont {Olivier},\ and\ \citenamefont
  {Greffet}}]{Greffet1997}%
  \BibitemOpen
  \bibfield  {author} {\bibinfo {author} {\bibfnamefont {J.}~\bibnamefont
  {Le~Gall}}, \bibinfo {author} {\bibfnamefont {M.}~\bibnamefont {Olivier}}, \
  and\ \bibinfo {author} {\bibfnamefont {J.-J.}\ \bibnamefont {Greffet}},\
  }\href {\doibase 10.1103/PhysRevB.55.10105} {\bibfield  {journal} {\bibinfo
  {journal} {Phys. Rev. B}\ }\textbf {\bibinfo {volume} {55}},\ \bibinfo
  {pages} {10105} (\bibinfo {year} {1997})}\BibitemShut {NoStop}%
\bibitem [{\citenamefont {Arnold}\ \emph {et~al.}(2012)\citenamefont {Arnold},
  \citenamefont {Marquier}, \citenamefont {Garin}, \citenamefont {Pardo},
  \citenamefont {Collin}, \citenamefont {Bardou}, \citenamefont {Pelouard},\
  and\ \citenamefont {Greffet}}]{Arnold2012}%
  \BibitemOpen
  \bibfield  {author} {\bibinfo {author} {\bibfnamefont {C.}~\bibnamefont
  {Arnold}}, \bibinfo {author} {\bibfnamefont {F.~m.~c.}\ \bibnamefont
  {Marquier}}, \bibinfo {author} {\bibfnamefont {M.}~\bibnamefont {Garin}},
  \bibinfo {author} {\bibfnamefont {F.}~\bibnamefont {Pardo}}, \bibinfo
  {author} {\bibfnamefont {S.}~\bibnamefont {Collin}}, \bibinfo {author}
  {\bibfnamefont {N.}~\bibnamefont {Bardou}}, \bibinfo {author} {\bibfnamefont
  {J.-L.}\ \bibnamefont {Pelouard}}, \ and\ \bibinfo {author} {\bibfnamefont
  {J.-J.}\ \bibnamefont {Greffet}},\ }\href {\doibase
  10.1103/PhysRevB.86.035316} {\bibfield  {journal} {\bibinfo  {journal} {Phys.
  Rev. B}\ }\textbf {\bibinfo {volume} {86}},\ \bibinfo {pages} {035316}
  (\bibinfo {year} {2012})}\BibitemShut {NoStop}%
\bibitem [{\citenamefont {Chalabi}\ \emph {et~al.}(2016)\citenamefont
  {Chalabi}, \citenamefont {Al\`u},\ and\ \citenamefont
  {Brongersma}}]{Chalabi2016}%
  \BibitemOpen
  \bibfield  {author} {\bibinfo {author} {\bibfnamefont {H.}~\bibnamefont
  {Chalabi}}, \bibinfo {author} {\bibfnamefont {A.}~\bibnamefont {Al\`u}}, \
  and\ \bibinfo {author} {\bibfnamefont {M.~L.}\ \bibnamefont {Brongersma}},\
  }\href {\doibase 10.1103/PhysRevB.94.094307} {\bibfield  {journal} {\bibinfo
  {journal} {Phys. Rev. B}\ }\textbf {\bibinfo {volume} {94}},\ \bibinfo
  {pages} {094307} (\bibinfo {year} {2016})}\BibitemShut {NoStop}%
\bibitem [{\citenamefont {Lu}\ \emph {et~al.}(2021{\natexlab{a}})\citenamefont
  {Lu}, \citenamefont {Tadjer}, \citenamefont {Caldwell},\ and\ \citenamefont
  {Folland}}]{Lu2021b}%
  \BibitemOpen
  \bibfield  {author} {\bibinfo {author} {\bibfnamefont {G.}~\bibnamefont
  {Lu}}, \bibinfo {author} {\bibfnamefont {M.}~\bibnamefont {Tadjer}}, \bibinfo
  {author} {\bibfnamefont {J.~D.}\ \bibnamefont {Caldwell}}, \ and\ \bibinfo
  {author} {\bibfnamefont {T.~G.}\ \bibnamefont {Folland}},\ }\href {\doibase
  10.1063/5.0048514} {\bibfield  {journal} {\bibinfo  {journal} {Applied
  Physics Letters}\ }\textbf {\bibinfo {volume} {118}},\ \bibinfo {pages}
  {141102} (\bibinfo {year} {2021}{\natexlab{a}})}\BibitemShut {NoStop}%
\bibitem [{\citenamefont {Gentle}\ and\ \citenamefont
  {Smith}(2010)}]{Gentle2010}%
  \BibitemOpen
  \bibfield  {author} {\bibinfo {author} {\bibfnamefont {A.~R.}\ \bibnamefont
  {Gentle}}\ and\ \bibinfo {author} {\bibfnamefont {G.~B.}\ \bibnamefont
  {Smith}},\ }\href {\doibase 10.1021/nl903271d} {\bibfield  {journal}
  {\bibinfo  {journal} {Nano Letters}\ }\textbf {\bibinfo {volume} {10}},\
  \bibinfo {pages} {373} (\bibinfo {year} {2010})},\ \bibinfo {note} {pMID:
  20055479}\BibitemShut {NoStop}%
\bibitem [{Dun(2018)}]{Dunkelberger2017}%
  \BibitemOpen
  \href {\doibase 10.1038/s41566-017-0069-0} {\bibfield  {journal} {\bibinfo
  {journal} {Nature Photonics}\ }\textbf {\bibinfo {volume} {12}},\ \bibinfo
  {pages} {50} (\bibinfo {year} {2018})}\BibitemShut {NoStop}%
\bibitem [{\citenamefont {Gubbin}\ \emph {et~al.}(2016)\citenamefont {Gubbin},
  \citenamefont {Martini}, \citenamefont {Politi}, \citenamefont {Maier},\ and\
  \citenamefont {De~Liberato}}]{Gubbin2016}%
  \BibitemOpen
  \bibfield  {author} {\bibinfo {author} {\bibfnamefont {C.~R.}\ \bibnamefont
  {Gubbin}}, \bibinfo {author} {\bibfnamefont {F.}~\bibnamefont {Martini}},
  \bibinfo {author} {\bibfnamefont {A.}~\bibnamefont {Politi}}, \bibinfo
  {author} {\bibfnamefont {S.~A.}\ \bibnamefont {Maier}}, \ and\ \bibinfo
  {author} {\bibfnamefont {S.}~\bibnamefont {De~Liberato}},\ }\href {\doibase
  10.1103/PhysRevLett.116.246402} {\bibfield  {journal} {\bibinfo  {journal}
  {Phys. Rev. Lett.}\ }\textbf {\bibinfo {volume} {116}},\ \bibinfo {pages}
  {246402} (\bibinfo {year} {2016})}\BibitemShut {NoStop}%
\bibitem [{\citenamefont {Lu}\ \emph {et~al.}(2021{\natexlab{b}})\citenamefont
  {Lu}, \citenamefont {Gubbin}, \citenamefont {Nolen}, \citenamefont {Folland},
  \citenamefont {Tadjer}, \citenamefont {De~Liberato},\ and\ \citenamefont
  {Caldwell}}]{Lu2021}%
  \BibitemOpen
  \bibfield  {author} {\bibinfo {author} {\bibfnamefont {G.}~\bibnamefont
  {Lu}}, \bibinfo {author} {\bibfnamefont {C.~R.}\ \bibnamefont {Gubbin}},
  \bibinfo {author} {\bibfnamefont {J.~R.}\ \bibnamefont {Nolen}}, \bibinfo
  {author} {\bibfnamefont {T.}~\bibnamefont {Folland}}, \bibinfo {author}
  {\bibfnamefont {M.~J.}\ \bibnamefont {Tadjer}}, \bibinfo {author}
  {\bibfnamefont {S.}~\bibnamefont {De~Liberato}}, \ and\ \bibinfo {author}
  {\bibfnamefont {J.~D.}\ \bibnamefont {Caldwell}},\ }\href {\doibase
  10.1021/acs.nanolett.0c04767} {\bibfield  {journal} {\bibinfo  {journal}
  {Nano Letters}\ }\textbf {\bibinfo {volume} {21}},\ \bibinfo {pages} {1831}
  (\bibinfo {year} {2021}{\natexlab{b}})},\ \bibinfo {note} {pMID:
  33587855}\BibitemShut {NoStop}%
\bibitem [{\citenamefont {Francoeur}\ \emph {et~al.}(2008)\citenamefont
  {Francoeur}, \citenamefont {Mengüç},\ and\ \citenamefont
  {Vaillon}}]{Francoeur2008}%
  \BibitemOpen
  \bibfield  {author} {\bibinfo {author} {\bibfnamefont {M.}~\bibnamefont
  {Francoeur}}, \bibinfo {author} {\bibfnamefont {M.~P.}\ \bibnamefont
  {Mengüç}}, \ and\ \bibinfo {author} {\bibfnamefont {R.}~\bibnamefont
  {Vaillon}},\ }\href {\doibase 10.1063/1.2963195} {\bibfield  {journal}
  {\bibinfo  {journal} {Applied Physics Letters}\ }\textbf {\bibinfo {volume}
  {93}},\ \bibinfo {pages} {043109} (\bibinfo {year} {2008})}\BibitemShut
  {NoStop}%
\bibitem [{\citenamefont {Pendry}(1999)}]{Pendry1999}%
  \BibitemOpen
  \bibfield  {author} {\bibinfo {author} {\bibfnamefont {J.~B.}\ \bibnamefont
  {Pendry}},\ }\href@noop {} {\bibfield  {journal} {\bibinfo  {journal}
  {Journal of Physics: Condensed Matter}\ }\textbf {\bibinfo {volume} {11}},\
  \bibinfo {pages} {6621} (\bibinfo {year} {1999})}\BibitemShut {NoStop}%
\bibitem [{\citenamefont {Hu}\ \emph {et~al.}(2008)\citenamefont {Hu},
  \citenamefont {Narayanaswamy}, \citenamefont {Chen},\ and\ \citenamefont
  {Chen}}]{Hu2008}%
  \BibitemOpen
  \bibfield  {author} {\bibinfo {author} {\bibfnamefont {L.}~\bibnamefont
  {Hu}}, \bibinfo {author} {\bibfnamefont {A.}~\bibnamefont {Narayanaswamy}},
  \bibinfo {author} {\bibfnamefont {X.}~\bibnamefont {Chen}}, \ and\ \bibinfo
  {author} {\bibfnamefont {G.}~\bibnamefont {Chen}},\ }\href {\doibase
  10.1063/1.2905286} {\bibfield  {journal} {\bibinfo  {journal} {Applied
  Physics Letters}\ }\textbf {\bibinfo {volume} {92}},\ \bibinfo {pages}
  {133106} (\bibinfo {year} {2008})}\BibitemShut {NoStop}%
\bibitem [{\citenamefont {Ghashami}\ \emph {et~al.}(2018)\citenamefont
  {Ghashami}, \citenamefont {Geng}, \citenamefont {Kim}, \citenamefont
  {Iacopino}, \citenamefont {Cho},\ and\ \citenamefont {Park}}]{Ghashami2018}%
  \BibitemOpen
  \bibfield  {author} {\bibinfo {author} {\bibfnamefont {M.}~\bibnamefont
  {Ghashami}}, \bibinfo {author} {\bibfnamefont {H.}~\bibnamefont {Geng}},
  \bibinfo {author} {\bibfnamefont {T.}~\bibnamefont {Kim}}, \bibinfo {author}
  {\bibfnamefont {N.}~\bibnamefont {Iacopino}}, \bibinfo {author}
  {\bibfnamefont {S.~K.}\ \bibnamefont {Cho}}, \ and\ \bibinfo {author}
  {\bibfnamefont {K.}~\bibnamefont {Park}},\ }\href@noop {} {\bibfield
  {journal} {\bibinfo  {journal} {Physical Review Letters}\ }\textbf {\bibinfo
  {volume} {120}},\ \bibinfo {pages} {175901} (\bibinfo {year}
  {2018})}\BibitemShut {NoStop}%
\bibitem [{\citenamefont {Tang}\ \emph {et~al.}(2020)\citenamefont {Tang},
  \citenamefont {DeSutter},\ and\ \citenamefont {Francoeur}}]{Tang2020}%
  \BibitemOpen
  \bibfield  {author} {\bibinfo {author} {\bibfnamefont {L.}~\bibnamefont
  {Tang}}, \bibinfo {author} {\bibfnamefont {J.}~\bibnamefont {DeSutter}}, \
  and\ \bibinfo {author} {\bibfnamefont {M.}~\bibnamefont {Francoeur}},\ }\href
  {\doibase 10.1021/acsphotonics.0c00404} {\bibfield  {journal} {\bibinfo
  {journal} {ACS Photonics}\ }\textbf {\bibinfo {volume} {7}},\ \bibinfo
  {pages} {1304} (\bibinfo {year} {2020})}\BibitemShut {NoStop}%
\bibitem [{\citenamefont {Fiorino}\ \emph {et~al.}(2018)\citenamefont
  {Fiorino}, \citenamefont {Zhu}, \citenamefont {Thompson}, \citenamefont
  {Mittapally}, \citenamefont {Reddy},\ and\ \citenamefont
  {Meyhofer}}]{Fiorino2018}%
  \BibitemOpen
  \bibfield  {author} {\bibinfo {author} {\bibfnamefont {A.}~\bibnamefont
  {Fiorino}}, \bibinfo {author} {\bibfnamefont {L.}~\bibnamefont {Zhu}},
  \bibinfo {author} {\bibfnamefont {D.}~\bibnamefont {Thompson}}, \bibinfo
  {author} {\bibfnamefont {R.}~\bibnamefont {Mittapally}}, \bibinfo {author}
  {\bibfnamefont {P.}~\bibnamefont {Reddy}}, \ and\ \bibinfo {author}
  {\bibfnamefont {E.}~\bibnamefont {Meyhofer}},\ }\href@noop {} {\bibfield
  {journal} {\bibinfo  {journal} {Nature Nanotechnology}\ }\textbf {\bibinfo
  {volume} {13}},\ \bibinfo {pages} {806} (\bibinfo {year} {2018})}\BibitemShut
  {NoStop}%
\bibitem [{\citenamefont {Narayanaswamy}\ and\ \citenamefont
  {Chen}(2003)}]{Narayanaswamy2003}%
  \BibitemOpen
  \bibfield  {author} {\bibinfo {author} {\bibfnamefont {A.}~\bibnamefont
  {Narayanaswamy}}\ and\ \bibinfo {author} {\bibfnamefont {G.}~\bibnamefont
  {Chen}},\ }\href {\doibase 10.1063/1.1575936} {\bibfield  {journal} {\bibinfo
   {journal} {Applied Physics Letters}\ }\textbf {\bibinfo {volume} {82}},\
  \bibinfo {pages} {3544} (\bibinfo {year} {2003})}\BibitemShut {NoStop}%
\bibitem [{\citenamefont {Laroche}\ \emph {et~al.}(2006)\citenamefont
  {Laroche}, \citenamefont {Carminati},\ and\ \citenamefont
  {Greffet}}]{Laroche2006}%
  \BibitemOpen
  \bibfield  {author} {\bibinfo {author} {\bibfnamefont {M.}~\bibnamefont
  {Laroche}}, \bibinfo {author} {\bibfnamefont {R.}~\bibnamefont {Carminati}},
  \ and\ \bibinfo {author} {\bibfnamefont {J.-J.}\ \bibnamefont {Greffet}},\
  }\href {\doibase 10.1063/1.2234560} {\bibfield  {journal} {\bibinfo
  {journal} {Journal of Applied Physics}\ }\textbf {\bibinfo {volume} {100}},\
  \bibinfo {pages} {063704} (\bibinfo {year} {2006})}\BibitemShut {NoStop}%
\bibitem [{\citenamefont {Song}\ \emph {et~al.}(2015)\citenamefont {Song},
  \citenamefont {Ganjeh}, \citenamefont {Sadat}, \citenamefont {Thompson},
  \citenamefont {Fiorino}, \citenamefont {Fern{\'a}ndez-Hurtado}, \citenamefont
  {Feist}, \citenamefont {Garcia-Vidal}, \citenamefont {Cuevas}, \citenamefont
  {Reddy},\ and\ \citenamefont {Meyhofer}}]{Song2015}%
  \BibitemOpen
  \bibfield  {author} {\bibinfo {author} {\bibfnamefont {B.}~\bibnamefont
  {Song}}, \bibinfo {author} {\bibfnamefont {Y.}~\bibnamefont {Ganjeh}},
  \bibinfo {author} {\bibfnamefont {S.}~\bibnamefont {Sadat}}, \bibinfo
  {author} {\bibfnamefont {D.}~\bibnamefont {Thompson}}, \bibinfo {author}
  {\bibfnamefont {A.}~\bibnamefont {Fiorino}}, \bibinfo {author} {\bibfnamefont
  {V.}~\bibnamefont {Fern{\'a}ndez-Hurtado}}, \bibinfo {author} {\bibfnamefont
  {J.}~\bibnamefont {Feist}}, \bibinfo {author} {\bibfnamefont {F.~J.}\
  \bibnamefont {Garcia-Vidal}}, \bibinfo {author} {\bibfnamefont {J.~C.}\
  \bibnamefont {Cuevas}}, \bibinfo {author} {\bibfnamefont {P.}~\bibnamefont
  {Reddy}}, \ and\ \bibinfo {author} {\bibfnamefont {E.}~\bibnamefont
  {Meyhofer}},\ }\href {\doibase 10.1038/nnano.2015.6} {\bibfield  {journal}
  {\bibinfo  {journal} {Nature Nanotechnology}\ }\textbf {\bibinfo {volume}
  {10}},\ \bibinfo {pages} {253} (\bibinfo {year} {2015})}\BibitemShut
  {NoStop}%
\bibitem [{\citenamefont {Tielrooij}\ \emph {et~al.}(2018)\citenamefont
  {Tielrooij}, \citenamefont {Hesp}, \citenamefont {Principi}, \citenamefont
  {Lundeberg}, \citenamefont {Pogna}, \citenamefont {Banszerus}, \citenamefont
  {Mics}, \citenamefont {Massicotte}, \citenamefont {Schmidt}, \citenamefont
  {Davydovskaya}, \citenamefont {Purdie}, \citenamefont {Goykhman},
  \citenamefont {Soavi}, \citenamefont {Lombardo}, \citenamefont {Watanabe},
  \citenamefont {Taniguchi}, \citenamefont {Bonn}, \citenamefont
  {Turchinovich}, \citenamefont {Stampfer}, \citenamefont {Ferrari},
  \citenamefont {Cerullo}, \citenamefont {Polini},\ and\ \citenamefont
  {Koppens}}]{Tielrooij2018}%
  \BibitemOpen
  \bibfield  {author} {\bibinfo {author} {\bibfnamefont {K.-J.}\ \bibnamefont
  {Tielrooij}}, \bibinfo {author} {\bibfnamefont {N.~C.~H.}\ \bibnamefont
  {Hesp}}, \bibinfo {author} {\bibfnamefont {A.}~\bibnamefont {Principi}},
  \bibinfo {author} {\bibfnamefont {M.~B.}\ \bibnamefont {Lundeberg}}, \bibinfo
  {author} {\bibfnamefont {E.~A.~A.}\ \bibnamefont {Pogna}}, \bibinfo {author}
  {\bibfnamefont {L.}~\bibnamefont {Banszerus}}, \bibinfo {author}
  {\bibfnamefont {Z.}~\bibnamefont {Mics}}, \bibinfo {author} {\bibfnamefont
  {M.}~\bibnamefont {Massicotte}}, \bibinfo {author} {\bibfnamefont
  {P.}~\bibnamefont {Schmidt}}, \bibinfo {author} {\bibfnamefont
  {D.}~\bibnamefont {Davydovskaya}}, \bibinfo {author} {\bibfnamefont {D.~G.}\
  \bibnamefont {Purdie}}, \bibinfo {author} {\bibfnamefont {I.}~\bibnamefont
  {Goykhman}}, \bibinfo {author} {\bibfnamefont {G.}~\bibnamefont {Soavi}},
  \bibinfo {author} {\bibfnamefont {A.}~\bibnamefont {Lombardo}}, \bibinfo
  {author} {\bibfnamefont {K.}~\bibnamefont {Watanabe}}, \bibinfo {author}
  {\bibfnamefont {T.}~\bibnamefont {Taniguchi}}, \bibinfo {author}
  {\bibfnamefont {M.}~\bibnamefont {Bonn}}, \bibinfo {author} {\bibfnamefont
  {D.}~\bibnamefont {Turchinovich}}, \bibinfo {author} {\bibfnamefont
  {C.}~\bibnamefont {Stampfer}}, \bibinfo {author} {\bibfnamefont {A.~C.}\
  \bibnamefont {Ferrari}}, \bibinfo {author} {\bibfnamefont {G.}~\bibnamefont
  {Cerullo}}, \bibinfo {author} {\bibfnamefont {M.}~\bibnamefont {Polini}}, \
  and\ \bibinfo {author} {\bibfnamefont {F.~H.~L.}\ \bibnamefont {Koppens}},\
  }\href {\doibase 10.1038/s41565-017-0008-8} {\bibfield  {journal} {\bibinfo
  {journal} {Nature Nanotechnology}\ }\textbf {\bibinfo {volume} {13}},\
  \bibinfo {pages} {41} (\bibinfo {year} {2018})}\BibitemShut {NoStop}%
\bibitem [{\citenamefont {Constant}\ \emph {et~al.}(2016)\citenamefont
  {Constant}, \citenamefont {Hornett}, \citenamefont {Chang},\ and\
  \citenamefont {Hendry}}]{Constant2016}%
  \BibitemOpen
  \bibfield  {author} {\bibinfo {author} {\bibfnamefont {T.~J.}\ \bibnamefont
  {Constant}}, \bibinfo {author} {\bibfnamefont {S.~M.}\ \bibnamefont
  {Hornett}}, \bibinfo {author} {\bibfnamefont {D.~E.}\ \bibnamefont {Chang}},
  \ and\ \bibinfo {author} {\bibfnamefont {E.}~\bibnamefont {Hendry}},\ }\href
  {\doibase 10.1038/nphys3545} {\bibfield  {journal} {\bibinfo  {journal}
  {Nature Physics}\ }\textbf {\bibinfo {volume} {12}},\ \bibinfo {pages} {124}
  (\bibinfo {year} {2016})}\BibitemShut {NoStop}%
\bibitem [{\citenamefont {Gubbin}\ and\ \citenamefont
  {De~Liberato}(2017)}]{Gubbin2017}%
  \BibitemOpen
  \bibfield  {author} {\bibinfo {author} {\bibfnamefont {C.~R.}\ \bibnamefont
  {Gubbin}}\ and\ \bibinfo {author} {\bibfnamefont {S.}~\bibnamefont
  {De~Liberato}},\ }\href {\doibase 10.1021/acsphotonics.7b00020} {\bibfield
  {journal} {\bibinfo  {journal} {ACS Photonics}\ }\textbf {\bibinfo {volume}
  {4}},\ \bibinfo {pages} {1381} (\bibinfo {year} {2017})}\BibitemShut
  {NoStop}%
\bibitem [{\citenamefont {Petrov}(2015)}]{Petrov2015}%
  \BibitemOpen
  \bibfield  {author} {\bibinfo {author} {\bibfnamefont {V.}~\bibnamefont
  {Petrov}},\ }\href {\doibase
  https://doi.org/10.1016/j.pquantelec.2015.04.001} {\bibfield  {journal}
  {\bibinfo  {journal} {Progress in Quantum Electronics}\ }\textbf {\bibinfo
  {volume} {42}},\ \bibinfo {pages} {1} (\bibinfo {year} {2015})}\BibitemShut
  {NoStop}%
\bibitem [{\citenamefont {Gubbin}\ and\ \citenamefont
  {De~Liberato}(2018)}]{Gubbin2018}%
  \BibitemOpen
  \bibfield  {author} {\bibinfo {author} {\bibfnamefont {C.~R.}\ \bibnamefont
  {Gubbin}}\ and\ \bibinfo {author} {\bibfnamefont {S.}~\bibnamefont
  {De~Liberato}},\ }\href {\doibase 10.1021/acsphotonics.7b00863} {\bibfield
  {journal} {\bibinfo  {journal} {ACS Photonics}\ }\textbf {\bibinfo {volume}
  {5}},\ \bibinfo {pages} {284} (\bibinfo {year} {2018})}\BibitemShut {NoStop}%
\bibitem [{\citenamefont {Vanderbilt}\ \emph {et~al.}(1986)\citenamefont
  {Vanderbilt}, \citenamefont {Louie},\ and\ \citenamefont
  {Cohen}}]{Vanderbilt1986}%
  \BibitemOpen
  \bibfield  {author} {\bibinfo {author} {\bibfnamefont {D.}~\bibnamefont
  {Vanderbilt}}, \bibinfo {author} {\bibfnamefont {S.~G.}\ \bibnamefont
  {Louie}}, \ and\ \bibinfo {author} {\bibfnamefont {M.~L.}\ \bibnamefont
  {Cohen}},\ }\href {\doibase 10.1103/PhysRevB.33.8740} {\bibfield  {journal}
  {\bibinfo  {journal} {Phys. Rev. B}\ }\textbf {\bibinfo {volume} {33}},\
  \bibinfo {pages} {8740} (\bibinfo {year} {1986})}\BibitemShut {NoStop}%
\bibitem [{\citenamefont {Teitelbaum}\ \emph {et~al.}(2018)\citenamefont
  {Teitelbaum}, \citenamefont {Henighan}, \citenamefont {Huang}, \citenamefont
  {Liu}, \citenamefont {Jiang}, \citenamefont {Zhu}, \citenamefont {Chollet},
  \citenamefont {Sato}, \citenamefont {Murray}, \citenamefont {Fahy},
  \citenamefont {O'Mahony}, \citenamefont {Bailey}, \citenamefont {Uher},
  \citenamefont {Trigo},\ and\ \citenamefont {Reis}}]{Teitelbaum2018}%
  \BibitemOpen
  \bibfield  {author} {\bibinfo {author} {\bibfnamefont {S.~W.}\ \bibnamefont
  {Teitelbaum}}, \bibinfo {author} {\bibfnamefont {T.}~\bibnamefont
  {Henighan}}, \bibinfo {author} {\bibfnamefont {Y.}~\bibnamefont {Huang}},
  \bibinfo {author} {\bibfnamefont {H.}~\bibnamefont {Liu}}, \bibinfo {author}
  {\bibfnamefont {M.~P.}\ \bibnamefont {Jiang}}, \bibinfo {author}
  {\bibfnamefont {D.}~\bibnamefont {Zhu}}, \bibinfo {author} {\bibfnamefont
  {M.}~\bibnamefont {Chollet}}, \bibinfo {author} {\bibfnamefont
  {T.}~\bibnamefont {Sato}}, \bibinfo {author} {\bibfnamefont {E.~D.}\
  \bibnamefont {Murray}}, \bibinfo {author} {\bibfnamefont {S.}~\bibnamefont
  {Fahy}}, \bibinfo {author} {\bibfnamefont {S.}~\bibnamefont {O'Mahony}},
  \bibinfo {author} {\bibfnamefont {T.~P.}\ \bibnamefont {Bailey}}, \bibinfo
  {author} {\bibfnamefont {C.}~\bibnamefont {Uher}}, \bibinfo {author}
  {\bibfnamefont {M.}~\bibnamefont {Trigo}}, \ and\ \bibinfo {author}
  {\bibfnamefont {D.~A.}\ \bibnamefont {Reis}},\ }\href {\doibase
  10.1103/PhysRevLett.121.125901} {\bibfield  {journal} {\bibinfo  {journal}
  {Phys. Rev. Lett.}\ }\textbf {\bibinfo {volume} {121}},\ \bibinfo {pages}
  {125901} (\bibinfo {year} {2018})}\BibitemShut {NoStop}%
\bibitem [{\citenamefont {Paarmann}\ \emph {et~al.}(2015)\citenamefont
  {Paarmann}, \citenamefont {Razdolski}, \citenamefont {Melnikov},
  \citenamefont {Gewinner}, \citenamefont {Schöllkopf},\ and\ \citenamefont
  {Wolf}}]{Paarmann2015}%
  \BibitemOpen
  \bibfield  {author} {\bibinfo {author} {\bibfnamefont {A.}~\bibnamefont
  {Paarmann}}, \bibinfo {author} {\bibfnamefont {I.}~\bibnamefont {Razdolski}},
  \bibinfo {author} {\bibfnamefont {A.}~\bibnamefont {Melnikov}}, \bibinfo
  {author} {\bibfnamefont {S.}~\bibnamefont {Gewinner}}, \bibinfo {author}
  {\bibfnamefont {W.}~\bibnamefont {Schöllkopf}}, \ and\ \bibinfo {author}
  {\bibfnamefont {M.}~\bibnamefont {Wolf}},\ }\href {\doibase
  10.1063/1.4929358} {\bibfield  {journal} {\bibinfo  {journal} {Applied
  Physics Letters}\ }\textbf {\bibinfo {volume} {107}},\ \bibinfo {pages}
  {081101} (\bibinfo {year} {2015})}\BibitemShut {NoStop}%
\bibitem [{\citenamefont {Paarmann}\ \emph {et~al.}(2016)\citenamefont
  {Paarmann}, \citenamefont {Razdolski}, \citenamefont {Gewinner},
  \citenamefont {Sch\"ollkopf},\ and\ \citenamefont {Wolf}}]{Paarmann2016}%
  \BibitemOpen
  \bibfield  {author} {\bibinfo {author} {\bibfnamefont {A.}~\bibnamefont
  {Paarmann}}, \bibinfo {author} {\bibfnamefont {I.}~\bibnamefont {Razdolski}},
  \bibinfo {author} {\bibfnamefont {S.}~\bibnamefont {Gewinner}}, \bibinfo
  {author} {\bibfnamefont {W.}~\bibnamefont {Sch\"ollkopf}}, \ and\ \bibinfo
  {author} {\bibfnamefont {M.}~\bibnamefont {Wolf}},\ }\href {\doibase
  10.1103/PhysRevB.94.134312} {\bibfield  {journal} {\bibinfo  {journal} {Phys.
  Rev. B}\ }\textbf {\bibinfo {volume} {94}},\ \bibinfo {pages} {134312}
  (\bibinfo {year} {2016})}\BibitemShut {NoStop}%
\bibitem [{\citenamefont {Kitade}\ \emph {et~al.}(2021)\citenamefont {Kitade},
  \citenamefont {Yamada}, \citenamefont {Morichika}, \citenamefont {Yabana},\
  and\ \citenamefont {Ashihara}}]{Kitade2021}%
  \BibitemOpen
  \bibfield  {author} {\bibinfo {author} {\bibfnamefont {S.}~\bibnamefont
  {Kitade}}, \bibinfo {author} {\bibfnamefont {A.}~\bibnamefont {Yamada}},
  \bibinfo {author} {\bibfnamefont {I.}~\bibnamefont {Morichika}}, \bibinfo
  {author} {\bibfnamefont {K.}~\bibnamefont {Yabana}}, \ and\ \bibinfo {author}
  {\bibfnamefont {S.}~\bibnamefont {Ashihara}},\ }\href {\doibase
  10.1021/acsphotonics.0c01680} {\bibfield  {journal} {\bibinfo  {journal} {ACS
  Photonics}\ }\textbf {\bibinfo {volume} {8}},\ \bibinfo {pages} {152}
  (\bibinfo {year} {2021})}\BibitemShut {NoStop}%
\bibitem [{\citenamefont {Passler}\ \emph {et~al.}(2017)\citenamefont
  {Passler}, \citenamefont {Razdolski}, \citenamefont {Gewinner}, \citenamefont
  {Schöllkopf}, \citenamefont {Wolf},\ and\ \citenamefont
  {Paarmann}}]{Passler2017}%
  \BibitemOpen
  \bibfield  {author} {\bibinfo {author} {\bibfnamefont {N.~C.}\ \bibnamefont
  {Passler}}, \bibinfo {author} {\bibfnamefont {I.}~\bibnamefont {Razdolski}},
  \bibinfo {author} {\bibfnamefont {S.}~\bibnamefont {Gewinner}}, \bibinfo
  {author} {\bibfnamefont {W.}~\bibnamefont {Schöllkopf}}, \bibinfo {author}
  {\bibfnamefont {M.}~\bibnamefont {Wolf}}, \ and\ \bibinfo {author}
  {\bibfnamefont {A.}~\bibnamefont {Paarmann}},\ }\href {\doibase
  10.1021/acsphotonics.7b00118} {\bibfield  {journal} {\bibinfo  {journal} {ACS
  Photonics}\ }\textbf {\bibinfo {volume} {4}},\ \bibinfo {pages} {1048}
  (\bibinfo {year} {2017})}\BibitemShut {NoStop}%
\bibitem [{\citenamefont {Passler}\ \emph {et~al.}(2019)\citenamefont
  {Passler}, \citenamefont {Razdolski}, \citenamefont {Katzer}, \citenamefont
  {Storm}, \citenamefont {Caldwell}, \citenamefont {Wolf},\ and\ \citenamefont
  {Paarmann}}]{Passler2019}%
  \BibitemOpen
  \bibfield  {author} {\bibinfo {author} {\bibfnamefont {N.~C.}\ \bibnamefont
  {Passler}}, \bibinfo {author} {\bibfnamefont {I.}~\bibnamefont {Razdolski}},
  \bibinfo {author} {\bibfnamefont {D.~S.}\ \bibnamefont {Katzer}}, \bibinfo
  {author} {\bibfnamefont {D.~F.}\ \bibnamefont {Storm}}, \bibinfo {author}
  {\bibfnamefont {J.~D.}\ \bibnamefont {Caldwell}}, \bibinfo {author}
  {\bibfnamefont {M.}~\bibnamefont {Wolf}}, \ and\ \bibinfo {author}
  {\bibfnamefont {A.}~\bibnamefont {Paarmann}},\ }\href {\doibase
  10.1021/acsphotonics.9b00290} {\bibfield  {journal} {\bibinfo  {journal} {ACS
  Photonics}\ }\textbf {\bibinfo {volume} {6}},\ \bibinfo {pages} {1365}
  (\bibinfo {year} {2019})}\BibitemShut {NoStop}%
\bibitem [{\citenamefont {Razdolski}\ \emph {et~al.}(2016)\citenamefont
  {Razdolski}, \citenamefont {Chen}, \citenamefont {Giles}, \citenamefont
  {Gewinner}, \citenamefont {Schöllkopf}, \citenamefont {Hong}, \citenamefont
  {Wolf}, \citenamefont {Giannini}, \citenamefont {Caldwell}, \citenamefont
  {Maier},\ and\ \citenamefont {Paarmann}}]{Razdolski2016}%
  \BibitemOpen
  \bibfield  {author} {\bibinfo {author} {\bibfnamefont {I.}~\bibnamefont
  {Razdolski}}, \bibinfo {author} {\bibfnamefont {Y.}~\bibnamefont {Chen}},
  \bibinfo {author} {\bibfnamefont {A.~J.}\ \bibnamefont {Giles}}, \bibinfo
  {author} {\bibfnamefont {S.}~\bibnamefont {Gewinner}}, \bibinfo {author}
  {\bibfnamefont {W.}~\bibnamefont {Schöllkopf}}, \bibinfo {author}
  {\bibfnamefont {M.}~\bibnamefont {Hong}}, \bibinfo {author} {\bibfnamefont
  {M.}~\bibnamefont {Wolf}}, \bibinfo {author} {\bibfnamefont {V.}~\bibnamefont
  {Giannini}}, \bibinfo {author} {\bibfnamefont {J.~D.}\ \bibnamefont
  {Caldwell}}, \bibinfo {author} {\bibfnamefont {S.~A.}\ \bibnamefont {Maier}},
  \ and\ \bibinfo {author} {\bibfnamefont {A.}~\bibnamefont {Paarmann}},\
  }\href {\doibase 10.1021/acs.nanolett.6b03014} {\bibfield  {journal}
  {\bibinfo  {journal} {Nano Letters}\ }\textbf {\bibinfo {volume} {16}},\
  \bibinfo {pages} {6954} (\bibinfo {year} {2016})},\ \bibinfo {note} {pMID:
  27766887}\BibitemShut {NoStop}%
\bibitem [{\citenamefont {Razdolski}\ \emph {et~al.}(2018)\citenamefont
  {Razdolski}, \citenamefont {Passler}, \citenamefont {Gubbin}, \citenamefont
  {Winta}, \citenamefont {Cernansky}, \citenamefont {Martini}, \citenamefont
  {Politi}, \citenamefont {Maier}, \citenamefont {Wolf}, \citenamefont
  {Paarmann},\ and\ \citenamefont {De~Liberato}}]{Razdolski2018}%
  \BibitemOpen
  \bibfield  {author} {\bibinfo {author} {\bibfnamefont {I.}~\bibnamefont
  {Razdolski}}, \bibinfo {author} {\bibfnamefont {N.~C.}\ \bibnamefont
  {Passler}}, \bibinfo {author} {\bibfnamefont {C.~R.}\ \bibnamefont {Gubbin}},
  \bibinfo {author} {\bibfnamefont {C.~J.}\ \bibnamefont {Winta}}, \bibinfo
  {author} {\bibfnamefont {R.}~\bibnamefont {Cernansky}}, \bibinfo {author}
  {\bibfnamefont {F.}~\bibnamefont {Martini}}, \bibinfo {author} {\bibfnamefont
  {A.}~\bibnamefont {Politi}}, \bibinfo {author} {\bibfnamefont {S.~A.}\
  \bibnamefont {Maier}}, \bibinfo {author} {\bibfnamefont {M.}~\bibnamefont
  {Wolf}}, \bibinfo {author} {\bibfnamefont {A.}~\bibnamefont {Paarmann}}, \
  and\ \bibinfo {author} {\bibfnamefont {S.}~\bibnamefont {De~Liberato}},\
  }\href {\doibase 10.1103/PhysRevB.98.125425} {\bibfield  {journal} {\bibinfo
  {journal} {Phys. Rev. B}\ }\textbf {\bibinfo {volume} {98}},\ \bibinfo
  {pages} {125425} (\bibinfo {year} {2018})}\BibitemShut {NoStop}%
\bibitem [{\citenamefont {Kiessling}\ \emph {et~al.}(2019)\citenamefont
  {Kiessling}, \citenamefont {Tong}, \citenamefont {Giles}, \citenamefont
  {Gewinner}, \citenamefont {Schöllkopf}, \citenamefont {Caldwell},
  \citenamefont {Wolf},\ and\ \citenamefont {Paarmann}}]{Kiessling2019}%
  \BibitemOpen
  \bibfield  {author} {\bibinfo {author} {\bibfnamefont {R.}~\bibnamefont
  {Kiessling}}, \bibinfo {author} {\bibfnamefont {Y.}~\bibnamefont {Tong}},
  \bibinfo {author} {\bibfnamefont {A.~J.}\ \bibnamefont {Giles}}, \bibinfo
  {author} {\bibfnamefont {S.}~\bibnamefont {Gewinner}}, \bibinfo {author}
  {\bibfnamefont {W.}~\bibnamefont {Schöllkopf}}, \bibinfo {author}
  {\bibfnamefont {J.~D.}\ \bibnamefont {Caldwell}}, \bibinfo {author}
  {\bibfnamefont {M.}~\bibnamefont {Wolf}}, \ and\ \bibinfo {author}
  {\bibfnamefont {A.}~\bibnamefont {Paarmann}},\ }\href {\doibase
  10.1021/acsphotonics.9b01335} {\bibfield  {journal} {\bibinfo  {journal} {ACS
  Photonics}\ }\textbf {\bibinfo {volume} {6}},\ \bibinfo {pages} {3017}
  (\bibinfo {year} {2019})}\BibitemShut {NoStop}%
\bibitem [{\citenamefont {De~Leon}\ \emph {et~al.}(2014)\citenamefont
  {De~Leon}, \citenamefont {Sipe},\ and\ \citenamefont {Boyd}}]{DeLeon2014}%
  \BibitemOpen
  \bibfield  {author} {\bibinfo {author} {\bibfnamefont {I.}~\bibnamefont
  {De~Leon}}, \bibinfo {author} {\bibfnamefont {J.~E.}\ \bibnamefont {Sipe}}, \
  and\ \bibinfo {author} {\bibfnamefont {R.~W.}\ \bibnamefont {Boyd}},\ }\href
  {\doibase 10.1103/PhysRevA.89.013855} {\bibfield  {journal} {\bibinfo
  {journal} {Phys. Rev. A}\ }\textbf {\bibinfo {volume} {89}},\ \bibinfo
  {pages} {013855} (\bibinfo {year} {2014})}\BibitemShut {NoStop}%
\bibitem [{\citenamefont {Ohtani}\ \emph {et~al.}(2019)\citenamefont {Ohtani},
  \citenamefont {Meng}, \citenamefont {Francki{\'e}}, \citenamefont {Bosco},
  \citenamefont {Ndebeka-Bandou}, \citenamefont {Beck},\ and\ \citenamefont
  {Faist}}]{Ohtani2019}%
  \BibitemOpen
  \bibfield  {author} {\bibinfo {author} {\bibfnamefont {K.}~\bibnamefont
  {Ohtani}}, \bibinfo {author} {\bibfnamefont {B.}~\bibnamefont {Meng}},
  \bibinfo {author} {\bibfnamefont {M.}~\bibnamefont {Francki{\'e}}}, \bibinfo
  {author} {\bibfnamefont {L.}~\bibnamefont {Bosco}}, \bibinfo {author}
  {\bibfnamefont {C.}~\bibnamefont {Ndebeka-Bandou}}, \bibinfo {author}
  {\bibfnamefont {M.}~\bibnamefont {Beck}}, \ and\ \bibinfo {author}
  {\bibfnamefont {J.}~\bibnamefont {Faist}},\ }\href {\doibase
  10.1126/sciadv.aau1632} {\bibfield  {journal} {\bibinfo  {journal} {Science
  Advances}\ }\textbf {\bibinfo {volume} {5}} (\bibinfo {year} {2019}),\
  10.1126/sciadv.aau1632}\BibitemShut {NoStop}%
\bibitem [{\citenamefont {Stewart}\ \emph {et~al.}(2020)\citenamefont
  {Stewart}, \citenamefont {Vella}, \citenamefont {Li}, \citenamefont {Fan},\
  and\ \citenamefont {Mikkelsen}}]{Stewart2020}%
  \BibitemOpen
  \bibfield  {author} {\bibinfo {author} {\bibfnamefont {J.~W.}\ \bibnamefont
  {Stewart}}, \bibinfo {author} {\bibfnamefont {J.~H.}\ \bibnamefont {Vella}},
  \bibinfo {author} {\bibfnamefont {W.}~\bibnamefont {Li}}, \bibinfo {author}
  {\bibfnamefont {S.}~\bibnamefont {Fan}}, \ and\ \bibinfo {author}
  {\bibfnamefont {M.~H.}\ \bibnamefont {Mikkelsen}},\ }\href {\doibase
  10.1038/s41563-019-0538-6} {\bibfield  {journal} {\bibinfo  {journal} {Nature
  Materials}\ }\textbf {\bibinfo {volume} {19}},\ \bibinfo {pages} {158}
  (\bibinfo {year} {2020})}\BibitemShut {NoStop}%
\bibitem [{\citenamefont {Kozen}\ \emph {et~al.}(2017)\citenamefont {Kozen},
  \citenamefont {Joress}, \citenamefont {Currie}, \citenamefont {Anderson},
  \citenamefont {Eddy},\ and\ \citenamefont {Wheeler}}]{Kozen2017}%
  \BibitemOpen
  \bibfield  {author} {\bibinfo {author} {\bibfnamefont {A.~C.}\ \bibnamefont
  {Kozen}}, \bibinfo {author} {\bibfnamefont {H.}~\bibnamefont {Joress}},
  \bibinfo {author} {\bibfnamefont {M.}~\bibnamefont {Currie}}, \bibinfo
  {author} {\bibfnamefont {V.~R.}\ \bibnamefont {Anderson}}, \bibinfo {author}
  {\bibfnamefont {C.~R.}\ \bibnamefont {Eddy}}, \ and\ \bibinfo {author}
  {\bibfnamefont {V.~D.}\ \bibnamefont {Wheeler}},\ }\href {\doibase
  10.1021/acs.jpcc.7b04682} {\bibfield  {journal} {\bibinfo  {journal} {The
  Journal of Physical Chemistry C}\ }\textbf {\bibinfo {volume} {121}},\
  \bibinfo {pages} {19341} (\bibinfo {year} {2017})}\BibitemShut {NoStop}%
\bibitem [{\citenamefont {Shao}\ \emph {et~al.}(2019)\citenamefont {Shao},
  \citenamefont {Guo}, \citenamefont {Zhu}, \citenamefont {Zhuang},
  \citenamefont {Fu},\ and\ \citenamefont {Cao}}]{Shao2019}%
  \BibitemOpen
  \bibfield  {author} {\bibinfo {author} {\bibfnamefont {D.}~\bibnamefont
  {Shao}}, \bibinfo {author} {\bibfnamefont {X.}~\bibnamefont {Guo}}, \bibinfo
  {author} {\bibfnamefont {Y.}~\bibnamefont {Zhu}}, \bibinfo {author}
  {\bibfnamefont {S.}~\bibnamefont {Zhuang}}, \bibinfo {author} {\bibfnamefont
  {Z.}~\bibnamefont {Fu}}, \ and\ \bibinfo {author} {\bibfnamefont
  {J.}~\bibnamefont {Cao}},\ }\href {\doibase 10.1088/1361-6463/aaeb77}
  {\bibfield  {journal} {\bibinfo  {journal} {Journal of Physics D: Applied
  Physics}\ }\textbf {\bibinfo {volume} {52}},\ \bibinfo {pages} {5} (\bibinfo
  {year} {2019})}\BibitemShut {NoStop}%
\bibitem [{\citenamefont {Wang}\ \emph {et~al.}(2018)\citenamefont {Wang},
  \citenamefont {Yoon}, \citenamefont {Kamboj}, \citenamefont {Petluru},
  \citenamefont {Zheng},\ and\ \citenamefont {Wasserman}}]{Wang2018}%
  \BibitemOpen
  \bibfield  {author} {\bibinfo {author} {\bibfnamefont {S.}~\bibnamefont
  {Wang}}, \bibinfo {author} {\bibfnamefont {N.}~\bibnamefont {Yoon}}, \bibinfo
  {author} {\bibfnamefont {A.}~\bibnamefont {Kamboj}}, \bibinfo {author}
  {\bibfnamefont {P.}~\bibnamefont {Petluru}}, \bibinfo {author} {\bibfnamefont
  {W.}~\bibnamefont {Zheng}}, \ and\ \bibinfo {author} {\bibfnamefont
  {D.}~\bibnamefont {Wasserman}},\ }\href {\doibase 10.1063/1.5017704}
  {\bibfield  {journal} {\bibinfo  {journal} {Applied Physics Letters}\
  }\textbf {\bibinfo {volume} {112}},\ \bibinfo {pages} {91104} (\bibinfo
  {year} {2018})}\BibitemShut {NoStop}%
\bibitem [{\citenamefont {Kim}\ and\ \citenamefont {Spitzer}(1979)}]{Kim1979}%
  \BibitemOpen
  \bibfield  {author} {\bibinfo {author} {\bibfnamefont {O.~K.}\ \bibnamefont
  {Kim}}\ and\ \bibinfo {author} {\bibfnamefont {W.~G.}\ \bibnamefont
  {Spitzer}},\ }\href {\doibase 10.1063/1.326422} {\bibfield  {journal}
  {\bibinfo  {journal} {Journal of Applied Physics}\ }\textbf {\bibinfo
  {volume} {50}},\ \bibinfo {pages} {4362} (\bibinfo {year}
  {1979})}\BibitemShut {NoStop}%
\bibitem [{\citenamefont {Gubbin}\ and\ \citenamefont
  {De~Liberato}(2020{\natexlab{a}})}]{Gubbin2020}%
  \BibitemOpen
  \bibfield  {author} {\bibinfo {author} {\bibfnamefont {C.~R.}\ \bibnamefont
  {Gubbin}}\ and\ \bibinfo {author} {\bibfnamefont {S.}~\bibnamefont
  {De~Liberato}},\ }\href {\doibase 10.1103/PhysRevX.10.021027} {\bibfield
  {journal} {\bibinfo  {journal} {Phys. Rev. X}\ }\textbf {\bibinfo {volume}
  {10}},\ \bibinfo {pages} {021027} (\bibinfo {year}
  {2020}{\natexlab{a}})}\BibitemShut {NoStop}%
\bibitem [{\citenamefont {Roldán}\ \emph {et~al.}(1997)\citenamefont
  {Roldán}, \citenamefont {Gámiz}, \citenamefont {López~Villanueva},\ and\
  \citenamefont {Caetujo}}]{Roldan1997}%
  \BibitemOpen
  \bibfield  {author} {\bibinfo {author} {\bibfnamefont {J.~B.}\ \bibnamefont
  {Roldán}}, \bibinfo {author} {\bibfnamefont {F.}~\bibnamefont {Gámiz}},
  \bibinfo {author} {\bibfnamefont {J.~A.}\ \bibnamefont {López~Villanueva}},
  \ and\ \bibinfo {author} {\bibfnamefont {P.}~\bibnamefont {Caetujo}},\ }\href
  {\doibase 10.1007/s11664-997-0151-3} {\bibfield  {journal} {\bibinfo
  {journal} {Journal of Electronic Materials}\ }\textbf {\bibinfo {volume}
  {26}},\ \bibinfo {pages} {203} (\bibinfo {year} {1997})}\BibitemShut
  {NoStop}%
\bibitem [{\citenamefont {{Jing Chen}}\ and\ \citenamefont
  {Khurgin}(2003)}]{Chen2003}%
  \BibitemOpen
  \bibfield  {author} {\bibinfo {author} {\bibnamefont {{Jing Chen}}}\ and\
  \bibinfo {author} {\bibfnamefont {J.}~\bibnamefont {Khurgin}},\ }\href
  {\doibase 10.1109/JQE.2003.809326} {\bibfield  {journal} {\bibinfo  {journal}
  {IEEE Journal of Quantum Electronics}\ }\textbf {\bibinfo {volume} {39}},\
  \bibinfo {pages} {600} (\bibinfo {year} {2003})}\BibitemShut {NoStop}%
\bibitem [{\citenamefont {Bluet}\ \emph {et~al.}(1999)\citenamefont {Bluet},
  \citenamefont {Chourou}, \citenamefont {Anikin},\ and\ \citenamefont
  {Madar}}]{Bluet1999}%
  \BibitemOpen
  \bibfield  {author} {\bibinfo {author} {\bibfnamefont {J.}~\bibnamefont
  {Bluet}}, \bibinfo {author} {\bibfnamefont {K.}~\bibnamefont {Chourou}},
  \bibinfo {author} {\bibfnamefont {M.}~\bibnamefont {Anikin}}, \ and\ \bibinfo
  {author} {\bibfnamefont {R.}~\bibnamefont {Madar}},\ }\href {\doibase
  https://doi.org/10.1016/S0921-5107(98)00504-2} {\bibfield  {journal}
  {\bibinfo  {journal} {Materials Science and Engineering: B}\ }\textbf
  {\bibinfo {volume} {61-62}},\ \bibinfo {pages} {212} (\bibinfo {year}
  {1999})}\BibitemShut {NoStop}%
\bibitem [{Gub()}]{Gubbin2016b}%
  \BibitemOpen
  \href@noop {} {\ }\BibitemShut {NoStop}%
\bibitem [{\citenamefont {Gubbin}\ and\ \citenamefont
  {De~Liberato}(2020{\natexlab{b}})}]{Gubbin2020b}%
  \BibitemOpen
  \bibfield  {author} {\bibinfo {author} {\bibfnamefont {C.~R.}\ \bibnamefont
  {Gubbin}}\ and\ \bibinfo {author} {\bibfnamefont {S.}~\bibnamefont
  {De~Liberato}},\ }\href {\doibase 10.1103/PhysRevB.102.235301} {\bibfield
  {journal} {\bibinfo  {journal} {Phys. Rev. B}\ }\textbf {\bibinfo {volume}
  {102}},\ \bibinfo {pages} {235301} (\bibinfo {year}
  {2020}{\natexlab{b}})}\BibitemShut {NoStop}%
\bibitem [{\citenamefont {Ciraci}\ \emph {et~al.}(2012)\citenamefont {Ciraci},
  \citenamefont {Hill}, \citenamefont {Mock}, \citenamefont {Urzhumov},
  \citenamefont {Fernandez-Dominguez}, \citenamefont {Maier}, \citenamefont
  {Pendry}, \citenamefont {Chilkoti},\ and\ \citenamefont
  {Smith}}]{Ciraci2012}%
  \BibitemOpen
  \bibfield  {author} {\bibinfo {author} {\bibfnamefont {C.}~\bibnamefont
  {Ciraci}}, \bibinfo {author} {\bibfnamefont {R.~T.}\ \bibnamefont {Hill}},
  \bibinfo {author} {\bibfnamefont {J.~J.}\ \bibnamefont {Mock}}, \bibinfo
  {author} {\bibfnamefont {Y.}~\bibnamefont {Urzhumov}}, \bibinfo {author}
  {\bibfnamefont {A.~I.}\ \bibnamefont {Fernandez-Dominguez}}, \bibinfo
  {author} {\bibfnamefont {S.~A.}\ \bibnamefont {Maier}}, \bibinfo {author}
  {\bibfnamefont {J.~B.}\ \bibnamefont {Pendry}}, \bibinfo {author}
  {\bibfnamefont {A.}~\bibnamefont {Chilkoti}}, \ and\ \bibinfo {author}
  {\bibfnamefont {D.~R.}\ \bibnamefont {Smith}},\ }\href@noop {} {\bibfield
  {journal} {\bibinfo  {journal} {Science}\ }\textbf {\bibinfo {volume}
  {337}},\ \bibinfo {pages} {1072} (\bibinfo {year} {2012})}\BibitemShut
  {NoStop}%
\bibitem [{\citenamefont {Rajabali}\ \emph {et~al.}(2021)\citenamefont
  {Rajabali}, \citenamefont {Cortese}, \citenamefont {Beck}, \citenamefont {{De
  Liberato}}, \citenamefont {Faist},\ and\ \citenamefont
  {Scalari}}]{Rajabali2021}%
  \BibitemOpen
  \bibfield  {author} {\bibinfo {author} {\bibfnamefont {S.}~\bibnamefont
  {Rajabali}}, \bibinfo {author} {\bibfnamefont {E.}~\bibnamefont {Cortese}},
  \bibinfo {author} {\bibfnamefont {M.}~\bibnamefont {Beck}}, \bibinfo {author}
  {\bibfnamefont {S.}~\bibnamefont {{De Liberato}}}, \bibinfo {author}
  {\bibfnamefont {J.}~\bibnamefont {Faist}}, \ and\ \bibinfo {author}
  {\bibfnamefont {G.}~\bibnamefont {Scalari}},\ }\href@noop {} {\enquote
  {\bibinfo {title} {Polaritonic nonlocality in light-matter interaction},}\ }
  (\bibinfo {year} {2021}),\ \Eprint {http://arxiv.org/abs/2101.08709}
  {arXiv:2101.08709 [cond-mat.mes-hall]} \BibitemShut {NoStop}%
\bibitem [{\citenamefont {Gubbin}\ and\ \citenamefont
  {De~Liberato}(2021)}]{Gubbin2021}%
  \BibitemOpen
  \bibfield  {author} {\bibinfo {author} {\bibfnamefont {C.~R.}\ \bibnamefont
  {Gubbin}}\ and\ \bibinfo {author} {\bibfnamefont {S.}~\bibnamefont
  {De~Liberato}},\ }\href@noop {} {\bibfield  {journal} {\bibinfo  {journal}
  {arXiv}\ ,\ \bibinfo {pages} {2105.13219}} (\bibinfo {year}
  {2021})}\BibitemShut {NoStop}%
\bibitem [{\citenamefont {Gubbin}\ and\ \citenamefont
  {De~Liberato}(2020{\natexlab{c}})}]{Gubbin2020c}%
  \BibitemOpen
  \bibfield  {author} {\bibinfo {author} {\bibfnamefont {C.~R.}\ \bibnamefont
  {Gubbin}}\ and\ \bibinfo {author} {\bibfnamefont {S.}~\bibnamefont
  {De~Liberato}},\ }\href {\doibase 10.1103/PhysRevB.102.201302} {\bibfield
  {journal} {\bibinfo  {journal} {Phys. Rev. B}\ }\textbf {\bibinfo {volume}
  {102}},\ \bibinfo {pages} {201302} (\bibinfo {year}
  {2020}{\natexlab{c}})}\BibitemShut {NoStop}%
\bibitem [{\citenamefont {Khan}\ \emph {et~al.}(2020)\citenamefont {Khan},
  \citenamefont {Fang}, \citenamefont {Palei}, \citenamefont {Lu},
  \citenamefont {Nordin}, \citenamefont {Simmons}, \citenamefont {Dominguez},
  \citenamefont {Islam}, \citenamefont {Xing}, \citenamefont {Jena},
  \citenamefont {Podolskiy}, \citenamefont {Wasserman},\ and\ \citenamefont
  {Hoffman}}]{Khan2020}%
  \BibitemOpen
  \bibfield  {author} {\bibinfo {author} {\bibfnamefont {I.}~\bibnamefont
  {Khan}}, \bibinfo {author} {\bibfnamefont {Z.}~\bibnamefont {Fang}}, \bibinfo
  {author} {\bibfnamefont {M.}~\bibnamefont {Palei}}, \bibinfo {author}
  {\bibfnamefont {J.}~\bibnamefont {Lu}}, \bibinfo {author} {\bibfnamefont
  {L.}~\bibnamefont {Nordin}}, \bibinfo {author} {\bibfnamefont {E.~L.}\
  \bibnamefont {Simmons}}, \bibinfo {author} {\bibfnamefont {O.}~\bibnamefont
  {Dominguez}}, \bibinfo {author} {\bibfnamefont {S.~M.}\ \bibnamefont
  {Islam}}, \bibinfo {author} {\bibfnamefont {H.~G.}\ \bibnamefont {Xing}},
  \bibinfo {author} {\bibfnamefont {D.}~\bibnamefont {Jena}}, \bibinfo {author}
  {\bibfnamefont {V.~A.}\ \bibnamefont {Podolskiy}}, \bibinfo {author}
  {\bibfnamefont {D.}~\bibnamefont {Wasserman}}, \ and\ \bibinfo {author}
  {\bibfnamefont {A.~J.}\ \bibnamefont {Hoffman}},\ }\href {\doibase
  10.1364/OE.401733} {\bibfield  {journal} {\bibinfo  {journal} {Optics
  Express}\ }\textbf {\bibinfo {volume} {28}},\ \bibinfo {pages} {28590}
  (\bibinfo {year} {2020})}\BibitemShut {NoStop}%
\bibitem [{\citenamefont {Ng}\ \emph {et~al.}(2007)\citenamefont {Ng},
  \citenamefont {Hassan},\ and\ \citenamefont {{Abu Hassan}}}]{Ng2007}%
  \BibitemOpen
  \bibfield  {author} {\bibinfo {author} {\bibfnamefont {S.~S.}\ \bibnamefont
  {Ng}}, \bibinfo {author} {\bibfnamefont {Z.}~\bibnamefont {Hassan}}, \ and\
  \bibinfo {author} {\bibfnamefont {H.}~\bibnamefont {{Abu Hassan}}},\ }\href
  {\doibase 081902 10.1063/645970} {\bibfield  {journal} {\bibinfo  {journal}
  {Applied Physics Letters}\ }\textbf {\bibinfo {volume} {90}},\ \bibinfo
  {pages} {81902} (\bibinfo {year} {2007})}\BibitemShut {NoStop}%
\bibitem [{\citenamefont {Tsao}\ \emph {et~al.}(2018)\citenamefont {Tsao},
  \citenamefont {Chowdhury}, \citenamefont {Hollis}, \citenamefont {Jena},
  \citenamefont {Johnson}, \citenamefont {Jones}, \citenamefont {Kaplar},
  \citenamefont {Rajan}, \citenamefont {{Van de Walle}}, \citenamefont
  {Bellotti}, \citenamefont {Chua}, \citenamefont {Collazo}, \citenamefont
  {Coltrin}, \citenamefont {Cooper}, \citenamefont {Evans}, \citenamefont
  {Graham}, \citenamefont {Grotjohn}, \citenamefont {Heller}, \citenamefont
  {Higashiwaki}, \citenamefont {Islam}, \citenamefont {Juodawlkis},
  \citenamefont {Khan}, \citenamefont {Koehler}, \citenamefont {Leach},
  \citenamefont {Mishra}, \citenamefont {Nemanich}, \citenamefont
  {Pilawa-Podgurski}, \citenamefont {Shealy}, \citenamefont {Sitar},
  \citenamefont {Tadjer}, \citenamefont {Witulski}, \citenamefont {Wraback},\
  and\ \citenamefont {Simmons}}]{Tsao2018}%
  \BibitemOpen
  \bibfield  {author} {\bibinfo {author} {\bibfnamefont {J.~Y.}\ \bibnamefont
  {Tsao}}, \bibinfo {author} {\bibfnamefont {S.}~\bibnamefont {Chowdhury}},
  \bibinfo {author} {\bibfnamefont {M.~A.}\ \bibnamefont {Hollis}}, \bibinfo
  {author} {\bibfnamefont {D.}~\bibnamefont {Jena}}, \bibinfo {author}
  {\bibfnamefont {N.~M.}\ \bibnamefont {Johnson}}, \bibinfo {author}
  {\bibfnamefont {K.~A.}\ \bibnamefont {Jones}}, \bibinfo {author}
  {\bibfnamefont {R.~J.}\ \bibnamefont {Kaplar}}, \bibinfo {author}
  {\bibfnamefont {S.}~\bibnamefont {Rajan}}, \bibinfo {author} {\bibfnamefont
  {C.~G.}\ \bibnamefont {{Van de Walle}}}, \bibinfo {author} {\bibfnamefont
  {E.}~\bibnamefont {Bellotti}}, \bibinfo {author} {\bibfnamefont {C.~L.}\
  \bibnamefont {Chua}}, \bibinfo {author} {\bibfnamefont {R.}~\bibnamefont
  {Collazo}}, \bibinfo {author} {\bibfnamefont {M.~E.}\ \bibnamefont
  {Coltrin}}, \bibinfo {author} {\bibfnamefont {J.~A.}\ \bibnamefont {Cooper}},
  \bibinfo {author} {\bibfnamefont {K.~R.}\ \bibnamefont {Evans}}, \bibinfo
  {author} {\bibfnamefont {S.}~\bibnamefont {Graham}}, \bibinfo {author}
  {\bibfnamefont {T.~A.}\ \bibnamefont {Grotjohn}}, \bibinfo {author}
  {\bibfnamefont {E.~R.}\ \bibnamefont {Heller}}, \bibinfo {author}
  {\bibfnamefont {M.}~\bibnamefont {Higashiwaki}}, \bibinfo {author}
  {\bibfnamefont {M.~S.}\ \bibnamefont {Islam}}, \bibinfo {author}
  {\bibfnamefont {P.~W.}\ \bibnamefont {Juodawlkis}}, \bibinfo {author}
  {\bibfnamefont {M.~A.}\ \bibnamefont {Khan}}, \bibinfo {author}
  {\bibfnamefont {A.~D.}\ \bibnamefont {Koehler}}, \bibinfo {author}
  {\bibfnamefont {J.~H.}\ \bibnamefont {Leach}}, \bibinfo {author}
  {\bibfnamefont {U.~K.}\ \bibnamefont {Mishra}}, \bibinfo {author}
  {\bibfnamefont {R.~J.}\ \bibnamefont {Nemanich}}, \bibinfo {author}
  {\bibfnamefont {R.~C.~N.}\ \bibnamefont {Pilawa-Podgurski}}, \bibinfo
  {author} {\bibfnamefont {J.~B.}\ \bibnamefont {Shealy}}, \bibinfo {author}
  {\bibfnamefont {Z.}~\bibnamefont {Sitar}}, \bibinfo {author} {\bibfnamefont
  {M.~J.}\ \bibnamefont {Tadjer}}, \bibinfo {author} {\bibfnamefont {A.~F.}\
  \bibnamefont {Witulski}}, \bibinfo {author} {\bibfnamefont {M.}~\bibnamefont
  {Wraback}}, \ and\ \bibinfo {author} {\bibfnamefont {J.~A.}\ \bibnamefont
  {Simmons}},\ }\href {\doibase https://doi.org/10.1002/aelm.201600501}
  {\bibfield  {journal} {\bibinfo  {journal} {Advanced Electronic Materials}\
  }\textbf {\bibinfo {volume} {4}},\ \bibinfo {pages} {1600501} (\bibinfo
  {year} {2018})}\BibitemShut {NoStop}%
\bibitem [{\citenamefont {L{\"{a}}hnemann}\ \emph {et~al.}(2017)\citenamefont
  {L{\"{a}}hnemann}, \citenamefont {Ajay}, \citenamefont {{Den Hertog}},\ and\
  \citenamefont {Monroy}}]{Lahnemann2017}%
  \BibitemOpen
  \bibfield  {author} {\bibinfo {author} {\bibfnamefont {J.}~\bibnamefont
  {L{\"{a}}hnemann}}, \bibinfo {author} {\bibfnamefont {A.}~\bibnamefont
  {Ajay}}, \bibinfo {author} {\bibfnamefont {M.~I.}\ \bibnamefont {{Den
  Hertog}}}, \ and\ \bibinfo {author} {\bibfnamefont {E.}~\bibnamefont
  {Monroy}},\ }\href {\doibase 10.1021/acs.nanolett.7b03414} {\bibfield
  {journal} {\bibinfo  {journal} {Nano Letters}\ }\textbf {\bibinfo {volume}
  {17}},\ \bibinfo {pages} {6954} (\bibinfo {year} {2017})}\BibitemShut
  {NoStop}%
\bibitem [{\citenamefont {Durmaz}\ \emph {et~al.}(2016)\citenamefont {Durmaz},
  \citenamefont {Nothern}, \citenamefont {Brummer}, \citenamefont {Moustakas},\
  and\ \citenamefont {Paiella}}]{Durmaz2016}%
  \BibitemOpen
  \bibfield  {author} {\bibinfo {author} {\bibfnamefont {H.}~\bibnamefont
  {Durmaz}}, \bibinfo {author} {\bibfnamefont {D.}~\bibnamefont {Nothern}},
  \bibinfo {author} {\bibfnamefont {G.}~\bibnamefont {Brummer}}, \bibinfo
  {author} {\bibfnamefont {T.~D.}\ \bibnamefont {Moustakas}}, \ and\ \bibinfo
  {author} {\bibfnamefont {R.}~\bibnamefont {Paiella}},\ }\href {\doibase
  10.1063/1.4950852} {\bibfield  {journal} {\bibinfo  {journal} {Applied
  Physics Letters}\ }\textbf {\bibinfo {volume} {108}},\ \bibinfo {pages}
  {201102} (\bibinfo {year} {2016})}\BibitemShut {NoStop}%
\bibitem [{\citenamefont {Falge}\ and\ \citenamefont {Otto}(1973)}]{Falge1973}%
  \BibitemOpen
  \bibfield  {author} {\bibinfo {author} {\bibfnamefont {H.~J.}\ \bibnamefont
  {Falge}}\ and\ \bibinfo {author} {\bibfnamefont {A.}~\bibnamefont {Otto}},\
  }\href {\doibase https://doi.org/10.1002/pssb.2220560213} {\bibfield
  {journal} {\bibinfo  {journal} {physica status solidi (b)}\ }\textbf
  {\bibinfo {volume} {56}},\ \bibinfo {pages} {523} (\bibinfo {year}
  {1973})}\BibitemShut {NoStop}%
\bibitem [{\citenamefont {Schlom}\ \emph {et~al.}(2008)\citenamefont {Schlom},
  \citenamefont {Chen}, \citenamefont {Pan}, \citenamefont {Schmehl},\ and\
  \citenamefont {Zurbuchen}}]{Schlom2008}%
  \BibitemOpen
  \bibfield  {author} {\bibinfo {author} {\bibfnamefont {D.~G.}\ \bibnamefont
  {Schlom}}, \bibinfo {author} {\bibfnamefont {L.-Q.}\ \bibnamefont {Chen}},
  \bibinfo {author} {\bibfnamefont {X.}~\bibnamefont {Pan}}, \bibinfo {author}
  {\bibfnamefont {A.}~\bibnamefont {Schmehl}}, \ and\ \bibinfo {author}
  {\bibfnamefont {M.~A.}\ \bibnamefont {Zurbuchen}},\ }\href {\doibase
  https://doi.org/10.1111/j.1551-2916.2008.02556.x} {\bibfield  {journal}
  {\bibinfo  {journal} {Journal of the American Ceramic Society}\ }\textbf
  {\bibinfo {volume} {91}},\ \bibinfo {pages} {2429} (\bibinfo {year}
  {2008})}\BibitemShut {NoStop}%
\bibitem [{\citenamefont {Stepanov}\ \emph {et~al.}(2016)\citenamefont
  {Stepanov}, \citenamefont {Nikolaev}, \citenamefont {Bougrov},\ and\
  \citenamefont {Romanov}}]{Stepanov2016}%
  \BibitemOpen
  \bibfield  {author} {\bibinfo {author} {\bibfnamefont {S.~I.}\ \bibnamefont
  {Stepanov}}, \bibinfo {author} {\bibfnamefont {V.~I.}\ \bibnamefont
  {Nikolaev}}, \bibinfo {author} {\bibfnamefont {V.~E.}\ \bibnamefont
  {Bougrov}}, \ and\ \bibinfo {author} {\bibfnamefont {A.~E.}\ \bibnamefont
  {Romanov}},\ }\href@noop {} {\bibfield  {journal} {\bibinfo  {journal}
  {Reviews on Advanced Materials Science}\ }\textbf {\bibinfo {volume} {44}},\
  \bibinfo {pages} {63} (\bibinfo {year} {2016})}\BibitemShut {NoStop}%
\bibitem [{\citenamefont {Beechem}\ \emph {et~al.}(2018)\citenamefont
  {Beechem}, \citenamefont {Goldflam}, \citenamefont {Sinclair}, \citenamefont
  {Peters}, \citenamefont {McDonald}, \citenamefont {Paisley}, \citenamefont
  {Kitahara}, \citenamefont {Drury}, \citenamefont {Burckel}, \citenamefont
  {Finnegan}, \citenamefont {Kim}, \citenamefont {Choi}, \citenamefont {Ryan},\
  and\ \citenamefont {Ihlefeld}}]{Beechem2018}%
  \BibitemOpen
  \bibfield  {author} {\bibinfo {author} {\bibfnamefont {T.~E.}\ \bibnamefont
  {Beechem}}, \bibinfo {author} {\bibfnamefont {M.~D.}\ \bibnamefont
  {Goldflam}}, \bibinfo {author} {\bibfnamefont {M.~B.}\ \bibnamefont
  {Sinclair}}, \bibinfo {author} {\bibfnamefont {D.~W.}\ \bibnamefont
  {Peters}}, \bibinfo {author} {\bibfnamefont {A.~E.}\ \bibnamefont
  {McDonald}}, \bibinfo {author} {\bibfnamefont {E.~A.}\ \bibnamefont
  {Paisley}}, \bibinfo {author} {\bibfnamefont {A.~R.}\ \bibnamefont
  {Kitahara}}, \bibinfo {author} {\bibfnamefont {D.~E.}\ \bibnamefont {Drury}},
  \bibinfo {author} {\bibfnamefont {D.~B.}\ \bibnamefont {Burckel}}, \bibinfo
  {author} {\bibfnamefont {P.~S.}\ \bibnamefont {Finnegan}}, \bibinfo {author}
  {\bibfnamefont {J.-W.}\ \bibnamefont {Kim}}, \bibinfo {author} {\bibfnamefont
  {Y.}~\bibnamefont {Choi}}, \bibinfo {author} {\bibfnamefont {P.~J.}\
  \bibnamefont {Ryan}}, \ and\ \bibinfo {author} {\bibfnamefont {J.~F.}\
  \bibnamefont {Ihlefeld}},\ }\href {\doibase
  https://doi.org/10.1002/adom.201870094} {\bibfield  {journal} {\bibinfo
  {journal} {Advanced Optical Materials}\ }\textbf {\bibinfo {volume} {6}},\
  \bibinfo {pages} {1870094} (\bibinfo {year} {2018})}\BibitemShut {NoStop}%
\bibitem [{\citenamefont {Taboada-Guti{\'{e}}rrez}\ \emph
  {et~al.}(2020)\citenamefont {Taboada-Guti{\'{e}}rrez}, \citenamefont
  {{\'{A}}lvarez-P{\'{e}}rez}, \citenamefont {Duan}, \citenamefont {Ma},
  \citenamefont {Crowley}, \citenamefont {Prieto}, \citenamefont {Bylinkin},
  \citenamefont {Autore}, \citenamefont {Volkova}, \citenamefont {Kimura},
  \citenamefont {Kimura}, \citenamefont {Berger}, \citenamefont {Li},
  \citenamefont {Bao}, \citenamefont {Gao}, \citenamefont {Errea},
  \citenamefont {Nikitin}, \citenamefont {Hillenbrand}, \citenamefont
  {Mart{\'{i}}n-S{\'{a}}nchez},\ and\ \citenamefont
  {Alonso-Gonz{\'{a}}lez}}]{Taboada-Gutierrez2020}%
  \BibitemOpen
  \bibfield  {author} {\bibinfo {author} {\bibfnamefont {J.}~\bibnamefont
  {Taboada-Guti{\'{e}}rrez}}, \bibinfo {author} {\bibfnamefont
  {G.}~\bibnamefont {{\'{A}}lvarez-P{\'{e}}rez}}, \bibinfo {author}
  {\bibfnamefont {J.}~\bibnamefont {Duan}}, \bibinfo {author} {\bibfnamefont
  {W.}~\bibnamefont {Ma}}, \bibinfo {author} {\bibfnamefont {K.}~\bibnamefont
  {Crowley}}, \bibinfo {author} {\bibfnamefont {I.}~\bibnamefont {Prieto}},
  \bibinfo {author} {\bibfnamefont {A.}~\bibnamefont {Bylinkin}}, \bibinfo
  {author} {\bibfnamefont {M.}~\bibnamefont {Autore}}, \bibinfo {author}
  {\bibfnamefont {H.}~\bibnamefont {Volkova}}, \bibinfo {author} {\bibfnamefont
  {K.}~\bibnamefont {Kimura}}, \bibinfo {author} {\bibfnamefont
  {T.}~\bibnamefont {Kimura}}, \bibinfo {author} {\bibfnamefont {M.-H.}\
  \bibnamefont {Berger}}, \bibinfo {author} {\bibfnamefont {S.}~\bibnamefont
  {Li}}, \bibinfo {author} {\bibfnamefont {Q.}~\bibnamefont {Bao}}, \bibinfo
  {author} {\bibfnamefont {X.~P.~A.}\ \bibnamefont {Gao}}, \bibinfo {author}
  {\bibfnamefont {I.}~\bibnamefont {Errea}}, \bibinfo {author} {\bibfnamefont
  {A.~Y.}\ \bibnamefont {Nikitin}}, \bibinfo {author} {\bibfnamefont
  {R.}~\bibnamefont {Hillenbrand}}, \bibinfo {author} {\bibfnamefont
  {J.}~\bibnamefont {Mart{\'{i}}n-S{\'{a}}nchez}}, \ and\ \bibinfo {author}
  {\bibfnamefont {P.}~\bibnamefont {Alonso-Gonz{\'{a}}lez}},\ }\href {\doibase
  10.1038/s41563-020-0665-0} {\bibfield  {journal} {\bibinfo  {journal} {Nature
  Materials}\ }\textbf {\bibinfo {volume} {19}},\ \bibinfo {pages} {964}
  (\bibinfo {year} {2020})}\BibitemShut {NoStop}%
\bibitem [{\citenamefont {Schubert}\ \emph {et~al.}(2019)\citenamefont
  {Schubert}, \citenamefont {Mock}, \citenamefont {Korlacki}, \citenamefont
  {Knight}, \citenamefont {Galazka}, \citenamefont {Wagner}, \citenamefont
  {Wheeler}, \citenamefont {Tadjer}, \citenamefont {Goto},\ and\ \citenamefont
  {Darakchieva}}]{Schubert2019}%
  \BibitemOpen
  \bibfield  {author} {\bibinfo {author} {\bibfnamefont {M.}~\bibnamefont
  {Schubert}}, \bibinfo {author} {\bibfnamefont {A.}~\bibnamefont {Mock}},
  \bibinfo {author} {\bibfnamefont {R.}~\bibnamefont {Korlacki}}, \bibinfo
  {author} {\bibfnamefont {S.}~\bibnamefont {Knight}}, \bibinfo {author}
  {\bibfnamefont {Z.}~\bibnamefont {Galazka}}, \bibinfo {author} {\bibfnamefont
  {G.}~\bibnamefont {Wagner}}, \bibinfo {author} {\bibfnamefont
  {V.}~\bibnamefont {Wheeler}}, \bibinfo {author} {\bibfnamefont
  {M.}~\bibnamefont {Tadjer}}, \bibinfo {author} {\bibfnamefont
  {K.}~\bibnamefont {Goto}}, \ and\ \bibinfo {author} {\bibfnamefont
  {V.}~\bibnamefont {Darakchieva}},\ }\href {\doibase 10.1063/1.5089145}
  {\bibfield  {journal} {\bibinfo  {journal} {Applied Physics Letters}\
  }\textbf {\bibinfo {volume} {114}},\ \bibinfo {pages} {102102} (\bibinfo
  {year} {2019})}\BibitemShut {NoStop}%
\bibitem [{\citenamefont {Kojima}(2018)}]{Kojima2018}%
  \BibitemOpen
  \bibfield  {author} {\bibinfo {author} {\bibfnamefont {S.}~\bibnamefont
  {Kojima}},\ }\href {\doibase 10.3390/photonics5040055} {\enquote {\bibinfo
  {title} {{Broadband Terahertz Spectroscopy of Phonon-Polariton Dispersion in
  Ferroelectrics}},}\ } (\bibinfo {year} {2018})\BibitemShut {NoStop}%
\bibitem [{\citenamefont {Ferro}(2015)}]{Ferro2015}%
  \BibitemOpen
  \bibfield  {author} {\bibinfo {author} {\bibfnamefont {G.}~\bibnamefont
  {Ferro}},\ }\href {\doibase 10.1080/10408436.2014.940440} {\bibfield
  {journal} {\bibinfo  {journal} {Critical Reviews in Solid State and Materials
  Sciences}\ }\textbf {\bibinfo {volume} {40}},\ \bibinfo {pages} {56}
  (\bibinfo {year} {2015})}\BibitemShut {NoStop}%
\bibitem [{\citenamefont {Howes}\ \emph {et~al.}(2020)\citenamefont {Howes},
  \citenamefont {Nolen}, \citenamefont {Caldwell},\ and\ \citenamefont
  {Valentine}}]{Howes2020}%
  \BibitemOpen
  \bibfield  {author} {\bibinfo {author} {\bibfnamefont {A.}~\bibnamefont
  {Howes}}, \bibinfo {author} {\bibfnamefont {J.~R.}\ \bibnamefont {Nolen}},
  \bibinfo {author} {\bibfnamefont {J.~D.}\ \bibnamefont {Caldwell}}, \ and\
  \bibinfo {author} {\bibfnamefont {J.}~\bibnamefont {Valentine}},\ }\href
  {\doibase 10.1002/adom.201901470} {\bibfield  {journal} {\bibinfo  {journal}
  {Advanced Optical Materials}\ }\textbf {\bibinfo {volume} {8}},\ \bibinfo
  {pages} {1901470} (\bibinfo {year} {2020})}\BibitemShut {NoStop}%
\bibitem [{\citenamefont {Kumah}\ \emph {et~al.}(2020)\citenamefont {Kumah},
  \citenamefont {Ngai},\ and\ \citenamefont {Kornblum}}]{Kumah2020}%
  \BibitemOpen
  \bibfield  {author} {\bibinfo {author} {\bibfnamefont {D.~P.}\ \bibnamefont
  {Kumah}}, \bibinfo {author} {\bibfnamefont {J.~H.}\ \bibnamefont {Ngai}}, \
  and\ \bibinfo {author} {\bibfnamefont {L.}~\bibnamefont {Kornblum}},\ }\href
  {\doibase https://doi.org/10.1002/adfm.201901597} {\bibfield  {journal}
  {\bibinfo  {journal} {Advanced Functional Materials}\ }\textbf {\bibinfo
  {volume} {30}},\ \bibinfo {pages} {1901597} (\bibinfo {year}
  {2020})}\BibitemShut {NoStop}%
\bibitem [{\citenamefont {Roelkens}\ \emph {et~al.}(2007)\citenamefont
  {Roelkens}, \citenamefont {{Van Campenhout}}, \citenamefont {Brouckaert},
  \citenamefont {{Van Thourhout}}, \citenamefont {Baets}, \citenamefont
  {Romeo}, \citenamefont {Regreny}, \citenamefont {Kazmierczak}, \citenamefont
  {Seassal}, \citenamefont {Letartre}, \citenamefont {Hollinger}, \citenamefont
  {Fedeli}, \citenamefont {{Di Cioccio}},\ and\ \citenamefont
  {Lagahe-Blanchard}}]{Roelkens2007}%
  \BibitemOpen
  \bibfield  {author} {\bibinfo {author} {\bibfnamefont {G.}~\bibnamefont
  {Roelkens}}, \bibinfo {author} {\bibfnamefont {J.}~\bibnamefont {{Van
  Campenhout}}}, \bibinfo {author} {\bibfnamefont {J.}~\bibnamefont
  {Brouckaert}}, \bibinfo {author} {\bibfnamefont {D.}~\bibnamefont {{Van
  Thourhout}}}, \bibinfo {author} {\bibfnamefont {R.}~\bibnamefont {Baets}},
  \bibinfo {author} {\bibfnamefont {P.~R.}\ \bibnamefont {Romeo}}, \bibinfo
  {author} {\bibfnamefont {P.}~\bibnamefont {Regreny}}, \bibinfo {author}
  {\bibfnamefont {A.}~\bibnamefont {Kazmierczak}}, \bibinfo {author}
  {\bibfnamefont {C.}~\bibnamefont {Seassal}}, \bibinfo {author} {\bibfnamefont
  {X.}~\bibnamefont {Letartre}}, \bibinfo {author} {\bibfnamefont
  {G.}~\bibnamefont {Hollinger}}, \bibinfo {author} {\bibfnamefont {J.~M.}\
  \bibnamefont {Fedeli}}, \bibinfo {author} {\bibfnamefont {L.}~\bibnamefont
  {{Di Cioccio}}}, \ and\ \bibinfo {author} {\bibfnamefont {C.}~\bibnamefont
  {Lagahe-Blanchard}},\ }\href {\doibase
  https://doi.org/10.1016/S1369-7021(07)70178-5} {\bibfield  {journal}
  {\bibinfo  {journal} {Materials Today}\ }\textbf {\bibinfo {volume} {10}},\
  \bibinfo {pages} {36} (\bibinfo {year} {2007})}\BibitemShut {NoStop}%
\bibitem [{\citenamefont {Higurashi}\ \emph {et~al.}(2015)\citenamefont
  {Higurashi}, \citenamefont {Okumura}, \citenamefont {Nakasuji},\ and\
  \citenamefont {Suga}}]{Higurashi2015}%
  \BibitemOpen
  \bibfield  {author} {\bibinfo {author} {\bibfnamefont {E.}~\bibnamefont
  {Higurashi}}, \bibinfo {author} {\bibfnamefont {K.}~\bibnamefont {Okumura}},
  \bibinfo {author} {\bibfnamefont {K.}~\bibnamefont {Nakasuji}}, \ and\
  \bibinfo {author} {\bibfnamefont {T.}~\bibnamefont {Suga}},\ }\href {\doibase
  10.7567/jjap.54.030207} {\bibfield  {journal} {\bibinfo  {journal} {Japanese
  Journal of Applied Physics}\ }\textbf {\bibinfo {volume} {54}},\ \bibinfo
  {pages} {30207} (\bibinfo {year} {2015})}\BibitemShut {NoStop}%
\bibitem [{\citenamefont {Baehr-Jones}\ \emph {et~al.}(2010)\citenamefont
  {Baehr-Jones}, \citenamefont {Spott}, \citenamefont {Ilic}, \citenamefont
  {Spott}, \citenamefont {Penkov}, \citenamefont {Asher},\ and\ \citenamefont
  {Hochberg}}]{Baehr-Jones2010}%
  \BibitemOpen
  \bibfield  {author} {\bibinfo {author} {\bibfnamefont {T.}~\bibnamefont
  {Baehr-Jones}}, \bibinfo {author} {\bibfnamefont {A.}~\bibnamefont {Spott}},
  \bibinfo {author} {\bibfnamefont {R.}~\bibnamefont {Ilic}}, \bibinfo {author}
  {\bibfnamefont {A.}~\bibnamefont {Spott}}, \bibinfo {author} {\bibfnamefont
  {B.}~\bibnamefont {Penkov}}, \bibinfo {author} {\bibfnamefont
  {W.}~\bibnamefont {Asher}}, \ and\ \bibinfo {author} {\bibfnamefont
  {M.}~\bibnamefont {Hochberg}},\ }\href {\doibase 10.1364/OE.18.012127}
  {\bibfield  {journal} {\bibinfo  {journal} {Optics Express}\ }\textbf
  {\bibinfo {volume} {18}},\ \bibinfo {pages} {12127} (\bibinfo {year}
  {2010})}\BibitemShut {NoStop}%
\bibitem [{\citenamefont {Beliaev}\ \emph {et~al.}(2021)\citenamefont
  {Beliaev}, \citenamefont {Shkondin}, \citenamefont {Lavrinenko},\ and\
  \citenamefont {Takayama}}]{Beliaev2021}%
  \BibitemOpen
  \bibfield  {author} {\bibinfo {author} {\bibfnamefont {L.~Y.}\ \bibnamefont
  {Beliaev}}, \bibinfo {author} {\bibfnamefont {E.}~\bibnamefont {Shkondin}},
  \bibinfo {author} {\bibfnamefont {A.~V.}\ \bibnamefont {Lavrinenko}}, \ and\
  \bibinfo {author} {\bibfnamefont {O.}~\bibnamefont {Takayama}},\ }\href
  {\doibase 10.1116/6.0000884} {\bibfield  {journal} {\bibinfo  {journal}
  {Journal of Vacuum Science \& Technology A}\ }\textbf {\bibinfo {volume}
  {39}},\ \bibinfo {pages} {43408} (\bibinfo {year} {2021})}\BibitemShut
  {NoStop}%
\bibitem [{\citenamefont {Castilla}\ \emph {et~al.}(2020)\citenamefont
  {Castilla}, \citenamefont {Vangelidis}, \citenamefont {Pusapati},
  \citenamefont {Goldstein}, \citenamefont {Autore}, \citenamefont
  {Slipchenko}, \citenamefont {Rajendran}, \citenamefont {Kim}, \citenamefont
  {Watanabe}, \citenamefont {Taniguchi}, \citenamefont {Mart{\'{i}}n-Moreno},
  \citenamefont {Englund}, \citenamefont {Tielrooij}, \citenamefont
  {Hillenbrand}, \citenamefont {Lidorikis},\ and\ \citenamefont
  {Koppens}}]{Castilla2020}%
  \BibitemOpen
  \bibfield  {author} {\bibinfo {author} {\bibfnamefont {S.}~\bibnamefont
  {Castilla}}, \bibinfo {author} {\bibfnamefont {I.}~\bibnamefont
  {Vangelidis}}, \bibinfo {author} {\bibfnamefont {V.-V.}\ \bibnamefont
  {Pusapati}}, \bibinfo {author} {\bibfnamefont {J.}~\bibnamefont {Goldstein}},
  \bibinfo {author} {\bibfnamefont {M.}~\bibnamefont {Autore}}, \bibinfo
  {author} {\bibfnamefont {T.}~\bibnamefont {Slipchenko}}, \bibinfo {author}
  {\bibfnamefont {K.}~\bibnamefont {Rajendran}}, \bibinfo {author}
  {\bibfnamefont {S.}~\bibnamefont {Kim}}, \bibinfo {author} {\bibfnamefont
  {K.}~\bibnamefont {Watanabe}}, \bibinfo {author} {\bibfnamefont
  {T.}~\bibnamefont {Taniguchi}}, \bibinfo {author} {\bibfnamefont
  {L.}~\bibnamefont {Mart{\'{i}}n-Moreno}}, \bibinfo {author} {\bibfnamefont
  {D.}~\bibnamefont {Englund}}, \bibinfo {author} {\bibfnamefont {K.-J.}\
  \bibnamefont {Tielrooij}}, \bibinfo {author} {\bibfnamefont {R.}~\bibnamefont
  {Hillenbrand}}, \bibinfo {author} {\bibfnamefont {E.}~\bibnamefont
  {Lidorikis}}, \ and\ \bibinfo {author} {\bibfnamefont {F.~H.~L.}\
  \bibnamefont {Koppens}},\ }\href {\doibase 10.1038/s41467-020-18544-z}
  {\bibfield  {journal} {\bibinfo  {journal} {Nature Communications}\ }\textbf
  {\bibinfo {volume} {11}},\ \bibinfo {pages} {4872} (\bibinfo {year}
  {2020})}\BibitemShut {NoStop}%
\bibitem [{\citenamefont {Novoselov}\ \emph {et~al.}(2016)\citenamefont
  {Novoselov}, \citenamefont {Mishchenko}, \citenamefont {Carvalho},\ and\
  \citenamefont {{Castro Neto}}}]{Novoselov2016}%
  \BibitemOpen
  \bibfield  {author} {\bibinfo {author} {\bibfnamefont {K.~S.}\ \bibnamefont
  {Novoselov}}, \bibinfo {author} {\bibfnamefont {A.}~\bibnamefont
  {Mishchenko}}, \bibinfo {author} {\bibfnamefont {A.}~\bibnamefont
  {Carvalho}}, \ and\ \bibinfo {author} {\bibfnamefont {A.~H.}\ \bibnamefont
  {{Castro Neto}}},\ }\href@noop {} {\bibfield  {journal} {\bibinfo  {journal}
  {Science}\ }\textbf {\bibinfo {volume} {353}} (\bibinfo {year}
  {2016})}\BibitemShut {NoStop}%
\bibitem [{\citenamefont {Basov}\ \emph {et~al.}(2016)\citenamefont {Basov},
  \citenamefont {Fogler},\ and\ \citenamefont {{Garcia de Abajo}}}]{Basov2016}%
  \BibitemOpen
  \bibfield  {author} {\bibinfo {author} {\bibfnamefont {D.~N.}\ \bibnamefont
  {Basov}}, \bibinfo {author} {\bibfnamefont {M.~M.}\ \bibnamefont {Fogler}}, \
  and\ \bibinfo {author} {\bibfnamefont {F.~J.}\ \bibnamefont {{Garcia de
  Abajo}}},\ }\href {\doibase 10.1126/science.aag1992} {\bibfield  {journal}
  {\bibinfo  {journal} {Science}\ }\textbf {\bibinfo {volume} {354}},\ \bibinfo
  {pages} {aag1992} (\bibinfo {year} {2016})}\BibitemShut {NoStop}%
\bibitem [{\citenamefont {Song}\ and\ \citenamefont {Gabor}(2018)}]{Song2018}%
  \BibitemOpen
  \bibfield  {author} {\bibinfo {author} {\bibfnamefont {J.~C.~W.}\
  \bibnamefont {Song}}\ and\ \bibinfo {author} {\bibfnamefont {N.~M.}\
  \bibnamefont {Gabor}},\ }\href {\doibase 10.1038/s41565-018-0294-9}
  {\bibfield  {journal} {\bibinfo  {journal} {Nature Nanotechnology}\ }\textbf
  {\bibinfo {volume} {13}},\ \bibinfo {pages} {986} (\bibinfo {year}
  {2018})}\BibitemShut {NoStop}%
\bibitem [{\citenamefont {{Herzig Sheinfux}}\ and\ \citenamefont
  {Koppens}(2020)}]{HerzigSheinfux2020}%
  \BibitemOpen
  \bibfield  {author} {\bibinfo {author} {\bibfnamefont {H.}~\bibnamefont
  {{Herzig Sheinfux}}}\ and\ \bibinfo {author} {\bibfnamefont {F.~H.~L.}\
  \bibnamefont {Koppens}},\ }\href {\doibase 10.1021/acs.nanolett.0c03175}
  {\bibfield  {journal} {\bibinfo  {journal} {Nano Letters}\ }\textbf {\bibinfo
  {volume} {20}},\ \bibinfo {pages} {6935} (\bibinfo {year}
  {2020})}\BibitemShut {NoStop}%
\bibitem [{\citenamefont {Zebo}\ \emph {et~al.}(2018)\citenamefont {Zebo},
  \citenamefont {Jianing}, \citenamefont {Yu}, \citenamefont {Ximiao},
  \citenamefont {Xiaobo}, \citenamefont {Pengyi}, \citenamefont {Jianbin},
  \citenamefont {Weiguang}, \citenamefont {Huanjun}, \citenamefont {Shaozhi},\
  and\ \citenamefont {Ningsheng}}]{Zebo2018}%
  \BibitemOpen
  \bibfield  {author} {\bibinfo {author} {\bibfnamefont {Z.}~\bibnamefont
  {Zebo}}, \bibinfo {author} {\bibfnamefont {C.}~\bibnamefont {Jianing}},
  \bibinfo {author} {\bibfnamefont {W.}~\bibnamefont {Yu}}, \bibinfo {author}
  {\bibfnamefont {W.}~\bibnamefont {Ximiao}}, \bibinfo {author} {\bibfnamefont
  {C.}~\bibnamefont {Xiaobo}}, \bibinfo {author} {\bibfnamefont
  {L.}~\bibnamefont {Pengyi}}, \bibinfo {author} {\bibfnamefont
  {X.}~\bibnamefont {Jianbin}}, \bibinfo {author} {\bibfnamefont
  {X.}~\bibnamefont {Weiguang}}, \bibinfo {author} {\bibfnamefont
  {C.}~\bibnamefont {Huanjun}}, \bibinfo {author} {\bibfnamefont
  {D.}~\bibnamefont {Shaozhi}}, \ and\ \bibinfo {author} {\bibfnamefont
  {X.}~\bibnamefont {Ningsheng}},\ }\href {\doibase doi:10.1002/adma.201705318}
  {\bibfield  {journal} {\bibinfo  {journal} {Advanced Materials}\ }\textbf
  {\bibinfo {volume} {30}},\ \bibinfo {pages} {1705318} (\bibinfo {year}
  {2018})}\BibitemShut {NoStop}%
\bibitem [{\citenamefont {Zheng}\ \emph {et~al.}(2019)\citenamefont {Zheng},
  \citenamefont {Xu}, \citenamefont {Oscurato}, \citenamefont {Tamagnone},
  \citenamefont {Sun}, \citenamefont {Jiang}, \citenamefont {Ke}, \citenamefont
  {Chen}, \citenamefont {Huang}, \citenamefont {Wilson}, \citenamefont
  {Ambrosio}, \citenamefont {Deng},\ and\ \citenamefont {Chen}}]{Zheng2019}%
  \BibitemOpen
  \bibfield  {author} {\bibinfo {author} {\bibfnamefont {Z.}~\bibnamefont
  {Zheng}}, \bibinfo {author} {\bibfnamefont {N.}~\bibnamefont {Xu}}, \bibinfo
  {author} {\bibfnamefont {S.~L.}\ \bibnamefont {Oscurato}}, \bibinfo {author}
  {\bibfnamefont {M.}~\bibnamefont {Tamagnone}}, \bibinfo {author}
  {\bibfnamefont {F.}~\bibnamefont {Sun}}, \bibinfo {author} {\bibfnamefont
  {Y.}~\bibnamefont {Jiang}}, \bibinfo {author} {\bibfnamefont
  {Y.}~\bibnamefont {Ke}}, \bibinfo {author} {\bibfnamefont {J.}~\bibnamefont
  {Chen}}, \bibinfo {author} {\bibfnamefont {W.}~\bibnamefont {Huang}},
  \bibinfo {author} {\bibfnamefont {W.~L.}\ \bibnamefont {Wilson}}, \bibinfo
  {author} {\bibfnamefont {A.}~\bibnamefont {Ambrosio}}, \bibinfo {author}
  {\bibfnamefont {S.}~\bibnamefont {Deng}}, \ and\ \bibinfo {author}
  {\bibfnamefont {H.}~\bibnamefont {Chen}},\ }\href {\doibase
  10.1126/sciadv.aav8690} {\bibfield  {journal} {\bibinfo  {journal} {Science
  Advances}\ }\textbf {\bibinfo {volume} {5}},\ \bibinfo {pages} {eaav8690}
  (\bibinfo {year} {2019})}\BibitemShut {NoStop}%
\bibitem [{\citenamefont {Autore}\ \emph {et~al.}(2021)\citenamefont {Autore},
  \citenamefont {Dolado}, \citenamefont {Li}, \citenamefont {Esteban},
  \citenamefont {Alfaro-Mozaz}, \citenamefont {Atxabal}, \citenamefont {Liu},
  \citenamefont {Edgar}, \citenamefont {V{\'{e}}lez}, \citenamefont {Casanova},
  \citenamefont {Hueso}, \citenamefont {Aizpurua},\ and\ \citenamefont
  {Hillenbrand}}]{Autore2021}%
  \BibitemOpen
  \bibfield  {author} {\bibinfo {author} {\bibfnamefont {M.}~\bibnamefont
  {Autore}}, \bibinfo {author} {\bibfnamefont {I.}~\bibnamefont {Dolado}},
  \bibinfo {author} {\bibfnamefont {P.}~\bibnamefont {Li}}, \bibinfo {author}
  {\bibfnamefont {R.}~\bibnamefont {Esteban}}, \bibinfo {author} {\bibfnamefont
  {F.~J.}\ \bibnamefont {Alfaro-Mozaz}}, \bibinfo {author} {\bibfnamefont
  {A.}~\bibnamefont {Atxabal}}, \bibinfo {author} {\bibfnamefont
  {S.}~\bibnamefont {Liu}}, \bibinfo {author} {\bibfnamefont {J.~H.}\
  \bibnamefont {Edgar}}, \bibinfo {author} {\bibfnamefont {S.}~\bibnamefont
  {V{\'{e}}lez}}, \bibinfo {author} {\bibfnamefont {F.}~\bibnamefont
  {Casanova}}, \bibinfo {author} {\bibfnamefont {L.~E.}\ \bibnamefont {Hueso}},
  \bibinfo {author} {\bibfnamefont {J.}~\bibnamefont {Aizpurua}}, \ and\
  \bibinfo {author} {\bibfnamefont {R.}~\bibnamefont {Hillenbrand}},\ }\href
  {\doibase https://doi.org/10.1002/adom.202001958} {\bibfield  {journal}
  {\bibinfo  {journal} {Advanced Optical Materials}\ }\textbf {\bibinfo
  {volume} {9}},\ \bibinfo {pages} {2001958} (\bibinfo {year}
  {2021})}\BibitemShut {NoStop}%
\bibitem [{\citenamefont {Lee}\ \emph {et~al.}(2020)\citenamefont {Lee},
  \citenamefont {He}, \citenamefont {Zhang}, \citenamefont {Luo}, \citenamefont
  {Liu}, \citenamefont {Edgar}, \citenamefont {Wang}, \citenamefont {Avouris},
  \citenamefont {Low}, \citenamefont {Caldwell},\ and\ \citenamefont
  {Oh}}]{Lee2020}%
  \BibitemOpen
  \bibfield  {author} {\bibinfo {author} {\bibfnamefont {I.-H.}\ \bibnamefont
  {Lee}}, \bibinfo {author} {\bibfnamefont {M.}~\bibnamefont {He}}, \bibinfo
  {author} {\bibfnamefont {X.}~\bibnamefont {Zhang}}, \bibinfo {author}
  {\bibfnamefont {Y.}~\bibnamefont {Luo}}, \bibinfo {author} {\bibfnamefont
  {S.}~\bibnamefont {Liu}}, \bibinfo {author} {\bibfnamefont {J.~H.}\
  \bibnamefont {Edgar}}, \bibinfo {author} {\bibfnamefont {K.}~\bibnamefont
  {Wang}}, \bibinfo {author} {\bibfnamefont {P.}~\bibnamefont {Avouris}},
  \bibinfo {author} {\bibfnamefont {T.}~\bibnamefont {Low}}, \bibinfo {author}
  {\bibfnamefont {J.~D.}\ \bibnamefont {Caldwell}}, \ and\ \bibinfo {author}
  {\bibfnamefont {S.-H.}\ \bibnamefont {Oh}},\ }\href {\doibase
  10.1038/s41467-020-17424-w} {\bibfield  {journal} {\bibinfo  {journal}
  {Nature Communications}\ }\textbf {\bibinfo {volume} {11}},\ \bibinfo {pages}
  {3649} (\bibinfo {year} {2020})}\BibitemShut {NoStop}%
\bibitem [{\citenamefont {Sun}\ and\ \citenamefont {Chang}(2014)}]{Sun2014}%
  \BibitemOpen
  \bibfield  {author} {\bibinfo {author} {\bibfnamefont {Z.}~\bibnamefont
  {Sun}}\ and\ \bibinfo {author} {\bibfnamefont {H.}~\bibnamefont {Chang}},\
  }\href {\doibase 10.1021/nn500508c} {\bibfield  {journal} {\bibinfo
  {journal} {ACS Nano}\ }\textbf {\bibinfo {volume} {8}},\ \bibinfo {pages}
  {4133} (\bibinfo {year} {2014})}\BibitemShut {NoStop}%
\bibitem [{\citenamefont {De~Liberato}(2015)}]{DeLiberato2015}%
  \BibitemOpen
  \bibfield  {author} {\bibinfo {author} {\bibfnamefont {S.}~\bibnamefont
  {De~Liberato}},\ }\href {\doibase 10.1103/PhysRevB.92.125433} {\bibfield
  {journal} {\bibinfo  {journal} {Phys. Rev. B}\ }\textbf {\bibinfo {volume}
  {92}},\ \bibinfo {pages} {125433} (\bibinfo {year} {2015})}\BibitemShut
  {NoStop}%
\bibitem [{\citenamefont {Ju}\ \emph {et~al.}(2017)\citenamefont {Ju},
  \citenamefont {Wang}, \citenamefont {Cao}, \citenamefont {Taniguchi},
  \citenamefont {Watanabe}, \citenamefont {Louie}, \citenamefont {Rana},
  \citenamefont {Park}, \citenamefont {Hone}, \citenamefont {Wang},\ and\
  \citenamefont {McEuen}}]{Ju2017}%
  \BibitemOpen
  \bibfield  {author} {\bibinfo {author} {\bibfnamefont {L.}~\bibnamefont
  {Ju}}, \bibinfo {author} {\bibfnamefont {L.}~\bibnamefont {Wang}}, \bibinfo
  {author} {\bibfnamefont {T.}~\bibnamefont {Cao}}, \bibinfo {author}
  {\bibfnamefont {T.}~\bibnamefont {Taniguchi}}, \bibinfo {author}
  {\bibfnamefont {K.}~\bibnamefont {Watanabe}}, \bibinfo {author}
  {\bibfnamefont {S.~G.}\ \bibnamefont {Louie}}, \bibinfo {author}
  {\bibfnamefont {F.}~\bibnamefont {Rana}}, \bibinfo {author} {\bibfnamefont
  {J.}~\bibnamefont {Park}}, \bibinfo {author} {\bibfnamefont {J.}~\bibnamefont
  {Hone}}, \bibinfo {author} {\bibfnamefont {F.}~\bibnamefont {Wang}}, \ and\
  \bibinfo {author} {\bibfnamefont {P.~L.}\ \bibnamefont {McEuen}},\ }\href
  {\doibase 10.1126/science.aam9175} {\bibfield  {journal} {\bibinfo  {journal}
  {Science}\ }\textbf {\bibinfo {volume} {358}},\ \bibinfo {pages} {907}
  (\bibinfo {year} {2017})}\BibitemShut {NoStop}%
\bibitem [{\citenamefont {Bandurin}\ \emph {et~al.}(2018)\citenamefont
  {Bandurin}, \citenamefont {Svintsov}, \citenamefont {Gayduchenko},
  \citenamefont {Xu}, \citenamefont {Principi}, \citenamefont {Moskotin},
  \citenamefont {Tretyakov}, \citenamefont {Yagodkin}, \citenamefont {Zhukov},
  \citenamefont {Taniguchi}, \citenamefont {Watanabe}, \citenamefont
  {Grigorieva}, \citenamefont {Polini}, \citenamefont {Goltsman}, \citenamefont
  {Geim},\ and\ \citenamefont {Fedorov}}]{Bandurin2018}%
  \BibitemOpen
  \bibfield  {author} {\bibinfo {author} {\bibfnamefont {D.~A.}\ \bibnamefont
  {Bandurin}}, \bibinfo {author} {\bibfnamefont {D.}~\bibnamefont {Svintsov}},
  \bibinfo {author} {\bibfnamefont {I.}~\bibnamefont {Gayduchenko}}, \bibinfo
  {author} {\bibfnamefont {S.~G.}\ \bibnamefont {Xu}}, \bibinfo {author}
  {\bibfnamefont {A.}~\bibnamefont {Principi}}, \bibinfo {author}
  {\bibfnamefont {M.}~\bibnamefont {Moskotin}}, \bibinfo {author}
  {\bibfnamefont {I.}~\bibnamefont {Tretyakov}}, \bibinfo {author}
  {\bibfnamefont {D.}~\bibnamefont {Yagodkin}}, \bibinfo {author}
  {\bibfnamefont {S.}~\bibnamefont {Zhukov}}, \bibinfo {author} {\bibfnamefont
  {T.}~\bibnamefont {Taniguchi}}, \bibinfo {author} {\bibfnamefont
  {K.}~\bibnamefont {Watanabe}}, \bibinfo {author} {\bibfnamefont {I.~V.}\
  \bibnamefont {Grigorieva}}, \bibinfo {author} {\bibfnamefont
  {M.}~\bibnamefont {Polini}}, \bibinfo {author} {\bibfnamefont {G.~N.}\
  \bibnamefont {Goltsman}}, \bibinfo {author} {\bibfnamefont {A.~K.}\
  \bibnamefont {Geim}}, \ and\ \bibinfo {author} {\bibfnamefont
  {G.}~\bibnamefont {Fedorov}},\ }\href {\doibase 10.1038/s41467-018-07848-w}
  {\bibfield  {journal} {\bibinfo  {journal} {Nature Communications}\ }\textbf
  {\bibinfo {volume} {9}},\ \bibinfo {pages} {5392} (\bibinfo {year}
  {2018})}\BibitemShut {NoStop}%
\bibitem [{\citenamefont {Sunku}\ \emph {et~al.}(2021)\citenamefont {Sunku},
  \citenamefont {Halbertal}, \citenamefont {Stauber}, \citenamefont {Chen},
  \citenamefont {McLeod}, \citenamefont {Rikhter}, \citenamefont {Berkowitz},
  \citenamefont {Lo}, \citenamefont {Gonzalez-Acevedo}, \citenamefont {Hone},
  \citenamefont {Dean}, \citenamefont {Fogler},\ and\ \citenamefont
  {Basov}}]{Sunku2021}%
  \BibitemOpen
  \bibfield  {author} {\bibinfo {author} {\bibfnamefont {S.~S.}\ \bibnamefont
  {Sunku}}, \bibinfo {author} {\bibfnamefont {D.}~\bibnamefont {Halbertal}},
  \bibinfo {author} {\bibfnamefont {T.}~\bibnamefont {Stauber}}, \bibinfo
  {author} {\bibfnamefont {S.}~\bibnamefont {Chen}}, \bibinfo {author}
  {\bibfnamefont {A.~S.}\ \bibnamefont {McLeod}}, \bibinfo {author}
  {\bibfnamefont {A.}~\bibnamefont {Rikhter}}, \bibinfo {author} {\bibfnamefont
  {M.~E.}\ \bibnamefont {Berkowitz}}, \bibinfo {author} {\bibfnamefont
  {C.~F.~B.}\ \bibnamefont {Lo}}, \bibinfo {author} {\bibfnamefont {D.~E.}\
  \bibnamefont {Gonzalez-Acevedo}}, \bibinfo {author} {\bibfnamefont {J.~C.}\
  \bibnamefont {Hone}}, \bibinfo {author} {\bibfnamefont {C.~R.}\ \bibnamefont
  {Dean}}, \bibinfo {author} {\bibfnamefont {M.~M.}\ \bibnamefont {Fogler}}, \
  and\ \bibinfo {author} {\bibfnamefont {D.~N.}\ \bibnamefont {Basov}},\ }\href
  {\doibase 10.1038/s41467-021-21792-2} {\bibfield  {journal} {\bibinfo
  {journal} {Nature Communications}\ }\textbf {\bibinfo {volume} {12}},\
  \bibinfo {pages} {1641} (\bibinfo {year} {2021})}\BibitemShut {NoStop}%
\bibitem [{\citenamefont {Hesp}\ \emph {et~al.}(2021)\citenamefont {Hesp},
  \citenamefont {Torre}, \citenamefont {Barcons-Ruiz}, \citenamefont {{Herzig
  Sheinfux}}, \citenamefont {Watanabe}, \citenamefont {Taniguchi},
  \citenamefont {{Krishna Kumar}},\ and\ \citenamefont {Koppens}}]{Hesp2021}%
  \BibitemOpen
  \bibfield  {author} {\bibinfo {author} {\bibfnamefont {N.~C.~H.}\
  \bibnamefont {Hesp}}, \bibinfo {author} {\bibfnamefont {I.}~\bibnamefont
  {Torre}}, \bibinfo {author} {\bibfnamefont {D.}~\bibnamefont {Barcons-Ruiz}},
  \bibinfo {author} {\bibfnamefont {H.}~\bibnamefont {{Herzig Sheinfux}}},
  \bibinfo {author} {\bibfnamefont {K.}~\bibnamefont {Watanabe}}, \bibinfo
  {author} {\bibfnamefont {T.}~\bibnamefont {Taniguchi}}, \bibinfo {author}
  {\bibfnamefont {R.}~\bibnamefont {{Krishna Kumar}}}, \ and\ \bibinfo {author}
  {\bibfnamefont {F.~H.~L.}\ \bibnamefont {Koppens}},\ }\href {\doibase
  10.1038/s41467-021-21862-5} {\bibfield  {journal} {\bibinfo  {journal}
  {Nature Communications}\ }\textbf {\bibinfo {volume} {12}},\ \bibinfo {pages}
  {1640} (\bibinfo {year} {2021})}\BibitemShut {NoStop}%
\bibitem [{\citenamefont {Rogalski}\ \emph {et~al.}(2019)\citenamefont
  {Rogalski}, \citenamefont {Kopytko},\ and\ \citenamefont
  {Martyniuk}}]{Rogalski2019}%
  \BibitemOpen
  \bibfield  {author} {\bibinfo {author} {\bibfnamefont {A.}~\bibnamefont
  {Rogalski}}, \bibinfo {author} {\bibfnamefont {M.}~\bibnamefont {Kopytko}}, \
  and\ \bibinfo {author} {\bibfnamefont {P.}~\bibnamefont {Martyniuk}},\ }\href
  {\doibase 10.1063/1.5088578} {\bibfield  {journal} {\bibinfo  {journal}
  {Applied Physics Reviews}\ }\textbf {\bibinfo {volume} {6}},\ \bibinfo
  {pages} {21316} (\bibinfo {year} {2019})}\BibitemShut {NoStop}%
\bibitem [{\citenamefont {Yu}\ \emph {et~al.}(2018)\citenamefont {Yu},
  \citenamefont {Yu}, \citenamefont {Wu}, \citenamefont {Singh}, \citenamefont
  {Zeng}, \citenamefont {Lin}, \citenamefont {Zhou}, \citenamefont {Lin},
  \citenamefont {Suenaga}, \citenamefont {Liu},\ and\ \citenamefont
  {Wang}}]{Yu2018}%
  \BibitemOpen
  \bibfield  {author} {\bibinfo {author} {\bibfnamefont {X.}~\bibnamefont
  {Yu}}, \bibinfo {author} {\bibfnamefont {P.}~\bibnamefont {Yu}}, \bibinfo
  {author} {\bibfnamefont {D.}~\bibnamefont {Wu}}, \bibinfo {author}
  {\bibfnamefont {B.}~\bibnamefont {Singh}}, \bibinfo {author} {\bibfnamefont
  {Q.}~\bibnamefont {Zeng}}, \bibinfo {author} {\bibfnamefont {H.}~\bibnamefont
  {Lin}}, \bibinfo {author} {\bibfnamefont {W.}~\bibnamefont {Zhou}}, \bibinfo
  {author} {\bibfnamefont {J.}~\bibnamefont {Lin}}, \bibinfo {author}
  {\bibfnamefont {K.}~\bibnamefont {Suenaga}}, \bibinfo {author} {\bibfnamefont
  {Z.}~\bibnamefont {Liu}}, \ and\ \bibinfo {author} {\bibfnamefont {Q.~J.}\
  \bibnamefont {Wang}},\ }\href {\doibase 10.1038/s41467-018-03935-0}
  {\bibfield  {journal} {\bibinfo  {journal} {Nature Communications}\ }\textbf
  {\bibinfo {volume} {9}},\ \bibinfo {pages} {1545} (\bibinfo {year}
  {2018})}\BibitemShut {NoStop}%
\bibitem [{\citenamefont {Chakraborty}\ \emph {et~al.}(2016)\citenamefont
  {Chakraborty}, \citenamefont {Marshall}, \citenamefont {Folland},
  \citenamefont {Kim}, \citenamefont {Grigorenko},\ and\ \citenamefont
  {Novoselov}}]{Chakraborty2016}%
  \BibitemOpen
  \bibfield  {author} {\bibinfo {author} {\bibfnamefont {S.}~\bibnamefont
  {Chakraborty}}, \bibinfo {author} {\bibfnamefont {O.~P.}\ \bibnamefont
  {Marshall}}, \bibinfo {author} {\bibfnamefont {T.~G.}\ \bibnamefont
  {Folland}}, \bibinfo {author} {\bibfnamefont {Y.-J.}\ \bibnamefont {Kim}},
  \bibinfo {author} {\bibfnamefont {A.~N.}\ \bibnamefont {Grigorenko}}, \ and\
  \bibinfo {author} {\bibfnamefont {K.~S.}\ \bibnamefont {Novoselov}},\ }\href
  {\doibase 10.1126/science.aad2930} {\bibfield  {journal} {\bibinfo  {journal}
  {Science (New York, N.Y.)}\ }\textbf {\bibinfo {volume} {351}},\ \bibinfo
  {pages} {246} (\bibinfo {year} {2016})}\BibitemShut {NoStop}%
\bibitem [{\citenamefont {Mey}\ \emph {et~al.}(2019)\citenamefont {Mey},
  \citenamefont {Wall}, \citenamefont {Schneider}, \citenamefont
  {G{\"{u}}nder}, \citenamefont {Walla}, \citenamefont {Soltani}, \citenamefont
  {Roskos}, \citenamefont {Yao}, \citenamefont {Qing}, \citenamefont {Fang},\
  and\ \citenamefont {Rahimi-Iman}}]{Mey2019}%
  \BibitemOpen
  \bibfield  {author} {\bibinfo {author} {\bibfnamefont {O.}~\bibnamefont
  {Mey}}, \bibinfo {author} {\bibfnamefont {F.}~\bibnamefont {Wall}}, \bibinfo
  {author} {\bibfnamefont {L.~M.}\ \bibnamefont {Schneider}}, \bibinfo {author}
  {\bibfnamefont {D.}~\bibnamefont {G{\"{u}}nder}}, \bibinfo {author}
  {\bibfnamefont {F.}~\bibnamefont {Walla}}, \bibinfo {author} {\bibfnamefont
  {A.}~\bibnamefont {Soltani}}, \bibinfo {author} {\bibfnamefont
  {H.}~\bibnamefont {Roskos}}, \bibinfo {author} {\bibfnamefont
  {N.}~\bibnamefont {Yao}}, \bibinfo {author} {\bibfnamefont {P.}~\bibnamefont
  {Qing}}, \bibinfo {author} {\bibfnamefont {W.}~\bibnamefont {Fang}}, \ and\
  \bibinfo {author} {\bibfnamefont {A.}~\bibnamefont {Rahimi-Iman}},\ }\href
  {\doibase 10.1021/acsnano.8b09659} {\bibfield  {journal} {\bibinfo  {journal}
  {ACS Nano}\ }\textbf {\bibinfo {volume} {13}},\ \bibinfo {pages} {5259}
  (\bibinfo {year} {2019})}\BibitemShut {NoStop}%
\bibitem [{\citenamefont {Fan}\ \emph {et~al.}(2020{\natexlab{a}})\citenamefont
  {Fan}, \citenamefont {Vu}, \citenamefont {Tran}, \citenamefont {Adhikari},\
  and\ \citenamefont {Lee}}]{Fan2020}%
  \BibitemOpen
  \bibfield  {author} {\bibinfo {author} {\bibfnamefont {S.}~\bibnamefont
  {Fan}}, \bibinfo {author} {\bibfnamefont {Q.~A.}\ \bibnamefont {Vu}},
  \bibinfo {author} {\bibfnamefont {M.~D.}\ \bibnamefont {Tran}}, \bibinfo
  {author} {\bibfnamefont {S.}~\bibnamefont {Adhikari}}, \ and\ \bibinfo
  {author} {\bibfnamefont {Y.~H.}\ \bibnamefont {Lee}},\ }\href {\doibase
  10.1088/2053-1583/ab7629} {\bibfield  {journal} {\bibinfo  {journal} {2D
  Materials}\ }\textbf {\bibinfo {volume} {7}},\ \bibinfo {pages} {22005}
  (\bibinfo {year} {2020}{\natexlab{a}})}\BibitemShut {NoStop}%
\bibitem [{\citenamefont {Fan}\ \emph {et~al.}(2020{\natexlab{b}})\citenamefont
  {Fan}, \citenamefont {Vu}, \citenamefont {Tran}, \citenamefont {Adhikari},\
  and\ \citenamefont {Lee}}]{Fan2020b}%
  \BibitemOpen
  \bibfield  {author} {\bibinfo {author} {\bibfnamefont {S.}~\bibnamefont
  {Fan}}, \bibinfo {author} {\bibfnamefont {Q.~A.}\ \bibnamefont {Vu}},
  \bibinfo {author} {\bibfnamefont {M.~D.}\ \bibnamefont {Tran}}, \bibinfo
  {author} {\bibfnamefont {S.}~\bibnamefont {Adhikari}}, \ and\ \bibinfo
  {author} {\bibfnamefont {Y.~H.}\ \bibnamefont {Lee}},\ }\href {\doibase
  10.1088/2053-1583/ab7629} {\bibfield  {journal} {\bibinfo  {journal} {2D
  Materials}\ }\textbf {\bibinfo {volume} {7}},\ \bibinfo {pages} {22005}
  (\bibinfo {year} {2020}{\natexlab{b}})}\BibitemShut {NoStop}%
\bibitem [{\citenamefont {Lu}\ \emph {et~al.}(2020)\citenamefont {Lu},
  \citenamefont {Nolen}, \citenamefont {Folland}, \citenamefont {Tadjer},
  \citenamefont {Walker},\ and\ \citenamefont {Caldwell}}]{Lu2020}%
  \BibitemOpen
  \bibfield  {author} {\bibinfo {author} {\bibfnamefont {G.}~\bibnamefont
  {Lu}}, \bibinfo {author} {\bibfnamefont {J.~R.}\ \bibnamefont {Nolen}},
  \bibinfo {author} {\bibfnamefont {T.~G.}\ \bibnamefont {Folland}}, \bibinfo
  {author} {\bibfnamefont {M.~J.}\ \bibnamefont {Tadjer}}, \bibinfo {author}
  {\bibfnamefont {D.~G.}\ \bibnamefont {Walker}}, \ and\ \bibinfo {author}
  {\bibfnamefont {J.~D.}\ \bibnamefont {Caldwell}},\ }\href {\doibase
  10.1021/acsomega.0c00600} {\bibfield  {journal} {\bibinfo  {journal} {ACS
  Omega}\ }\textbf {\bibinfo {volume} {5}},\ \bibinfo {pages} {10900} (\bibinfo
  {year} {2020})}\BibitemShut {NoStop}%
\end{thebibliography}%

\end{document}